\documentclass{article}

\usepackage{arxiv}
\usepackage[T1]{fontenc}    % use 8-bit T1 fonts
\usepackage{amsmath, amssymb}
\usepackage{amsthm}
\usepackage{nicefrac}       % compact symbols for 1/2, etc.

\usepackage{graphicx}
\usepackage{booktabs}       % professional-quality tables
\usepackage{threeparttable}
\usepackage{pdflscape}
\usepackage{subcaption}
\usepackage{longtable}
\usepackage{tabularx}
\usepackage{multirow}       % セルを縦に結合するために使用
\usepackage{array}
\renewcommand{\arraystretch}{1.3}
\usepackage[table]{xcolor}
\usepackage{enumitem}

\usepackage[numbers]{natbib}
\usepackage{doi}

\usepackage{microtype}      % microtypography

\usepackage{hyperref}

\usepackage{epigraph}                       % 引用から始めるときに使用
\newtheoremstyle{spaceddef} % スタイルの名前
  {10pt plus 4pt minus 4pt} % ブロック上部の余白（柔軟に伸縮する設定）
  {10pt plus 4pt minus 4pt} % ブロック下部の余白（柔軟に伸縮する設定）
  {}                        % 本文のフォント（空白＝斜体にしない標準フォント）
  {}                        % 全体のインデント（空白＝インデントなし）
  {\bfseries}               % 見出しのフォント（太字）
  {.}                       % 見出しの後の句読点（ピリオド）
  {5pt}                     % 見出しと本文の間のスペース
  {}                        % 見出しのカスタムフォーマット（空白＝標準）
\theoremstyle{spaceddef}
\newtheorem{definition}{Definition}

\newtheorem{hypothesis}{Hypothesis}
\newtheorem{result}{Result}

\usepackage{url}            % simple URL typesetting
\usepackage{lipsum}         % Can be removed after putting your text content

\title{Do Matching Mechanisms Work with LLM Agents?}

\author{{Yukihiro Hoshino}\thanks{\texttt{y.hoshino@css.t.u-tokyo.ac.jp}}\\
    School of Engineering\\
	The University of Tokyo\\\\
	\And
    \href{https://orcid.org/0000-0002-4774-1506}{\includegraphics[scale=0.06]{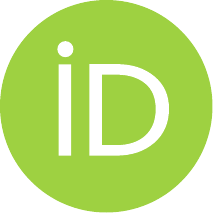}\hspace{1 mm}Ayato Kitadai} \\
    School of Engineering\\
	The University of Tokyo\\\\
    \And
	\href{https://orcid.org/0000-0002-6411-8716}{\includegraphics[scale=0.06]{orcid.pdf}\hspace{1 mm}Nariaki Nishino} \\
    School of Engineering\\
	The University of Tokyo\\
}

\renewcommand{\shorttitle}{Do Matching Mechanisms Work with LLM Agents?}

\hypersetup{
  pdftitle={Do Matching Mechanisms Work with LLM Agents?},
  pdfauthor={Yukihiro Hoshino, Ayato Kitadai, Nariaki Nishino},
  pdfsubject={Matching mechanisms and LLM agents},
  pdfkeywords={Large language models, LLM agents, matching theory, market design, mechanism design, strategy-proofness}
}

\begin{document}
\maketitle

\begin{abstract}
This study examines whether standard matching mechanisms function as intended in LLM-agent markets, where LLM agents make allocation-related decisions as delegated decision-makers.
We compare decentralized free-negotiation markets with centralized mechanism-based markets including several representative mechanisms.
Across controlled one-to-one matching environments, mechanism-based markets generally outperform free negotiation in terms of stability and efficiency.
We also find that LLM agents report preferences truthfully at substantially higher rates than human subjects in comparable DA and EADA environments.
However, truth-telling is not uniformly aligned with formal strategy-proofness across all mechanisms: TTC, despite being strategy-proof, does not always elicit higher truth-telling than EADA.
These results suggest that matching theory provides a useful but incomplete guide for designing institutions in LLM-agent markets.
\end{abstract}

% keywords can be removed

\section{Introduction}\label{sec:int}
\epigraph{%
``We will say that an agency relationship has arisen between two (or more) parties when one, designated as the agent, acts for, on behalf of, or as representative for the other, designated the principal, in a particular domain of decision problems.''%
}{Stephen A. Ross (1973)}%

%%%%% Meanning of agent  and its changes %%%%%
The word ``agent'' traces its etymology back to the Latin agere (to act), and in the contexts of law and economics, it has implied a proxy who carries out actions on behalf of another~\cite{OED-2024}.
In economics, law, and computer science, an agent is commonly understood as an entity that acts on behalf of another party.
Agency theory formalizes this relationship as one in which an agent makes decisions for a principal under informational and incentive constraints~\cite{Jensen-1976,Ross-1973}.
In computer science, the term has developed in a different technical tradition, but it retains the same core idea: an autonomous system that performs tasks on behalf of a user~\cite{Jennings-1996,Nwana-1996,Maes-1994,Wooldridge-1995}.
On the other hand, the concept of agents in engineering and computer science has undergone its own unique development, but it remains consistent with the original meaning of a proxy in that it autonomously executes tasks on behalf of a user~\cite{Jennings-1996,Nwana-1996}.
Even in the early definitions, an agent refers to a computer system that possesses autonomy and acts to achieve goals on behalf of its human owner~\cite{Maes-1994, Wooldridge-1995}.

Multi-agent simulation (MAS) has long been used to study how aggregate social outcomes emerge from interactions among many decision-making units~\cite{Axelrod-1997,Epstein-1996,Nagel-1992,Schelling-1971}.
However, traditional agents are typically governed by researcher-specified rules.
This has limited their ability to represent flexible communication, context-dependent reasoning, and strategic adaptation in institutionally rich environments~\cite{An-2012}.
These limitations are particularly consequential for market design, where the performance of an institution depends not only on the formal allocation rule but also on how participants understand, communicate, and strategically respond to that rule.

The recent emergence of large language models (LLMs) has substantially changed what agent-based simulations can represent~\cite{Gao-2024}.
LLMs, pre-trained on vast amounts of text data, acquired the ability to generate human-like reasoning and responses according to context~\cite{Aher-2023,Andreas-2022,Xi-2023}.
\citet{Park-2023,Park-2024} demonstrated that ``Generative Agents'' equipped with LLMs autonomously form memories, make plans, and cooperate with others in natural language within a sandbox environment.
These studies suggest that LLM-driven agents can generate forms of planning, memory use, and social interaction that are difficult to specify in advance using conventional rule-based agents.
Consequently, agent simulations are now entering a new phase, such as targeting larger-scale cities~\cite{Bougie-2025,Feng-2025}.

%%%%% LLM act as an proxy of an human %%%%%
The introduction of LLMs is not limited to reproducing macro-social phenomena; it is rapidly expanding its scope to approximate the micro-decision-making processes of individuals, and even to utilize them as human substitutes.
In recent years, a growing literature on ``silicon samples'' has examined whether LLMs can serve as synthetic respondents in surveys, behavioral experiments, and studies of opinion formation~\cite{Argyle-2023,Bisbee-2024,Cao-2025,Dominguezolmedo-2024,Horton-2023,Kapania-2024,Li-2024,Sinacola-2025,Sun-2024}.
These suggest the possibility that LLMs can act as virtual respondents with diverse demographic attributes, replacing or complementing surveys and experiments in the social sciences.

In order to treat an LLM agent as a subject that substitutes for an individual, it is necessary to sufficiently reflect the preferences, beliefs, and cognitive characteristics of that individual, rather than just average human likeness.
In this regard, verifications regarding the faithful reproduction of human behavior through persona assignment and the input of questionnaire response histories have progressed.
Specifically, research verifies to what extent social biases in social science experiments~\cite{Kolluri-2025,Lutz-2025}, cognitive biases and cognitive characteristics in psychology experiments~\cite{Kwok-2024,Toubia-2025}, and bounded rationality and heterogeneity in economic experiments can be imitated~\cite{Chen-2023-b,Phelps-2024,Toubia-2025}, examining the substitutability of humans.
In particular, regarding economic preference parameters such as risk preference and time preference, it has been shown that LLMs can capture structures equivalent to humans~\cite{Goli-2024,Kim-2024,Liu-2025-b}.

In addition to reproducing individual characteristics and preferences, the improvement of strategic decision-making abilities in interactions with others is also remarkable.
Recent studies have demonstrated that LLM agents can engage in strategic learning for cooperation, betrayal, or self-interest maximization in situations such as negotiations, auctions, and repeated games~\cite{Akata-2025,Chen-2024,Chiu-2025,Fu-2023,Guo-2024,Jia-2024}.
These studies show that agents are acquiring not only single-task processing abilities but also strategic abilities in social interactions.

Taken together, this literature suggests a shift in the role of LLM agents: they are no longer studied only as simulation tools, but increasingly as potential delegates that may act on behalf of individuals in decision-making environments.
As \citet{Meng-2024} points out, the condition that LLM faithfully reflects human behavior and is consistent with individual preferences is a prerequisite for humans to delegate their important decision-making to LLM.
In other words, current technological progress can be said to be enabling a return to the original meaning of an agent: a ``subject who carries out actions on behalf of another.''
In fact, while research is progressing on situations where humans delegate financial decision-making to LLM agents~\cite{Candrian-2022, Dvorak-2025}, research is also advancing on systems where LLM agents autonomously negotiate and trade as human proxies in specific economic activities, such as constructing stock portfolios, price negotiations in e-commerce, and condition negotiations in supply chains~\cite{Allouah-2025,Deng-2024,Kirshner-2025,Saha-2025,Xia-2024,Yu-2024,Zhu-2025}.
These trends point to the emergence of what we call an ``LLM-agent market'': an environment in which delegated LLM agents interact with one another, negotiate on behalf of users, and determine allocations or terms of exchange.

%%%%% Market design for LLM agent market %%%%%%
The rise of LLM-agent markets raises a question that is not addressed by the MAS literature alone.
Even if agents can communicate and negotiate in natural language, decentralized interaction need not generate stable or efficient allocations.
For many allocation problems, human societies have therefore relied not only on bilateral negotiation but also on explicitly designed market mechanisms.
This is the domain of mechanism design and market design, which study how rules can be constructed so that self-interested participants generate desirable social outcomes.

Mechanism design and market design provide a formal framework for studying how allocation rules shape outcomes when participants pursue their own objectives~\cite{Hurwicz-1973,Maskin-1999,Myerson-1981,Roth-2002}.
In their canonical form, these theories analyze strategic behavior under explicitly specified preferences, information structures, and rules, and characterize mechanisms that satisfy properties such as incentive compatibility, efficiency, and stability.
Its effectiveness has been verified in diverse fields, such as frequency auctions aimed at the efficient allocation of public resources~\cite{McAfee-1986,Milgrom-2000}, public goods provision mechanisms aiming to solve the free-rider problem~\cite{Clarke-1971,Groves-1973}, and furthermore, the design of matching markets like the assignment of medical residents or school choice~\cite{Abdulkadiroglu-2003,Roth-1984}.

Matching theory is a central area of market design, and its development has been closely connected to experimental and empirical work on how allocation mechanisms perform in practice.
Starting with the Deferred Acceptance (DA) mechanism~\cite{Gale-1962}, matching theory has clarified fundamental trade-offs among stability, efficiency, and strategy-proofness~\cite{Abdulkadiroglu-2020,Abdulkadiroglu-2003,Ergin-2006,Gale-1962,Morrill-2014,Roth-2002}.
At the same time, laboratory experiments have shown that theoretical incentive properties do not automatically translate into truthful or optimal behavior by human participants~\cite{Calsamiglia-2010,Chen-2006,Hakimov-2018,Pais-2008,Smith-1976}.
Even when truthful reporting is a weakly dominant strategy, participants may fail to recognize it, distrust the mechanism, or respond to the cognitive complexity of the rule~\cite{Ashlagi-2018,Guillen-2018,Guillen-2019,Li-2017}.
This has motivated the study of mechanisms that may perform well behaviorally even when they relax some canonical theoretical properties.
For example, Efficiency-Adjusted Deferred Acceptance (EADA) improves efficiency relative to DA and has been shown in recent experiments to elicit high levels of truthful reporting despite not satisfying standard strategy-proofness~\cite{Cerrone-2024,Kesten-2010,Tang-2014}.

This raises a new question for LLM-agent markets.
Existing matching mechanisms were developed in formal models of strategic behavior and have been empirically evaluated with human participants.
LLM agents fit neither category perfectly: they may follow formal instructions more consistently than humans, but their behavior is generated through probabilistic language models and may depend on prompts, context, and model-specific reasoning patterns.
Thus, the relevant issue is not simply whether LLM agents are more or less rational than humans, but whether the theoretical properties of matching mechanisms are behaviorally manifested when decisions are generated by LLM agents.
At the present moment, when social interactions in which LLM agents are delegated to make decisions are becoming common, determining what kind of institutional design should be applied to this new market is an urgent question.
As a starting point to answer this question, verifying whether existing matching mechanisms accumulated in human society are useful in the LLM agent market is extremely important from both an engineering and economic perspective.

%%%%% Objective %%%%%
This study examines whether standard matching mechanisms function as intended in an LLM-agent matching market, where LLM agents make allocation-related decisions as proxies for human participants.
Specifically, we compare decentralized free-negotiation markets with centralized mechanism-based markets in which several representative matching mechanisms are implemented.

The analysis focuses on three questions.
First, do mechanism-based markets generate more stable and efficient outcomes than decentralized free negotiation among LLM agents?
Second, are the theoretical properties associated with each mechanism, including stability, Pareto efficiency, and strategy-proofness, behaviorally realized when the participants are LLM agents?
Third, how do LLM agents differ from human subjects in their propensity to report preferences truthfully under mechanisms such as DA and EADA?
Because LLM behavior may depend on natural-language framing, we also examine whether institutional performance varies across substantively different but formally equivalent market contexts: labor-market matching, high-school admissions, and nursery-school allocation.

The remainder of the paper is organized as follows. 
Section~\ref{sec:lite1} reviews recent work on LLM agents and delegated decision-making.
Section~\ref{sec:lite2} introduces the relevant concepts from matching theory and summarizes experimental evidence on matching mechanisms with human participants.
Section~\ref{sec:mth} develops the hypotheses, and Section~\ref{sec:exp} describes the experimental design.
Section~\ref{sec:res} reports the results.
Section~\ref{sec:disc} discusses their implications for mechanism design in LLM-agent markets.
Section~\ref{sec:con} concludes.

\section{Agent Research Using Large Language Models}\label{sec:lite1}
\subsection{Large Language Model-Driven Agents and Social Simulation}\label{subsec:lite1_simu}
Emergent phenomena refer to the process where individual components comprising a system act based on local rules and interactions, thereby manifesting global behavior~\cite{Goldstein-1999,Ueda-2001}. Through a bidirectional dynamic process where this global behavior imposes new constraints on the behavior of the components, a globally ordered structure expressing new functions is formed. In the social sciences, understanding how the micro-level decision-making of individuals (such as purchasing, movement, and opinion formation) emerges into macro-social phenomena like market price fluctuations, residential segregation, or the polarization of public opinion is extremely important for policymaking and institutional design~\cite{Coleman-1987,Macy-2002,Merton-1936}.

MAS, which has developed as a powerful tool bridging this micro and macro gap, enables the bottom-up reproduction of nonlinear interactions among heterogeneous subjects that cannot be fully described by top-down mathematical models~\cite{Heath-2009}. MAS began as an analogy to particle models in physics~\cite{Reynolds-1987,Vicsek-1995},and has been applied in diverse fields such as sociology, economics, and ecology, contributing to the understanding of social systems as complex adaptive systems~\cite{Axelrod-1997,Axelrod-2006,Bonabeau-2002,Epstein-1996,Nagel-1992,Schelling-1971}.
% 一部例を削除しました

However, traditional classical MAS had fundamental limitations as an analytical method in the social sciences~\cite{Castelfranchi-1995}. First is the limitation in describing the behavioral rules of agents. In traditional MAS, researchers had to design the behavioral rules of agents in advance and explicitly, such as in an If-Then format. While this was justified as the KISS (Keep It Simple, Stupid) principle~\cite{Macy-2002} to maintain model transparency, it was extremely difficult to reflect the complex cognitive biases, context dependencies, emotions, and sophisticated communication using natural language that real humans possess~\cite{Edmonds-2005,Qiu-2020}. As a result, agents within the simulation were often overly simplified, failing to reproduce the complex dynamics of reality. Second is the lack of environmental adaptability and versatility. Classical agents could not respond to events outside their programmed scope and lacked the ability to generate emergent responses in unknown situations or perform advanced learning and reasoning based on past experiences~\cite{An-2012,Edmonds-2005,Gao-2024}.It has remained difficult to incorporate realistic reasoning into traditional agents, presenting a limitation in the predictive accuracy and persuasiveness of simulations.
% 一部例を削除しました

The integration of LLMs into MAS is fundamentally overturning these limitations by granting agents the ability to generate behavior rather than just having their behavior described. LLM-driven agents are expected to be capable of generating autonomous and human-like behavior according to the situation without having rules described in detail in advance, by being given roles, personalities, motives, and memories through natural language instructions~\cite{Aher-2023,Andreas-2022,Gao-2024,Xi-2023}. In a study by Park et al.~\cite{Park-2023}, a simulation was conducted where 25 agents (Generative Agents) lived in a virtual town called Smallville. What is particularly noteworthy in this simulation is that unprogrammed cooperative behavior occurred spontaneously. This demonstrates that LLMs possess the ability to understand social contexts and adjust behavior through interactions with others. Furthermore, from 2024 to 2025, large-scale social simulation platforms driving thousands to millions of LLM agents have been proposed one after another~\cite{Bougie-2025,Feng-2025,Liu-2025-c,Piao-2025,Yang-2025}.

In the field of urban planning as well, the sophistication of simulations using LLM agents is advancing. In projects like OpenCity~\cite{Yan-2024} and TravelAgent~\cite{Noyman-2024}, LLM agents are introduced into the simulation of urban activities. Unlike traditional agents that moved based on fixed cost functions such as shortest paths, LLM agents make flexible decisions based on context and individual states, such as deciding to walk through a park because the weather is nice today. Attempts are also progressing in the simulation of macroeconomic activities, such as EconAgent~\cite{Li-2024-b}. Here, individual agents read news articles and policy announcements regarding macro indicators like inflation and interest rate fluctuations, make future predictions, and determine consumption and investment behaviors. This makes it possible to capture the feedback loops that people's expectations and narratives impose on the market, which tended to be abstracted away in traditional economic models. The emergence of LLMs can be said to have evolved MAS from an execution environment for rules designed by researchers into an experimental environment of society woven by autonomous subjects with human-like flexible reasoning abilities.

\subsection{Imitation of Human Behavior by Large Language Models}\label{subsec:lite1_imitation}
The introduction of LLMs is not limited to reproducing macro-social phenomena; it has also created a new research trend in approximating the micro-decision-making processes of individuals. In particular, verification studies on the extent to which LLM agents can accurately imitate human behavior, preferences, and cognitive biases are actively being conducted~\cite{Dillion-2023,Grossmann-2023}. As verifications proceed regarding the validity of using LLM agents as proxies for humans across diverse tasks in broad academic fields such as economics, psychology, cognitive science, medicine, and political science, both positive and negative results are mixed.

In the fields of social and public opinion surveys, attempts to use LLMs as virtual respondents ("silicon samples") and verify whether they can reproduce the response distributions of actual humans are rapidly advancing, suggesting the possibility of replacing costly and time-consuming preliminary surveys with LLMs~\cite{Sarstedt-2024,Shrestha-2024,Sun-2024}. Spearheaded by Argyle et al.~\cite{Argyle-2023}, many studies report that by giving LLMs appropriate personas (demographic information such as age, gender, race, and political orientation), the opinion distributions of specific social groups can be reproduced with high accuracy. In verifications using the 2020 US election survey, LLMs not only predicted individual voting behavior and policy preferences but also showed high consistency with actual data in terms of opinion polarization and correlations at the group level~\cite{Jiang-2025,Miranda-2025}.

On the other hand, research from 2024 to 2025 also points out the inherent biases in LLMs and the limits of reproduction due to averaged opinions~\cite{Wang-2025}. Bisbee et al.~\cite{Bisbee-2024} and Santurkar et al.~\cite{Santurkar-2023} revealed that LLM responses tend to amplify stereotypes more than actual humans do. While LLMs excel at playing average personas, they underestimate the diversity and noise seen in real human opinions, and their responses tend to converge to the median~\cite{Bisbee-2024,Li-2025}. Furthermore, Dominguez-Olmedo et al.~\cite{Dominguezolmedo-2024} attempted to reproduce the American Community Survey (ACS) using 39 different LLMs, but reported that the response accuracy for certain racial and income groups was significantly low depending on the model. Additionally, it has been reported that LLMs contain WEIRD (Western, Educated, Industrialized, Rich, and Democratic) biases originating from the developers' cultural backgrounds and training data, as well as specific political biases, indicating that further verification is necessary regarding the practical use of silicon samples~\cite{Qu-2024,Rettenberger-2024,Rozado-2023,Shrestha-2024,Xu-2025}.
% 短縮検討

The reproduction of typical laboratory experiments in economics (such as the ultimatum game, dictator game, public goods game, and prisoner's dilemma) has become a major benchmark for verifying not only the rationality of LLMs but also social preferences for fairness and reciprocity~\cite{Mei-2024}. 
\citet{Horton-2023} and \citet{Aher-2023} observed that LLMs, like humans, recognize the trade-off between monetary gains and fairness, and exhibit behaviors such as rejecting unfair proposals. These suggest that LLMs can understand economic contexts and make decision-making similar to humans. However, according to a study by \citet{Kitadai-2025} and other verifications~\cite{Akata-2025,Chen-2023,Jia-2024}, it has been confirmed that LLMs prioritize logical consistency over emotions and cognitive biases, showing a strong tendency to behave akin to the rational economic man (homo economicus) assumed in traditional economic theory. To reproduce boundedly rational properties unique to humans, such as the status quo bias and loss aversion, detailed persona settings and adjustments to the model's hyperparameters may be necessary~\cite{Kitadai-2025,Kolluri-2025,Kwok-2024,Lutz-2025,Moon-2024,Toubia-2025}.
% 一部事例を削除

Even outside of economics, attempts to imitate humans with LLMs are progressing in a wide variety of fields. In efforts to realize digital twins of humans, there are attempts to predict decision-making as a specific individual by inputting the individual's past behavioral history, spoken data such as SNS posts and emails, Big Five personality traits, and values into the LLM as detailed context~\cite{Goethals-2025,Li-2025-b,Toubia-2025}.
Furthermore, whether LLMs can possess a Theory of Mind (ToM) to understand the beliefs and intentions of others is a major point of debate in cognitive science.
Strachan et al.~\cite{Strachan-2024} reported that LLMs recorded scores equal to or higher than humans in standard ToM tests such as the false belief task.
However, it is still unsettled whether this is based on true cognitive ability or merely an apparent understanding through pattern matching in the training data.
In the medical field, their use as virtual patients imitating patient symptoms, medical histories, and speaking styles is also advancing~\cite{Holderried-2024,Luo-2025,Yamamoto-2024}.

\subsection{Substitutability of Decision-Making by Large Language Models}\label{subsec:lite1_substitute}
Alongside the debate over whether LLM agents can faithfully imitate humans, research is also being conducted on areas where LLM agents possessing a certain degree of human likeness interact with humans to build consensus, or where LLM agents delegated by humans make decisions as proxies among themselves.
Here, we specifically focus on the qualitative changes in interactions in a society where humans and AI coexist, and organize research results on the substitutability of decision-making in highly human and emotional contexts such as romance and negotiation.

When LLM agents are integrated into social systems as decision-making subjects, how humans perceive and behave toward LLM agents is a crucial element that determines the success or failure of the system~\cite{Candrian-2022,Meng-2024}.
\citet{Dvorak-2025} conducted a large-scale experimental study on the social interactions between humans and LLM.
Targeting 3,552 participants, they conducted classical economic games such as the ultimatum game, the trust game, and the prisoner's dilemma, comparing behavioral changes when the opponent was a human versus an LLM.
The experimental results revealed that when humans recognize their opponent as an LLM, they significantly reduce trust, cooperation, fairness, and coordination compared to when the opponent is a human. On the other hand, in situations where it is not revealed that the opponent is an LLM, it has been shown that humans tend to delegate decision-making to the LLM.
% 輪読会で読んだ論文の事例も書いてよさそう

As the imitation of humans by LLM agents and the accompanying delegation of decision-making become a reality, discussions are also developing regarding the spaces where interactions between LLM agents and humans, or interactions among LLM agents, take place~\cite{Shah-2025}. 
Research is already underway regarding responses when delegating financial decision-making~\cite{Allouah-2025,Fish-2025,Saha-2025,Yu-2024}, negotiations\cite{Bianchi-2024,Zhu-2025}, and purchasing transactions~\cite{Deng-2024,Xia-2024} to LLM agents.
In an AI negotiation competition by \citet{Vaccaro-2026}, it was confirmed that giving an LLM agent a warm persona increased the subjective satisfaction of the opponent agent and improved the consensus-building rate, while simultaneously presenting a trade-off in which it was inferior to agents adopting dominant strategies in acquiring pure economic gains. 
In romance, which is a complex and personal area in building human relationships, a new approach has also been proposed.
\citet{Shang-2025} pointed out the limitations of traditional matching apps that rely on matching static profiles such as hobbies, income, and educational background, and proposed a dynamic compatibility diagnosis system using LLM agents. 
Here, two LLM agents possessing the personas of User A and User B freely converse and interact in scenarios of important events such as virtual dates, career conflicts, and family planning.
An evaluating LLM observes the content of the conversations, emotional transitions, and decision-making that occur in this process, and scores their compatibility by learning it as a reward model.
This approach showed significantly higher predictive accuracy for long-term relationship stability than traditional methods.

\subsection{The Position of This Study in the Field of LLM Agent Research}\label{subsec:lite1_position}
Recent LLM agent research can be broadly categorized into three trends.
The first is the attempt to integrate LLMs into MAS as behavior generators and observe emergent phenomena arising from interactions among subjects equipped with natural language, memory, and social reasoning. 
The second is research that positions LLMs as silicon samples and verifies whether they can reproduce the statistical properties of human response distributions and decision-making; both positive results and limitations have been reported regarding the reproducibility of social surveys and laboratory experiments.
The third is a trend dealing with whether decision-making can be delegated to LLM agents as human proxies, exploring how interactions between humans and LLM agents, or among LLM agents, transform decision-making and outcomes in scenarios such as financial decision-making, negotiations, and transactions.

However, in the knowledge accumulated by these three trends, an important gap remains when considering the phase where LLM agents participate in social institutions.
Looking at social institutions as a whole, there may be resource allocation problems that cannot be solved solely by the autonomous negotiations of individual agents.
Particularly in matching markets—typified by school admissions, job hunting, organ transplants, and personnel assignments—it is expected that a centralized mechanism intervenes to aggregate the preferences of multiple participants and determine a globally desirable allocation.
Currently, while there are sporadic studies where LLM agents act as subjects to vote~\cite{Yang-2024} or participate in auctions~\cite{Zhu-2024}, to the best of our knowledge, no study has systematically verified how LLM agents behave in matching markets and how existing mechanisms function in response to that behavior.

Therefore, this study constructs a matching market in which LLM agents delegated to make decisions as proxies for humans participate, and verifies the responses of agents under multiple representative matching mechanism-based markets.
This serves as a starting point for institutional design theory anticipating a society where the delegation of decision-making to LLM agents becomes commonplace.

\section{Matching Theory and Empirical Research on Mechanisms}\label{sec:lite2}
\subsection{Basic Model of Matching Theory}\label{subsec:lite2_basic}
Matching theory is a theoretical framework for realizing efficient resource allocation in markets without prices~\cite{Gale-1962,Roth-1984,Roth-2002}. Traditional microeconomics has developed based on the fundamental idea that the price mechanism adjusts supply and demand to achieve efficient resource allocation~\cite{Arrow-1954}. However, in the real world, there are broad areas where the price mechanism does not function or where making it function should be avoided from a socially accepted or legal perspective. A representative example is the exchange of donor organs for kidney transplants~\cite{Roth-2007,Roth-2005}, which is a system that matches pairs of willing living donors and patients who cannot undergo direct transplants due to blood type incompatibility or other reasons, allowing both to receive transplants by exchanging kidneys between pairs; this has developed in various countries since the mid-2000s. In this kidney exchange market, the buying and selling of organs for the purpose of allocation by money is ethically and legally prohibited in many countries, making it a typical example of a market without prices. The assignment of medical residents to hospitals, job hunting for companies, high school entrance exams, and admission to nursery schools are also considered markets where prices do not function, and matching theory studies the allocation of scarce resources (organs, employment quotas at hospitals and companies, admission quotas for high schools, admission quotas for nursery schools) in such markets without prices~\cite{Abdulkadiroglu-2003,Gong-2016,Kamada-2015,Kennes-2014,Okumura-2019,Reischmann-2021,Roth-1984,Terrier-2026}.

Matching theory~\cite{Hurwicz-1973,Maskin-1999}, which deals with such problems, is positioned as a branch of institutional design theory, and its purpose is to theoretically and mathematically design mechanisms (predetermined procedures) to realize desirable matchings (social states) in situations where multiple decision-making subjects hold preferences for each other and act in pursuit of their own self-interest. Here, the preference $\succsim$ held by a decision-making subject is a binary relation indicating the desired order the subject has over the entire set of outcomes $\mathcal{A}$, and it satisfies both completeness and transitivity shown below. In this definition, $x \succsim_i y$ means that subject $i$ prefers outcome $x \in \mathcal{A}$ at least as much as outcome $y \in \mathcal{A}$.

\begin{definition}[Completeness]
For any $x, y \in \mathcal{A}$, $x \succsim_i y$ or $y \succsim_i x$.
\end{definition}

\begin{definition}[Transitivity]
For any $x, y, z \in \mathcal{A}$, if $x \succsim_i y$ and $y \succsim_i z$, then $x \succsim_i z$.
\end{definition}

Furthermore, the symmetric and asymmetric parts of the preference are defined as follows. The symmetric part represents that $i$ likes $x$ and $y$ equally, and the asymmetric part represents that $i$ strictly prefers $x$ over $y$.

\begin{definition}[Symmetric Part]
For any $x, y \in \mathcal{A}$, $x \succ_i y \iff [x \succsim_i y \text{ and } y \succsim_i x]$.
\end{definition}

\begin{definition}[Asymmetric Part]
For any $x, y \in \mathcal{A}$, $x \succ_i y \iff [x \succsim_i y \text{ and not } y \succsim_i x]$.
\end{definition}

In addition, when a preference satisfying completeness and transitivity does not have a symmetric part (satisfies antisymmetry), it is called a strict preference, and this study consistently deals with this strict preference.

\begin{definition}[Antisymmetry]
For any $x, y \in \mathcal{A}$, $x = y \iff [x \succsim_i y \text{ and } y \succsim_i x]$.
\end{definition}

The mathematical models dealt with in matching theory abstract the diverse and complex market structures of reality, and the most basic model among them is called two-sided matching~\cite{Gale-1962}, which deals with problems such as how to combine decision-making subjects belonging to two different groups, or how to allocate the resources of one group to the decision-making subjects of the other group. Here, following Gale and Shapley, we formulate the basic model dealt with in matching theory, taking matching between subjects belonging to two groups—job seekers and companies—as an example. Let the proposing job seekers be $s \in \mathcal{S}$, and the accepting companies be $c \in \mathcal{C}$. Job seekers have strict preferences over each company, and companies have strict preferences over each job seeker. In addition, it is assumed that a job seeker has an outside option of not getting a job at any company, and this outcome is represented by the symbol $\varnothing$. That is, any job seeker $s \in \mathcal{S}$ has a strict preference over the set $\mathcal{C} \cup \{\varnothing\}$, and any company $c \in \mathcal{C}$ has a strict preference over the set $\mathcal{S} \cup \{\varnothing\}$. A company being acceptable to a job seeker is denoted as $c \succ_s \varnothing$, and similarly, $s \succ_c \varnothing$ means that a job seeker is acceptable to a company.In this situation, matching is represented by the following function $\mu$. Note that $\mu(s)$ represents the matched company for the job seeker, and $\mu(c)$ represents the matched job seeker for the company.
$$\mu: \mathcal{S} \cup \mathcal{C} \rightarrow \mathcal{S} \cup \mathcal{C} \cup \{\varnothing\}$$
In particular, $\mu$ being a one-to-one matching is equivalent to satisfying all three of the following conditions.
\[
\begin{aligned}
\textbf{1.}\;& \forall s \in \mathcal{S},\quad 
\mu(s) \in \mathcal{C} \cup \{\varnothing\} \\
\textbf{2.}\;& \forall c \in \mathcal{C},\quad 
\mu(c) \in \mathcal{S} \cup \{\varnothing\} \\
\textbf{3.}\;& \forall s \in \mathcal{S},\ \forall c \in \mathcal{C},\quad
\mu(s) = c \iff \mu(c) = s
\end{aligned}
\]

Conditions 1 and 2 mean that all job seekers are matched with some company or not matched with any company, and all companies are also matched with some job seeker or not matched with any job seeker. Condition 3 means that if a job seeker is matched with a company, that company is simultaneously matched with that job seeker. This study will consistently deal with this two-sided, one-to-one matching, which is the most fundamental model.

\subsection{Definition of Concepts Related to Desirable Matching}\label{subsec:lite2_def}
While the desirability of matching in the real world depends on complex constraints and contexts, matching theory abstracts this and evaluates it by defining several desirable properties.
Following \citet{Cerrone-2024}, this study defines desirable matching based on the stability and efficiency of matching outcomes, and the strategy-proofness of matching mechanisms.

First, stability refers to the property that there is no incentive for withdrawal or deviation by mutual agreement between parties in the matching outcome.
In other words, this requires the absence of a blocking pair, which is a pair of individuals who prefer each other to their current partners.
In addition, individual rationality—meaning that each decision-making subject prefers their current partner over being unmatched—must also be satisfied, otherwise the individual would leave.
Therefore, it is usually considered a prerequisite for stability.

\begin{definition}[Stability]
Consider a one-to-one matching $\mu$.
Assume the preferences of each subject are defined over a set including the outside option $\emptyset$.
A matching $\mu$ is stable if $\mu$ is individually rational and there is no blocking pair for $\mu$.

\begin{itemize}[label=, leftmargin=1.5em]
    \item (Individual Rationality) $\mu$ is said to be individually rational if it satisfies the following for any $s \in S$ and $c \in C$:
    $$\mu(s) \succsim_s \emptyset \text{ and } \mu(c) \succsim_c \emptyset$$
    
    \item (Blocking Pair) A pair $(s, c) \in S \times C$ is said to be a blocking pair for $\mu$ if it satisfies the following:
    $$c \succ_s \mu(s) \text{ and } s \succ_c \mu(c)$$
\end{itemize}
\end{definition}

In a matching with stability, no individual or pair is improved by other options compared to their current state, so as a result, no preemptive actions occur, enabling stable market operation.
For example, a stable solution in the stable marriage problem is a combination in which there is no pair where both prefer each other to their current partners among any male-female pair, and it has been shown that at least one such solution is always found by the mechanism of \citet{Gale-1962}.
A stable matching is also a solution belonging to the core of a cooperative game, and it is known that in a one-to-one two-sided matching model in particular, the set of stable matchings and the core set coincide. This means that stability is a strong property that prevents deviations not only by pairs but also by coalitions of groups. Moreover, stability is extremely important in practice; according to research by \citet{Roth-1984}, producing a stable outcome is almost essential for the success of a centralized matching mechanism.
It has been reported that in actual labor markets and school choice systems, after mechanisms satisfying stability were introduced, early unofficial offers and informal transactions decreased, realizing fair and predictable system operations~\cite{Abdulkadiroglu-2006,Niederle-2003,Roth-1984,Roth-1994}.

Next, efficiency refers to the property of having no waste from the perspective of resource allocation, and a representative example is Pareto efficiency.
A matching is defined as Pareto efficient if there is no other matching (Pareto improvement) where at least one subject obtains a more favorable outcome than the current one without anyone else being disadvantaged, compared to any other matching. The concept of efficiency is further subdivided: in addition to Pareto efficiency for all subjects, which represents efficiency for everyone including both job seekers and companies, there is one-sided Pareto efficiency.
One-sided Pareto efficiency represents a state where, looking only at the preferences of the job seekers, for example, it is impossible to achieve Pareto improvement through trades among job seekers. It is used when the interests of subjects on one side (e.g., student welfare) are prioritized as a matter of policy, such as in school choice~\cite{Abdulkadiroglu-2003,Kesten-2010}.

\begin{definition}[Pareto Efficiency for All Subjects]
Consider a one-to-one matching $\mu$.
Assume the preferences of each subject are defined over a set including the outside option $\emptyset$.
A matching $\mu$ is Pareto efficient for all subjects if there is no matching $\mu'$ that is a Pareto improvement over $\mu$.

\begin{itemize}[label=, leftmargin=1.5em]
    \item (Pareto Improvement) Another matching $\mu'$ is a Pareto improvement over $\mu$ if:
    $$\mu'(s) \succsim_s \mu(s) (\forall s \in S) \text{ and } \mu'(c) \succsim_c \mu(c) (\forall c \in C)$$
    holds, and for at least one subject $i \in S \cup C$, the following holds:
    $$\mu'(i) \succ_i \mu(i)$$

    \item (One-sided Pareto Efficiency: Proposing Side) A matching $\mu$ is Pareto efficient on the proposing side if, for any matching $\mu'$ such that
    $$\mu'(s) \succsim_s \mu(s) (\forall s \in S)$$
    there does not exist such a $\mu'$ satisfying the following for at least one $s' \in S$:
    $$\mu'(s') \succ_{s'} \mu(s')$$
\end{itemize}
\end{definition}

The one-sided Pareto efficiency of the proposing side, which has a trade-off relationship with stability, is defined as "efficiency" (Table \ref{tab:def_prop}) in this study. In general, Pareto efficiency and stability are not necessarily compatible and are in a trade-off relationship depending on preference profiles~\cite{Abdulkadiroglu-2003,Cerrone-2024,Roth-1982}.

Finally, strategy-proofness~\cite{Gibbard-1973} refers to the property that makes it the best strategy (weakly dominant strategy) for decision-making subjects to state their true preferences without falsification in a matching mechanism, regardless of the preferences or actions of others. Specifically, if no subject can obtain a better outcome by falsely stating their true preference than by stating the truth, the mechanism is said to be immune to strategic manipulation. If a mechanism is not strategy-proof, subjects who are well-versed in the system's structure and capable of highly strategic behavior will benefit, while honest subjects or those lacking information will suffer losses. A strategy-proof mechanism eliminates this disparity by providing a simple optimal strategy: for all subjects to state their preferences honestly~\cite{Pathak-2008}. This is an excellent property for decision-making subjects, as they do not need to perform complex calculations or gather information, reducing psychological and time costs~\cite{Abdulkadiroglu-2006,Pathak-2011}. It also has the advantage for institutional designers that the stated preference information becomes reliable, contributing to welfare analysis and policy decisions~\cite{Abdulkadiroglu-2009,Pathak-2011}.

\begin{definition}[One-to-one Matching]
Let a mechanism that outputs a one-to-one matching be denoted as
$$\varphi: \mathcal{P}^S \times \mathcal{P}^C \rightarrow M$$
Here, $\mathcal{P}$ is the set of preferences, including the outside option, and $M$ is the set of one-to-one matchings. The true preference of subject $i \in S \cup C$ is written as $\succsim_i$, the stated preference as $\hat{\succsim}_i$, and the profile of stated preferences as $\hat{\succsim} = (\hat{\succsim}_j)_{j \in S \cup C}$. Also, let $\varphi(\hat{\succsim})(i)$ be the partner (or $\emptyset$) assigned to subject $i$ under the stated profile $\hat{\succsim}$.
\begin{itemize}[label=, leftmargin=1.5em]
    \item (Strategy-proofness for All Subjects) A mechanism is strategy-proof for all subjects if, for any subject $i \in S \cup C$, any reports of others $\hat{\succsim}_{-i}$, and any misreport $\hat{\succsim}'_i$, the following holds:
    $$\varphi(\succsim_i, \hat{\succsim}_{-i})(i) \succsim_i \varphi(\hat{\succsim}'_i, \hat{\succsim}_{-i})(i)$$

    \item (One-sided Strategy-proofness: Proposing Side) A mechanism is strategy-proof on the proposing side if, for any $s \in S$, any reports of others $\hat{\succsim}_{-s}$, and any misreport $\hat{\succsim}'_s$, the following holds:
    $$\varphi(\succsim_s, \hat{\succsim}_{-s})(s) \succsim_s \varphi(\hat{\succsim}'_s, \hat{\succsim}_{-s})(s)$$
\end{itemize}
\end{definition}

In this study, the one-sided strategy-proofness of the proposing side is defined as ``strategy-proofness'' (Table \ref{tab:def_prop}). 
\citet{Roth-1982} presents an impossibility theorem that states there is no mechanism that simultaneously satisfies stability and strategy-proofness for all groups in a market.
On the other hand, it is known that one-sided strategy-proofness and stability can be compatible~\cite{Gale-1962}.

Furthermore, whether decision-makers actually report honestly even under a mechanism satisfying strategy-proofness is an important empirical question.
Strategy-proofness implies that truthful reporting is a weakly dominant strategy, but it does not imply that participants necessarily recognize, trust, or follow that strategy in practice.
Previous laboratory experiments have shown that human subjects sometimes misreport their preferences even under strategy-proof mechanisms, suggesting that the behavioral realization of strategy-proofness depends on participants' understanding of the mechanism and their strategic reasoning (e.g. \cite{Guillen-2019}).
Therefore, in this study, the proportion of subjects who state their preferences honestly among all decision-making subjects on the proposing side is defined as the truth-telling rate, which is used as an indicator of whether decision-making subjects actually report their preferences honestly.

\begin{definition}[Truth-telling Rate]
The truth-telling rate $TR$ is defined as follows. Note that $|S|$ represents the total number of subjects.
$$TR := \frac{1}{|S|} |\{s \in S \mid \hat{\succsim}_s = \succsim_s\}|$$
\end{definition}

\begin{table}[htbp]
    \centering
    \caption{Definitions of desirable matching properties in this study}
    \label{tab:def_prop}
    \begin{tabular}{@{}ll@{}}
        \toprule % 表の一番上の太い線
        \textbf{Term} & \textbf{Definition in this study} \\
        \midrule % ヘッダーと本体を分ける中間の線
        Stability & Absence of blocking pairs and being individually rational \\
        Efficiency & Satisfies one-sided Pareto efficiency based on the proposing side's preferences \\
        Strategy-proofness & The proposing side cannot benefit from misreporting \\
        \bottomrule % 表の一番下の太い線
    \end{tabular}
\end{table}

\subsection{Theory and Empirical Research on Existing Matching Mechanisms}\label{subsec:lite2_mecha}
Regarding representative mechanisms in matching theory, this section outlines their procedures, theoretical properties, and findings obtained from subject experiments, based on the definitions in Table \ref{tab:def_prop}. Specifically, it covers Deferred Acceptance (DA), Efficiency-Adjusted Deferred Acceptance (EADA), Boston Matching (Boston), Random Serial Dictatorship (RSD), and Top Trading Cycle (TTC).

These five mechanisms are in a trade-off relationship where they sacrifice one of stability, efficiency, or strategy-proofness to prioritize another property~\cite{Abdulkadiroglu-2003,Gale-1962,Kesten-2010,Roth-1982}.
An overview of the properties theoretically satisfied by each mechanism is shown in Table~\ref{tab:mecha_prop}.
The stability and efficiency in the table represent the properties satisfied by the matching outcome when all decision-making subjects on the proposing side report honestly.
DA is the only mechanism that achieves both stability and strategy-proofness, but efficiency is not guaranteed.
In contrast, while EADA, Boston, and TTC satisfy efficiency if all participants submit their true preferences, these mechanisms do not satisfy stability.
Regarding strategy-proofness, DA, RSD, and TTC satisfy it, whereas EADA and Boston involve the possibility that strategic misreporting becomes advantageous.
Below, details of the theoretical properties of each mechanism and related empirical studies will be described.

%%% DA %%%
The Deferred Acceptance mechanism is a matching mechanism proposed by \citet{Gale-1962}, and is known as a mechanism that constructs the optimal stable matching.
Typically, the proposing side applies to the accepting side in order of preference, and the accepting side tentatively holds the applicants with the highest priority in order, rejecting others when capacity is exceeded.
By repeating this proposal and rejection, a stable matching is eventually obtained in which there is no pair in any combination where both prefer each other over their current partners~\cite{Gale-1962}.

The DA mechanism theoretically satisfies stability and strategy-proofness, but does not satisfy efficiency.
It is known that the outcome produced by DA is the most favorable combination for the proposing side and the most unfavorable for the accepting side among the entire set of stable matchings, and this is called the proposing-side optimal stable matching~\cite{Gale-1962}.
As theoretically desirable properties, DA satisfies the absence of blocking pairs~\cite{Gale-1962}, as well as individual rationality and one-sided strategy-proofness for the proposing side~\cite{Dubins-1981,Roth-1982}.
On the other hand, there are constraints regarding efficiency, and matching by DA does not necessarily satisfy Pareto efficiency based on all preferences in general~\cite{Abdulkadiroglu-2003}.
The target handled in this study is one-to-one two-sided matching, and it is known that Pareto efficiency for all subjects is satisfied in this model~\cite{Roth-1990}.
However, when limiting the target to the proposing side's preferences, there may be other matchings that provide Pareto improvements by compromising stability.
In fact, in the school choice in New York City where students used proposing-side DA, it was reported that approximately 5\% of the students could have been assigned to schools with a higher preference ranking~\cite{Abdulkadiroglu-2009}.

In subject experiments, while DA theoretically makes honest reporting of preference rankings a weakly dominant strategy for the proposing side, in reality, a tendency for some participants to attempt strategic manipulation has been observed. An experiment by \citet{Chen-2006} showed that the truth-telling rate of participants under the DA mechanism was around 72\%.
In the experiment by \citet{Cerrone-2024}, the subjects who reported honestly in DA remained between 45\% and 55\%, pointing out the possibility of insufficient understanding or trust in its strategic properties. 
Furthermore, in a review paper by \citet{Hakimov-2018}, it was also stated that empirical experiments of the DA mechanism targeting humans show truth-telling rates ranging from 30\% to 70\%~\cite{Bo-2020,Chen-2019,Pais-2008,Rees-Jones-2018}.

%%% EADA %%%
The Efficiency-Adjusted Deferred Acceptance (EADA) mechanism is a mechanism devised by \citet{Kesten-2010}, and its main focus is to increase the efficiency of the proposing side by partially relaxing the stability of DA.
The procedure is based on the same proposal-acceptance process as DA, but it is characterized by detecting and removing specific pairs called interrupters during the execution of DA.
According to the explanation by \citet{Cerrone-2024}, an interrupter refers to a pair consisting of job seeker A, who was temporarily put on hold, and company X, which temporarily put them on hold, in a situation where ``a certain job seeker A (proposing side) proposes acceptance to a certain company X (accepting side) and is put on hold, causing that company X to reject another job seeker B, but ultimately job seeker A themselves is also rejected by that company X.''
In EADA, such pairs of proposing and accepting sides are identified, the proposing side is made to waive its priority, the accepting side candidate is removed from the proposing side's preference, and DA is repeated anew.
By iteratively identifying interrupters, removing the relevant priority claims, and re-running DA, EADA can improve the proposing side’s welfare relative to standard DA by relaxing some of the efficiency losses associated with stability.
Following \citet{Cerrone-2024}, there are three variants of EADA: (1) EADA Consent, in which priority waivers are implemented only for proposing-side agents who consent to waive them; (2) EADA Object, in which priority waivers are implemented unless the relevant proposing-side agent objects; and (3) EADA Enforced, in which priority waivers are implemented without an additional consent or objection stage. In this study, we use EADA Enforced as the EADA condition in order to keep the decision task comparable across mechanisms: as in DA, Boston, RSD, and TTC, proposing agents submit only a preference ranking, and the matching outcome is then determined by the mechanism.

The EADA mechanism theoretically satisfies efficiency, but for stability and strategy-proofness, it only satisfies weakened versions of each property. While the matching outcome by EADA is one-sided Pareto efficient for the proposing side, the stability of the matching is not guaranteed because it eliminates interrupters~\cite{Kesten-2010}. In addition, regarding strategy-proofness, EADA has no guarantee of eliminating misreporting for the proposing side (one-sided strategy-proofness), and there is a possibility that strategically falsely reporting preference rankings could be advantageous depending on the preference profile~\cite{Cerrone-2024,Kesten-2010}.
However, even if a blocking pair exists, the matching outcome of EADA possesses Reasonable Stability~\cite{Kesten-2010} in the sense that there is no other matching outcome that becomes more stable upon the realization of that blocking pair. 
Furthermore, it possesses the property of Regret-free Truth-telling~\cite{Cerrone-2024}, meaning that under incomplete information (a state where the preferences of others are unknown) and only with feedback of the results, it is not certain that strategic misreporting will yield a more advantageous outcome than honest reporting.
Both of these are properties that relax stability and one-sided strategy-proofness for the proposing side.

Empirical findings also support the characteristics of honest reporting promoted by EADA. In a subject experiment by \citet{Cerrone-2024}, 65\% to 70\% of participants honestly reported their preferred schools when EADA was adopted, which significantly exceeded the truth-telling rate of 45\% to 55\% in DA.
That is, it has been reported that EADA is more likely to elicit honest behavior from real human subjects than DA. Furthermore, the same study confirmed efficiency improvements by EADA when compared with DA as a baseline.
These results suggest the possibility that a mechanism that initially appears inferior in strategic aspects may actually bring about better properties in practice, and it is noteworthy that EADA has an effect of inducing honest reporting that surpasses DA at the laboratory level.
%Its introduction into actual institutional design is also progressing, with a new school allocation system based on EADA scheduled to be introduced in the Flanders region of Belgium.

%%% Boston %%%
The Boston Matching (Boston) mechanism is an admission quota mechanism traditionally used in public school selection in Boston, USA, and is a matching mechanism where acceptance is immediately finalized for each preference rank of the proposers. Typically, the proposing side applies to the accepting side in order of preference, and the accepting side finalizes the acceptance from the applicants with the highest priority, rejecting others when capacity is exceeded. Pairs once finalized are not overturned in this matching, and by repeating this proposal and rejection, the final matching outcome is obtained.

The Boston mechanism theoretically satisfies efficiency but does not satisfy stability or strategy-proofness.
Under the sequential finalization method, if the first choice is popular, students who are rejected may face a situation where the capacity is already filled at their second choice, resulting in rejection there as well.
Therefore, from the perspective of stability, there is a possibility that blocking pairs may arise in the Boston outcome~\cite{Abdulkadiroglu-2003}.
On the other hand, in terms of efficiency, it has been shown that if an appropriate equilibrium strategy is achieved under common knowledge, the matching outcome produced by Boston can achieve one-sided Pareto improvements for the proposing side over DA~\cite{Abdulkadiroglu-2011}.
This is deeply related to Boston's failure to satisfy one-sided strategy-proofness, stemming from the fact that falsely reporting preference rankings can be advantageous for the proposers.

In fact, experimental studies by \citet{Chen-2006} and \citet{Pais-2008} have confirmed that participants perform large-scale rank manipulation under the Boston mechanism condition.
\citet{Pathak-2008} point out the existence of a strategic disparity in the Boston mechanism, where participants who report honestly are at a disadvantage compared to subjects who report strategically.
However, according to empirical verification by \citet{Featherstone-2016}, it is not easy for decision-making subjects to actually learn and coordinate on such counterintuitive Bayesian equilibria of misreporting, and behaviors that deviate significantly from the theoretical equilibrium are observed even after repeated practice in simple environments.
While subjects reported their true preference rankings with high probability under the DA condition, under the Boston condition, although many avoided honest reporting, they could not reach the optimal misreporting predicted by theory, and often made preference reports that deviated from optimal responses.

%%% RSD %%%
The Random Serial Dictatorship (RSD) mechanism is a mechanism using a random order based solely on the preference rankings of one side. In this mechanism, the proposing side is given the right to choose in a randomly determined order, and matching is performed by repeating the process where the proposing side selects the option with the highest preference rank among the options currently available according to that order. While RSD deterministically produces a unique outcome, all subjects on the proposing side are randomly assigned to any order with equal probability, so it functions as an ex-ante fair lottery mechanism.

The RSD mechanism theoretically satisfies strategy-proofness and ex-post efficiency, but does not satisfy stability.
This mechanism, also called the random priority mechanism, does not consider the priorities of the accepting side, so it generally does not satisfy stability~\cite{Abdulkadiroglu-1998,Svensson-1999}.
For example, if job seeker A, who has high priority for company X, cannot enter their first-choice company X due to a disadvantage in the lottery, and instead job seeker B with lower priority enters that company X, job seeker A and company X could become a more preferred combination for each other than their current ones.
Because such blocking pairs exist, RSD is not stable. On the other hand, RSD generally does not satisfy one-sided Pareto efficiency regarding the proposing side's preferences, but ex-post it realizes a one-sided Pareto efficient matching~\cite{Bogomolnaia-2001,Svensson-1999}.
Ex-post Pareto efficiency refers to Pareto efficiency in a state where the order of choice has been determined.
Furthermore, as a theoretical property, RSD is known to possess one-sided strategy-proofness for the proposing side~\cite{Abdulkadiroglu-1998,Svensson-1999}.
This has also been proven by \citet{Bogomolnaia-2001}, who state that it satisfies one-sided strategy-proofness in a situation where the distribution regarding the order of choice is fixed.
Typically, the order handled in RSD is random, and because participants are assigned to any order with equal probability, it can be said to satisfy strategy-proofness.

In subject experiments, it has been suggested that even in RSD, where truth-telling is a weakly dominant strategy, people do not necessarily always state their preferences honestly.
Regarding the truth-telling rate, it has been reported to be 60\% in a study by \citet{Li-2017} and 40\% in a study by \citet{Kloosterman-2020}.
These findings indicate that even mechanisms with straightforward incentive properties do not automatically induce truthful reporting in human subjects.

%%% TTC %%%
The Top Trading Cycle (TTC) mechanism is an application of Shapley and Scarf's housing market model~\cite{Shapley-1974} to matching. This method, introduced by Abdulkadiro\u{g}lu and S\"{o}nmez~\cite{Abdulkadiroglu-2003}, is based on a design philosophy that prioritizes efficiency over stability. TTC functions as a market mechanism that regards the priorities of the accepting side as "pseudo-property rights" and allows the proposing side to exchange those rights among themselves. The proposing side points to the accepting side ranked first on their own preference list. Similarly, the accepting side points to the proposing side ranked highest on their own preference list. Since this structure is a finite set, there must always exist at least one "closed cycle." Specifically, a closed cycle takes a circular structure such as Job Seeker A $\rightarrow$ Company X $\rightarrow$ Job Seeker A, or Job Seeker A $\rightarrow$ Company X $\rightarrow$ Job Seeker B $\rightarrow$ Company Y $\rightarrow$ Job Seeker A. For all proposing sides included in the cycle, the accepting side that the proposing side pointed to is assigned to them, and the acceptance is finalized. Pairs once finalized are not overturned in this matching, and the final matching outcome is obtained by repeating this pointing and cycle identification.

The TTC mechanism theoretically satisfies strategy-proofness and efficiency, but does not satisfy stability. TTC is considered not to satisfy stability because blocking pairs arise during the process of trading pseudo-property rights~\cite{Abdulkadiroglu-2020,Abdulkadiroglu-2003}. For example, if a cycle like School X $\rightarrow$ Student A $\rightarrow$ School A $\rightarrow$ Student B $\rightarrow$ School X occurs, a student with high priority for School X waives their right to School X and goes to School A, and as a result, Student B with lower priority enters School A. At this time, Student C, who has an intermediate priority between A and B and desired School X, and School X form a blocking pair. On the other hand, TTC is considered to satisfy one-sided Pareto efficiency for the proposing side~\cite{Abdulkadiroglu-2003,Abdulkadiroglu-2020}. Because TTC is not bound by stability constraints, it allows students to exchange property rights and mutually increase their utility. Therefore, under this mechanism, an allocation is achieved where no one else's utility can be improved without sacrificing someone's welfare. In addition, TTC satisfies one-sided strategy-proofness for the proposing side. According to the theorem of Abdulkadiro\u{g}lu et al.~\cite{Abdulkadiroglu-2020}, TTC satisfies the minimality of Justified Envy among mechanisms that are Pareto efficient and strategy-proof. Justified Envy here refers to the envy $(s, (c, s'))$ harbored by a job seeker $s$ toward a specific job seeker $s'$ assigned to a company when $s \succ_c s'$, and it is used as an indicator regarding stability.

In many experimental studies, TTC has recorded higher truth-telling rates than DA. A study by Pais and Pint\'{e}r~\cite{Pais-2008} reported a truth-telling rate of 87\% to 96\%, stating that the exchange logic of TTC is more intuitively understandable than the holding logic of DA, and that its high efficiency induces honest reporting. Similarly, a study by Chen and S\"{o}nmez~\cite{Chen-2006} reported a truth-telling rate of 56\% to 72\%.

\begin{table}[htbp]
    \centering
    \begin{threeparttable}
    \caption{Overview of theoretical properties satisfied by each matching mechanism}
    \label{tab:mecha_prop}
    \begin{tabular}{@{}lccc@{}}
        \toprule
        \textbf{Mechanism} & \textbf{Stability} & \textbf{Efficiency} & \textbf{Strategy-proofness} \\
        \midrule
        Deferred Acceptance (DA) & $\checkmark$ & $\times$ & $\checkmark$  \\
        Efficiency-Adjusted DA (EADA) & $\triangle$ & $\checkmark$ & $\triangle$  \\
        Boston Matching (Boston) & $\times$ & $\checkmark$ & $\times$  \\
        Random Serial Dictatorship (RSD) & $\times$ & $\triangle$ & $\checkmark$  \\
        Top Trading Cycle (TTC) & $\times$ & $\checkmark$ & $\checkmark$  \\
        \bottomrule
    \end{tabular}
    
    % ここから注釈環境
    \begin{tablenotes}
        \footnotesize
        \item Note: $\checkmark$ means the property is satisfied, $\times$ means it is not satisfied, and $\triangle$ means the property is satisfied under a relaxed definition.
    \end{tablenotes}
    
    \end{threeparttable}
\end{table}

\subsection{The Position of This Study in the Field of Matching Theory}\label{subsec:lite2_position}
In the lineage of matching theory and experimental economics, it has been assumed that decision-making subjects are homo economicus, and the properties satisfied by mechanisms such as DA, EADA, Boston, RSD, and TTC have been organized mathematically. At the same time, it has been repeatedly shown that real humans do not report honestly as theory dictates, and that the degree to which they do so varies greatly depending on the mechanism. However, these findings have been interpreted primarily against the backdrop of humans' bounded rationality and cognitive load limits, and it is not self-evident whether similar empirical tendencies hold when decision-making subjects are replaced by LLM-driven agents. While LLM agents may behave more rationally and strategically than humans, they may also exhibit behavioral changes depending on the uncertainty associated with probabilistic generation and natural language contexts, so the applicability of traditional theory is an object that needs to be re-verified.

Based on the above, the originality of this study lies in evaluating the effectiveness of matching mechanisms in an LLM agent market, using existing theoretical properties and findings from subject experiments as a basis for comparison. Specifically, a decentralized market where matching is formed through free negotiation is set as a baseline, and on top of that, multiple representative matching mechanisms (DA, EADA, Boston, RSD, TTC) are introduced into the same one-to-one matching scenario and experimentally compared using LLM agents as decision-making subjects.

\section{Methodology}\label{sec:mth}
\subsection{Hypothesis}\label{subsec:mth_hypo}
%%%%% ここの最初の文章修正したい %%%%%
In this study, free negotiation markets where agents form matchings through natural language dialogue is set as the baseline for comparative verification. Against this, we verify whether mechanism-based markets applying multiple matching mechanisms yields better or theoretically valid outcomes compared to the baseline (free negotiation markets). Note that the hypotheses, experimental methods, and main analysis plan of this study were pre-registered on the Open Science Framework (OSF) prior to data collection and analysis\footnote{\url{https://osf.io/cnmz5/overview?view_only=11a0fc2229db4f32b4f177a7a41fac6b}}. The hypotheses and experimental items below are based on the pre-registered content.

First, regarding whether intervention by mechanisms improves market quality compared to free negotiation, Hypotheses ~\ref{hyp:h1} and ~\ref{hyp:h2} were formulated from the perspectives of stability and efficiency. As shown in Section ~\ref{sec:lite2}, existing matching mechanisms each possess properties they should satisfy, and if LLM agents aim to maximize their own utility and act rationally, it is highly likely that they will satisfy these properties consistently with the theory. Previous studies in the context of human behavior imitation by LLM agents report that LLMs behave more rationally and strategically than humans~\cite{Chen-2023,Jia-2024,Kitadai-2025}. Therefore, in this study as well, it is expected that the LLM agents' responses will reproduce the theoretically expected properties of each mechanism. On the other hand, the free negotiation markets are markets without mechanisms—that is, a decentralized market where matchings are constructed by interacting decision-making subjects without a centralized system. While some studies suggest that stability and efficiency are achievable in decentralized matching targeting humans, it has been pointed out that LLM agents are not well-suited for interactions involving coordination~\cite{Akata-2025,Echenique-2024,Raykov-2017}. Based on the above, using the free negotiation markets as baseline, we hypothesized that desirable matching outcomes reflecting the properties of the mechanisms would be obtained in the mechanism-based markets.

\begin{hypothesis}\label{hyp:h1}
The proportion of yielding stable matching outcomes is higher in mechanism-Based markets that theoretically satisfy stability compared to the free negotiation market.
\end{hypothesis}

\begin{hypothesis}\label{hyp:h2}
The proportion of yielding Pareto efficient matching outcomes is higher in mechanism-based markets that theoretically satisfy Pareto efficiency compared to the free negotiation market.
\end{hypothesis}

Next, regarding whether the theoretical predictions of differences in properties (stability, efficiency, strategy-proofness) among matching mechanisms are reproduced even in a proxy matching market by LLM agents, the following three hypotheses were formulated. As mentioned above, if LLM agents act rationally and strategically, it is inferred that the differences in properties among matching mechanisms will also manifest consistently with theory.

\begin{hypothesis}\label{hyp:h3}
The proportion of yielding stable matching outcomes is higher in mechanism-based markets that theoretically satisfy stability compared to those that do not.
\end{hypothesis}

\begin{hypothesis}\label{hyp:h4}
The proportion of yielding Pareto efficient matching outcomes is higher in mechanism-based markets that theoretically satisfy Pareto efficiency compared to those that do not.
\end{hypothesis}

\begin{hypothesis}\label{hyp:h5}
Mechanism-based markets that satisfy strategy-proofness show higher truth-telling rates compared to those that do not.
\end{hypothesis}

Finally, we verify whether there are strategic behaviors specific to the matching market with LLM agents as subjects, compared to a market where humans are the subjects.
Focusing on the truth-telling rate, Hypotheses \ref{hyp:h6} and \ref{hyp:h7} were formulated.
In Cerrone et al.~\cite{Cerrone-2024}, contrary to theoretical predictions, results from human subject experiments showed that the truth-telling rate was higher for EADA, which does not satisfy strategy-proofness, than for the DA mechanism, which does.
However, if LLM agents act more rationally and strategically than humans, mechanisms are considered to function consistently with theory.
In other words, we predict that a higher truth-telling rate will be observed under the DA mechanism-based market in this study, and a lower truth-telling rate will be observed under the EADA mechanism-based market.

\begin{hypothesis}\label{hyp:h6}
The truth-telling rate is higher in the DA mechanism-based market, which satisfies strategy-proofness, than in empirical experiments with human subjects.
\end{hypothesis}

\begin{hypothesis}\label{hyp:h7}
The truth-telling rate is lower in the EADA mechanism-based market, which does not satisfy strategy-proofness, than in empirical experiments with human subjects.
\end{hypothesis}

\subsection{Configuration of Experimental Conditions}\label{subsec:mth_conditions}
To verify the hypotheses in the previous section, this study established (1) market environment settings, (2) market scenario settings, and (3) preference profile settings, which serve as the main objects of comparison. Across all combinations of settings, we observed the stability and efficiency of matching outcomes and the truth-telling rates for the matching mechanisms.

(1) The market environment is the environment where LLM agents perform matching. We constructed two types of free negotiation markets (passive free negotiation market, active free negotiation market) and five types of mechanism-based markets (DA, EADA, Boston, RSD, TTC mechanism-based markets).
(2) The market scenario refers to the roles assigned to LLM agents and the name of the market. Experiments were conducted in three scenarios: the labor market (matching job seekers and companies), high school entrance exams (matching students and high schools), and nursery school selection (matching parents and nursery schools).
(3) The preference profile is the set of preferences given to LLM agents. We prepared five types of preference profiles that differ in the properties theoretically satisfied by the matching mechanisms.

An experiment consisting of a configured market environment, market scenario, and preference profile is defined as one set, and 100 independent matching trials were conducted for each set. In all experiments, we consistently dealt with one-to-one matching with 5 proposing agents and 5 accepting agents.

Regarding the information structure, following Cerrone et al.~\cite{Cerrone-2024}, we conducted the experiments under complete information. Complete information here refers to a situation where all agents know the preference information of any proposing or accepting agent, the capacity of the accepting side, and the matching mechanism. Complete information can be considered the most disadvantageous condition for verifying a mechanism, as it is the easiest situation for agents to strategically misreport. In this study, complete information was chosen to verify that matching mechanisms function even under the most disadvantageous situations.

\section{Experimental Design}\label{sec:exp}
\subsection{Market Environments}\label{subsec:exp_env}
\begin{figure}[htbp]
    \centering
    \includegraphics[width=0.7\linewidth]{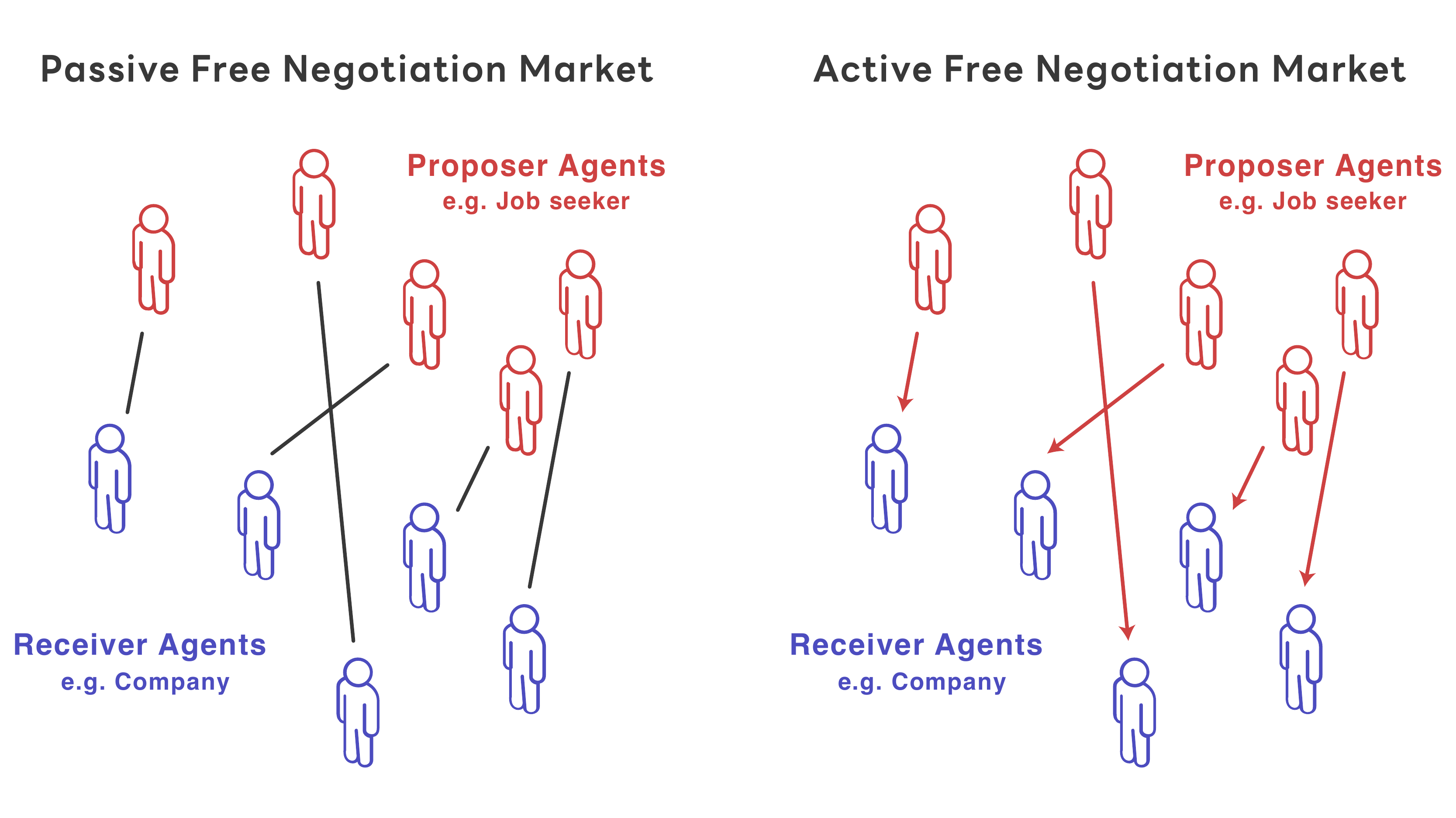}
    \caption{Passive free negotiation vs. Active free negotiation}
    \label{fig:passive_active}
\end{figure}

In the free negotiation market, proposing agents and accepting agents can negotiate freely through natural language dialogue. Within this, we constructed a passive free negotiation market where dialogue partners are randomly paired in a centralized manner, and an active free negotiation market where the proposing side selects their dialogue partners. Conceptual diagrams for both are shown in Figure \ref{fig:passive_active}.

Both free negotiation markets use a round system, and in each round, one message from the proposing side and one message from the accepting side are exchanged. By including tags in their messages, the proposing side explicitly indicates their actions in each round and can formally apply for acceptance at any time. The accepting side, having received a formal application, also replies regarding their acceptance or rejection using tags. By repeating this, the matching process concludes either when all agents are matched or at the end of the 30th round.

\begin{figure}[htbp]
    \centering
    \includegraphics[width=0.9\linewidth]{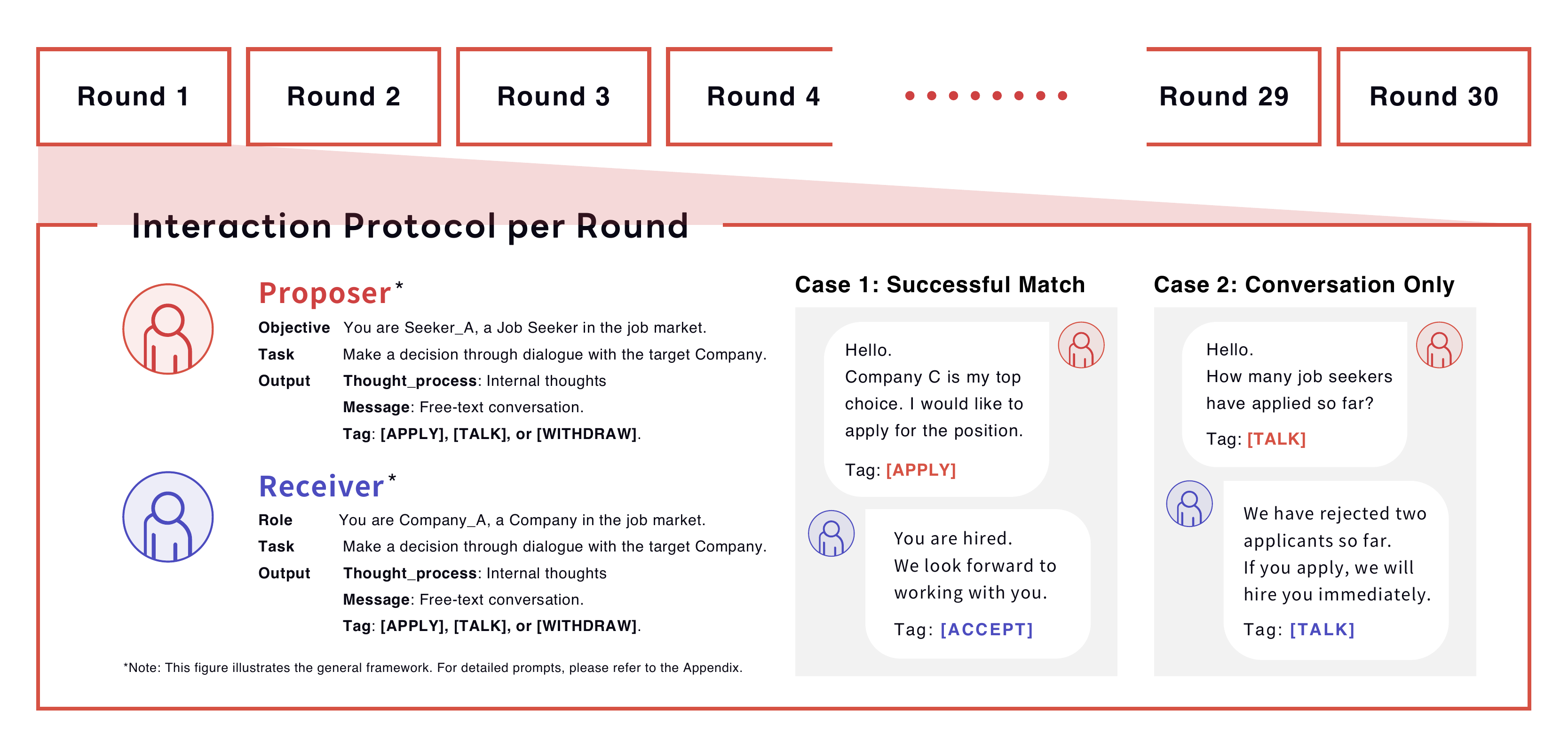}
    \caption{Protocol of passive free negotiation markets}
    \label{fig:free_nego}
\end{figure}

In the passive free negotiation market (Figure \ref{fig:free_nego}), in each round, 5 non-overlapping pairs of proposing and accepting agents are randomly matched, and the proposing side initiates the dialogue. The proposing side has three types of tags: [APPLY] (formally apply for acceptance), [TALK] (engage in conversation such as asking questions or making an appeal), and [WITHDRAW] (decline the dialogue). The proposing side sends one of these tags along with their message to their paired accepting agent. If the proposing side selects [APPLY], the accepting side decides whether to accept or reject using either the [ACCEPT] or [REJECT] tag. If the tag is [TALK], the accepting side also replies with a [TALK] tag, and if the tag is [WITHDRAW], the dialogue for that round ends at that point. Example prompts are provided in Appendix A: Table \ref{tab:free_passive_proposer_prompt} for the proposing side and Table \ref{tab:free_passive_accepter_prompt} for the accepting side.

On the other hand, in the active free negotiation market, in each round, the proposing side selects one accepting agent to dialogue with (overlaps among proposing agents are allowed), and the proposing side initiates the dialogue. The proposing side has two types of tags: [APPLY] and [TALK]. The proposing side sends one of these tags along with their message to their paired accepting agent. If an accepting agent receives messages from multiple opponents, they receive all messages at once for that round. If the proposing side selects [APPLY], the accepting side decides whether to accept or reject using either the [ACCEPT] or [REJECT] tag. If the tag is [TALK], the accepting side also replies with a [TALK] tag. Example prompts for the active free negotiation market are provided in Appendix A: Table \ref{tab:free_active_proposer} for the proposing side and Table \ref{tab:free_active_accepter} for the accepting side.

The mechanism-based markets apply each of the matching mechanisms discussed in Section 3.4 (DA, EADA, Boston, RSD, TTC), and matching is performed by the proposing side submitting a list of preference rankings. The true preference list given to the proposing side does not necessarily have to match the choice ranking list ultimately submitted by the proposing side, and when both lists match, it is considered that a truthful report was made. Unlike free negotiation, there is no round system, and the matching outcome is determined by a single decision from each proposing agent.

Table \ref{tab:DA_prompt} in Appendix A is an example prompt for the proposing side in the DA mechanism-based market. The only difference from the prompts of other matching mechanism-based markets is the explanation of the matching market, which are listed in Tables \ref{tab:EADA_prompt} to \ref{tab:TTC_prompt}. For the execution procedures of the DA and EADA mechanisms, to compare truth-telling rates, we quoted the instructions from the subject experiment by Cerrone et al. For the execution procedures of other mechanisms, we described them in a format consistent with DA and EADA, as long as the mechanism names and the presence or absence of dominant strategies were not explicitly stated as text.

\subsection{Market Scenarios}\label{subsec:exp_scn}
It is argued that the strategic responses of LLM agents depend not only on mathematical game structures but also on scenarios and contexts~\cite{Lore-2023,Robinson-2025}. If this tendency is reproduced in proxy matching markets by LLM agents as well, differences in matching outcomes and strategic behaviors among scenarios should be verified. In this study, we conducted verifications using three different scenarios typical of matchings: the labor market~\cite{Gong-2016,Roth-1984,Roth-1994}, high school entrance exams~\cite{Abdulkadiroglu-2005,Abdulkadiroglu-2009,Terrier-2026,Wang-2020}, and nursery school selection~\cite{Kennes-2014,Okumura-2019,Reischmann-2021}.

In each scenario, we set (1) the name of the proposing subject, (2) the name of the accepting subject, and (3) the name of the market, which are included in the prompt input to the LLM agent. Table \ref{tab:scenario_names} shows the names set for each scenario.

\begin{table}[htbp]
    \centering
    \caption{Terminology for each market scenario}
    \label{tab:scenario_names}
    \begin{tabular}{@{}llll@{}}
        \toprule
        \textbf{Scenario} & \textbf{Proposing Side} & \textbf{Receiving Side} & \textbf{Market Name} \\
        \midrule
        Job  & Job Seeker & Company & Job Market \\
        School  & Student & High School & High School Entrance Exam Market \\
        Nursery & Parent & Nursery & Nursery School Allocation Market \\
        \bottomrule
    \end{tabular}
\end{table}

\subsection{Preference Profiles}\label{subsec:exp_pref}
In this study, we prepared five types of preference profiles for the proposing agents and accepting agents (Table \ref{tab:type_of_preferences}):(1) Preferences with uniform priorities on the accepting side;(2) Preferences with heterogeneous priorities on the accepting side, with no interrupters in the EADA mechanism and unmanipulable;(3) Preferences with heterogeneous priorities on the accepting side, with interrupters in the EADA mechanism but unmanipulable;(4) Preferences with heterogeneous priorities on the accepting side, with no interrupters in the EADA mechanism but manipulable; and(5) Preferences with heterogeneous priorities on the accepting side, with interrupters in the EADA mechanism and manipulable.Here, uniform priorities on the accepting side mean that all accepting agents have the same priority order. The presence of an interrupter in EADA means that the preference profile contains an interrupter during the DA executed first within EADA. Furthermore, being manipulable in EADA means that there is a preference profile where a specific proposing agent can gain an advantage over reporting truthfully by strategically misreporting their preference rankings. Let these be denoted as Preference 1 to Preference 5, respectively. Preferences 3, 4, and 5 adopt the preference profiles themed in Cerrone et al~\cite{Cerrone-2024}. Preference 3 corresponds to the "Non-Manipulable Market with three interrupters", Preference 4 to the "Manipulable Market without interrupter", and Preference 5 to the "Manipulable Market with three interrupters".

% tab:pref1〜5はappendix行き→文章修正した方がいいね→修正しました
Each preference profile differs in the properties satisfied when the mechanism functions consistently with theory. For example, Preference 1 has uniform priorities on the accepting side; thus, when all agents report truthfully, the matching outcome by the DA mechanism satisfies one-sided Pareto efficiency, and the matching outcome by the TTC mechanism satisfies stability. Preference 4 has no interrupters, but it features a characteristic where a specific agent can obtain a more favorable outcome than DA only when other agents report truthfully by making a misreport in EADA. In Appendix B, Tables \ref{tab:preference1} to \ref{tab:preference5} show the specific preference profiles.

Finally, Table \ref{tab:preferences_characteristics} summarizes the properties satisfied by the matching outcomes derived when all proposing agents report truthfully under each matching mechanism, organized by preference profile. Note that stability and efficiency here are premised on truthful reporting, not the matching outcomes in an equilibrium where all agents express their preferences rationally.

\begin{table}[htbp]
  \centering
  \renewcommand{\arraystretch}{1.2}
  \caption{Overview of preference profile settings}
  \label{tab:type_of_preferences}
  \begin{tabular}{@{}llllll@{}} 
    \toprule
    & Pref. 1 & Pref. 2 & Pref. 3 & Pref. 4 & Pref. 5 \\ 
    \midrule
    Receiving Side's Priorities & Uniform & Non-uniform & Non-uniform & Non-uniform & Non-uniform \\
    Interrupter & Absent & Absent & Present & Absent & Present \\
    Manipulability & Impossible & Impossible & Impossible & Possible & Possible \\ 
    \bottomrule
  \end{tabular}
\end{table}

\begin{table}[htbp]
  \centering
  \begin{threeparttable}
    \renewcommand{\arraystretch}{1.2}
    \caption{Properties satisfied by matching outcomes when all proposing agents report truthfully}
    \label{tab:preferences_characteristics}
    \begin{tabular}{@{}l cc cc cc cc cc@{}}
      \toprule
      & \multicolumn{2}{c}{DA} & \multicolumn{2}{c}{EADA} & \multicolumn{2}{c}{Boston}
      & \multicolumn{2}{c}{RSD} & \multicolumn{2}{c}{TTC} \\
      \cmidrule(lr){2-3}\cmidrule(lr){4-5}\cmidrule(lr){6-7}\cmidrule(lr){8-9}\cmidrule(lr){10-11}
      Profile & Stable & Efficient & Stable & Efficient & Stable & Efficient & Stable & Efficient & Stable & Efficient \\
      \midrule
      Pref. 1 & $\checkmark$ & $\checkmark$ & $\checkmark$ & $\checkmark$ & $\checkmark$ & $\checkmark$ & $\checkmark$ & $\checkmark$ & $\checkmark$ & $\checkmark$ \\
      Pref. 2 & $\checkmark$ & $\checkmark$ & $\checkmark$ & $\checkmark$ & $\checkmark$ & $\checkmark$ & $\times$ & $\checkmark$ & $\checkmark$ & $\checkmark$ \\
      Pref. 3 & $\checkmark$ & $\times$ & $\triangle$ & $\checkmark$ & $\times$ & $\checkmark$ & $\times$ & $\checkmark$ & $\times$ & $\checkmark$ \\
      Pref. 4 & $\checkmark$ & $\checkmark$ & $\triangle$ & $\checkmark$ & $\times$ & $\checkmark$ & $\times$ & $\checkmark$ & $\times$ & $\checkmark$ \\
      Pref. 5 & $\checkmark$ & $\times$ & $\triangle$ & $\checkmark$ & $\times$ & $\checkmark$ & $\times$ & $\checkmark$ & $\times$ & $\checkmark$ \\
      \bottomrule
    \end{tabular}
    
    \begin{tablenotes}
      \footnotesize
      \item Note: $\checkmark$ indicates the property is satisfied, and $\times$ indicates it is not satisfied. $\triangle$ under EADA represents Reasonable Stability.
    \end{tablenotes}
  \end{threeparttable}
\end{table}

\subsection{LLM Models}\label{subsec:exp_llm}
We employed \texttt{gpt-5.2-2025-12-11} and \texttt{gemini-2.5-flash-preview-09-2025} as the underlying LLMs (hereafter referred to as GPT and Gemini, respectively). To adjust the randomness when agents make decisions, the hyperparameter temperature was utilized. A higher value for this parameter generates more random outputs. In this study, following several previous studies~\cite{Argyle-2023,Lee-2024,Strachan-2024}, the temperature was set to 0.7. Note that the list of configurable hyperparameters including temperature and the main values set in this study are shown in Table \ref{tab:params_gpt} and \ref{tab:params_gemini}.

\begin{table}[htbp]
    \centering
    % 左側の表（幅をテキスト幅の48%に設定し、上端[t]で揃える）
    \begin{minipage}[t]{0.48\textwidth}
        \centering
        \caption{Hyperparameters for GPT model}
        \label{tab:params_gpt}
        \begin{tabular}{@{}ll@{}}
            \toprule
            Parameter (GPT) & Value \\
            \midrule
            reasoning.effort & none (default) \\
            text.verbosity & medium (default) \\
            max\_output\_tokens & - \\
            temperature & 0.7 \\
            top\_p & 1 (default) \\
            logprobs & - (default) \\
            \bottomrule
        \end{tabular}
    \end{minipage} % \hfill 
    \begin{minipage}[t]{0.48\textwidth}
        \centering
        \caption{Hyperparameters for Gemini model}
        \label{tab:params_gemini}
        \begin{tabular}{@{}ll@{}}
            \toprule
            Parameter (Gemini) & Value \\
            \midrule
            thinkingBudget & -1 (default) \\
            temperature & 0.7 \\
            topP & 0.95 (default) \\
            topK & 64 (fixed) \\
            candidateCount & 1 (default) \\
            maxOutputTokens & - \\
            stopSequences & - \\
            presencePenalty & - \\
            frequencyPenalty & - \\
            seed & - \\
            responseMimeType & application/json \\
            responseSchema & - \\
            \bottomrule
        \end{tabular}
    \end{minipage}
\end{table}

\section{Results}\label{sec:res}
\subsection{Free Negotiation vs. Mechanism-Based Markets}\label{subsec:res_env1}
\begin{figure}[htbp]
    \centering
    \includegraphics[width=0.8\linewidth]{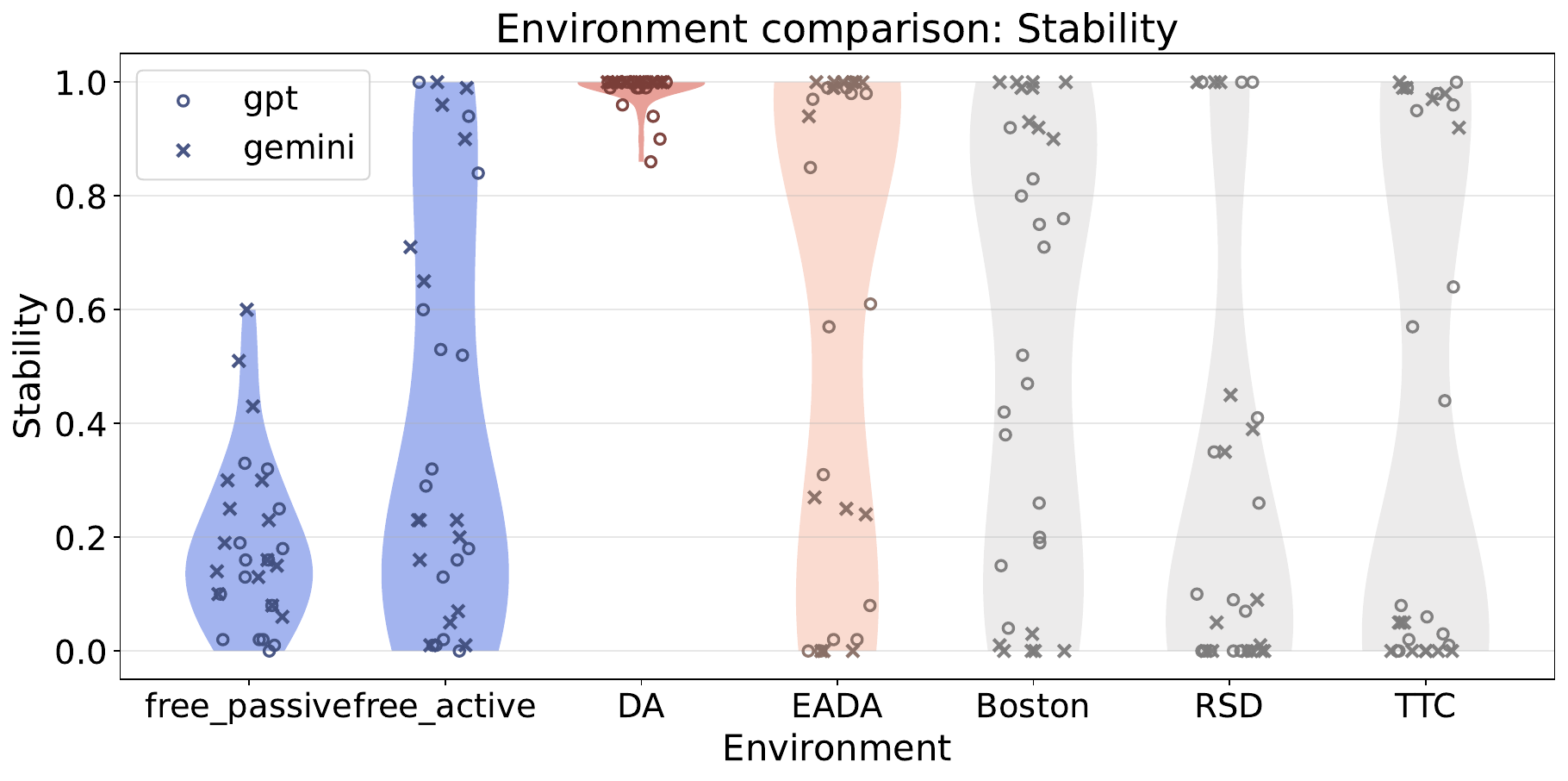}
    \caption{Stability among markets environments}
    \label{fig:violin_stability}
\end{figure}

Figure~\ref{fig:violin_stability} plots the proportion $P_{stability}$ of stable matching results in the free negotiation market and the mechanism-based market for each experimental set (100 trials). Each column of the graph corresponds to a market environment (passive free negotiation, active free negotiation, and the DA, EADA, Boston, RSD, and TTC mechanism-based markets), and the vertical axis represents the proportion $P_{stability}$ of stable matching results in one experimental set. Each plot corresponds to a different experimental set, and the $\circ$/× plots correspond to GPT and Gemini, respectively. The data for each market environment column includes 30 sets of matching results, covering three market scenarios, five preference profiles, and two LLM models. Note that Figure~\ref{fig:violin_stability} is intended to provide an overview of qualitative trends and is not a distribution map representing the probability of obtaining stable results.

In the DA mechanism-based market, which satisfies stability in all preference profiles, $P_{stability}$ was at least 86\% (market scenario: labor market, preference profile: Preference 5), whereas in the passive free negotiation market (free\_passive), it was at most 60\% (market scenario: nursery school selection, preference profile: Preference 2). Even in the active free negotiation market (free\_active), the center of gravity of the overall plots is concentrated below 70\%, indicating that the DA mechanism-based market achieves more desirable results in terms of stability.

\begin{figure}[htbp]
    \centering
    \includegraphics[width=0.80\linewidth]{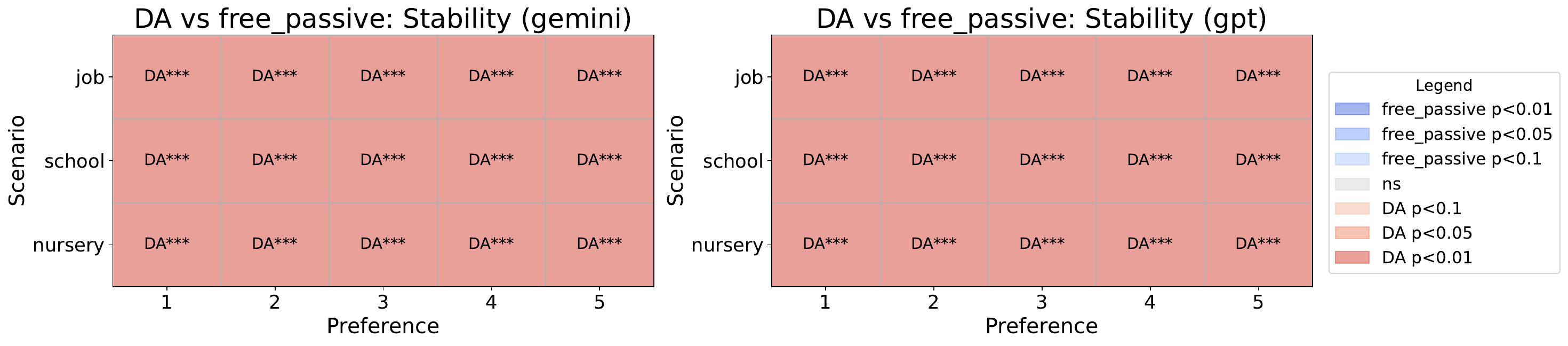}
    \includegraphics[width=0.80\linewidth]{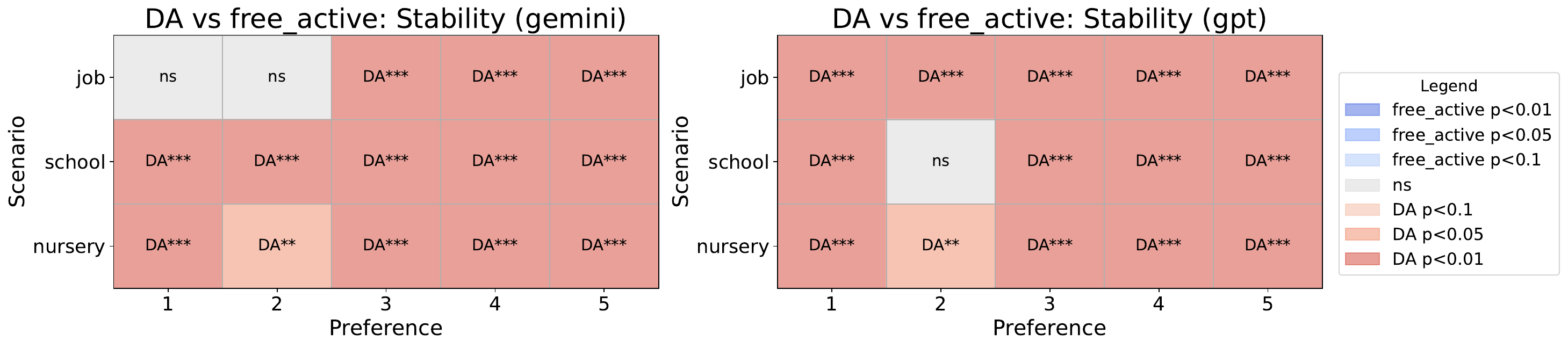}
    \caption{DA vs. free negotiation markets (Stability)}
    \label{fig:stability_da_free}
\end{figure}

Next, we verify whether the $P_{stability}$ of the stability-satisfying mechanism-based market is statistically significantly higher than that of the free negotiation markets. In this study, stability is defined as 0 (not stable) or 1 (stable) for each matching result, which can be treated as a Bernoulli distribution. Therefore, by using the Chi-square test~\cite{Pearson-1900}, we were able to verify whether there is a significant difference in the population proportions that yield stable matchings between any two market environments. Figure~\ref{fig:stability_da_free} shows the p-values estimated using the Chi-square test (two-sided) to compare the $P_{stability}$ of the DA mechanism with the passive and active free negotiation markets. The top two panels show the comparison with the passive free negotiation market, and the bottom two panels show the comparison with the active free negotiation market. Each cell represents whether the difference in the population proportions yielding stable matchings is significant between the DA mechanism-based market and the free negotiation markets when all conditions other than the market environment (same market scenario, same preference profile, and same LLM) are identical. The significance levels are color-coded into three levels: $\alpha = 0.1, 0.05, 0.01$. Gray cells (ns) indicate that there was no statistically significant difference.

The $P_{stability}$ between the DA mechanism-based market and the passive free negotiation market shows a significant difference across all experimental settings, demonstrating that the DA mechanism-based market is superior. Furthermore, in comparison with the active free negotiation market, the $P_{stability}$ of the DA mechanism-based market is significantly higher in most cases. Cases where no significant difference occurred were only for Gemini in the labor market with Preferences 1 and 2, and for GPT in high school entrance exams with Preference 2. However, this is due to the fact that $P_{stability}$ in these settings reached 0.99, 1.0, and 1.0, respectively, indicating that the active free negotiation market functioned in a desirable manner, rather than the DA mechanism failing to function appropriately.

\begin{figure}[htbp]
    \centering
    \includegraphics[width=0.8\linewidth]{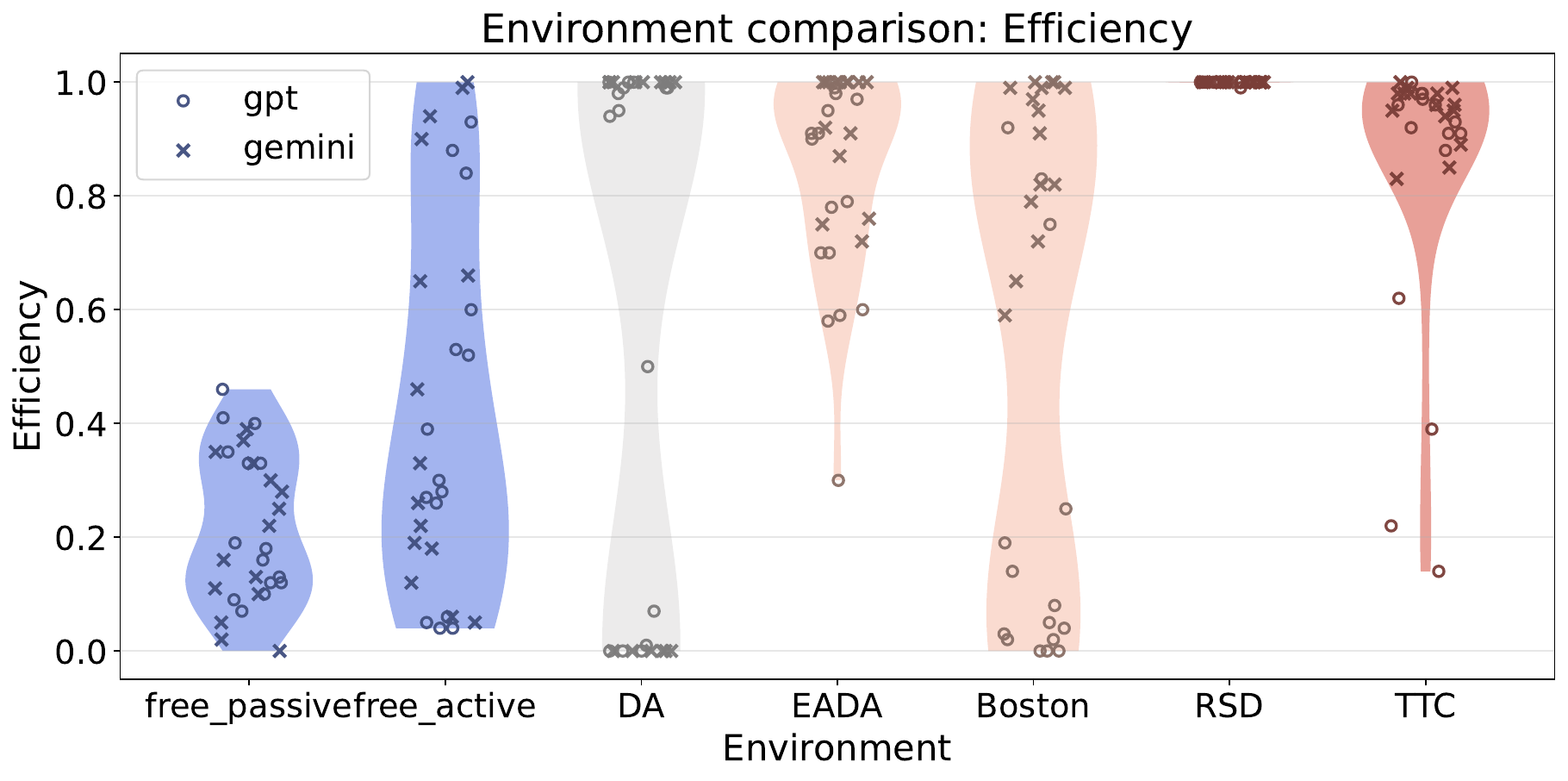}
    \caption{Efficiency among matching environments}
    \label{fig:violin_efficiency}
\end{figure}

Figure~\ref{fig:violin_efficiency} plots the proportion $P_{efficiency}$ of efficient matching results in the free negotiation markets and the mechanism-based markets for each experimental set (100 trials). Each element in Figure~\ref{fig:violin_efficiency}, other than the vertical axis, is common to Figure~\ref{fig:violin_stability} regarding stability, and like before, it is not a distribution map representing the probability of obtaining efficient results.

In the RSD mechanism-based market, which satisfies efficiency in all preference profiles, $P_{efficiency}$ was at least 99\% (market scenario: labor market, preference profile: Preference 5), whereas in the passive free negotiation environment, it was at most 46\% (market scenario: high school entrance exams, preference profile: Preference 3). Even in the active negotiation environment, the center of gravity of the overall plots is concentrated below 70\%, indicating that the RSD mechanism-based environment achieves more desirable results in terms of efficiency. On the other hand, although the TTC mechanism also satisfies efficiency in all preference profiles, results were obtained where $P_{efficiency}$ fell below 50\% in some condition settings (market scenario: labor market, preference profiles: Preferences 3, 4, and 5) in this environment. However, the center of gravity of the overall plots for the TTC mechanism-based environment is concentrated between 80\% and 100\%, suggesting that it has superior properties for achieving efficient matching compared to the passive or active free negotiation environments. Furthermore, in the EADA mechanism-based environment, the plots are concentrated between 60\% and 100\%, and although it is not a mechanism that strictly satisfies efficiency because it does not satisfy strategy-proofness, a tendency was observed for it to achieve a higher proportion of efficient matching results compared to either free negotiation environment.

\begin{figure}[htbp]
    \centering
    \includegraphics[width=0.8\linewidth]{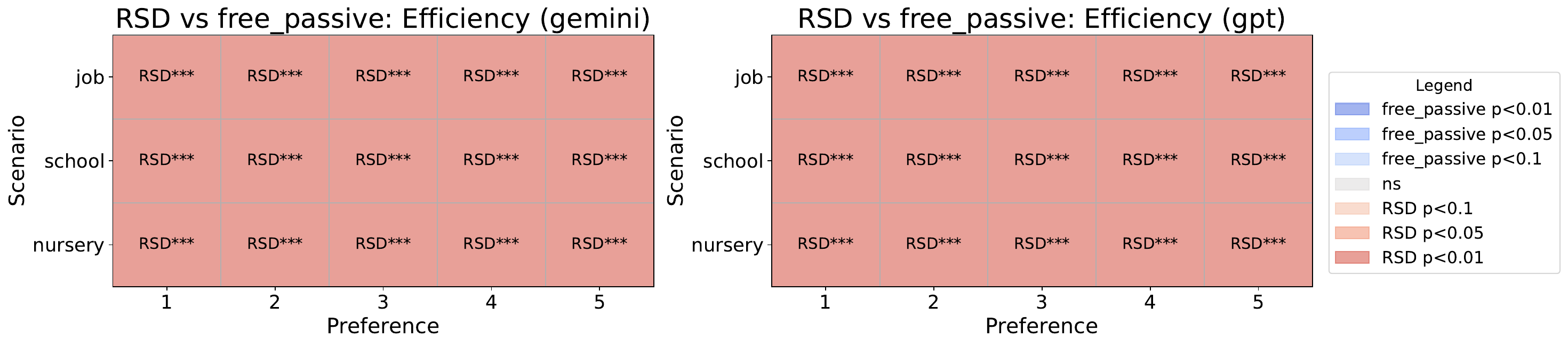}
    \includegraphics[width=0.8 \linewidth]{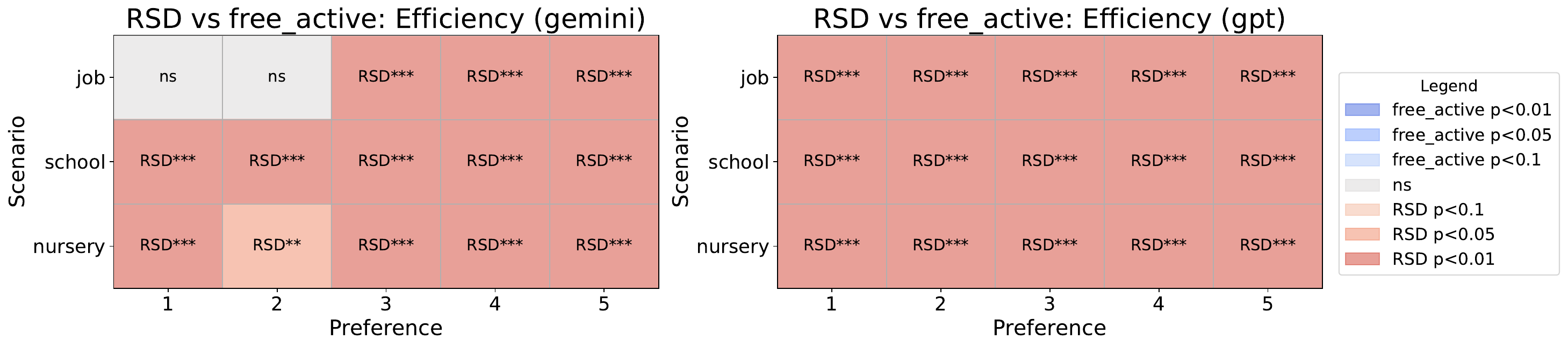}
    \caption{RSD vs. free negotiation markets (Efficiency)}
    \label{fig:efficiency_rsd_free}
    
    \centering
    \includegraphics[width=0.8\linewidth]{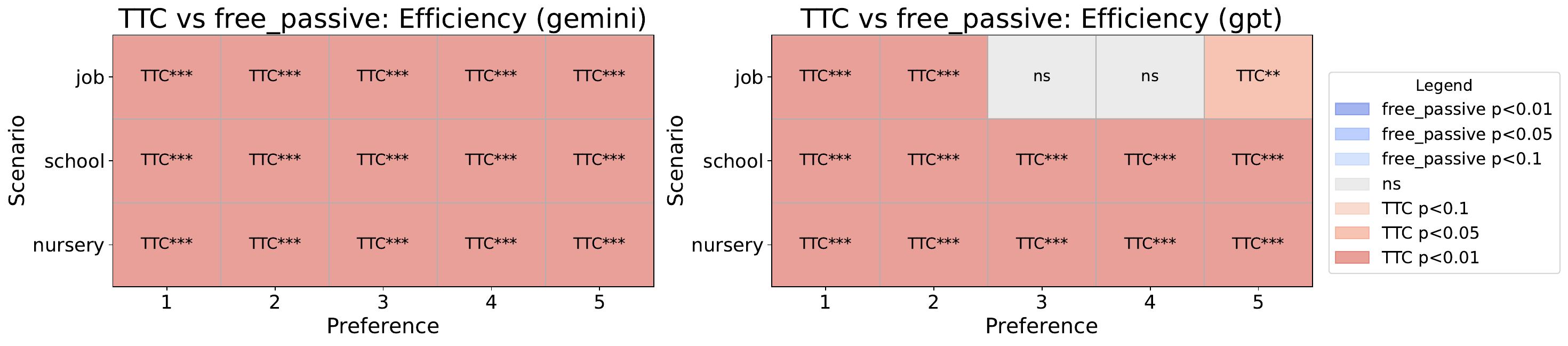}
    \includegraphics[width=0.8\linewidth]{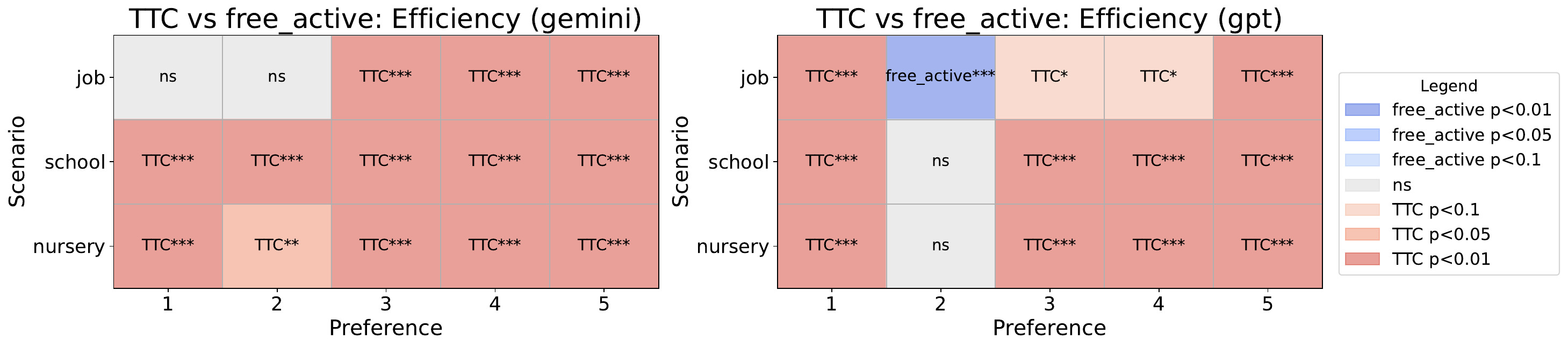}
    \caption{TTC vs. free negotiation markets (Efficiency)}
    \label{fig:efficiency_ttc_free}

    \centering
    \includegraphics[width=0.8\linewidth]{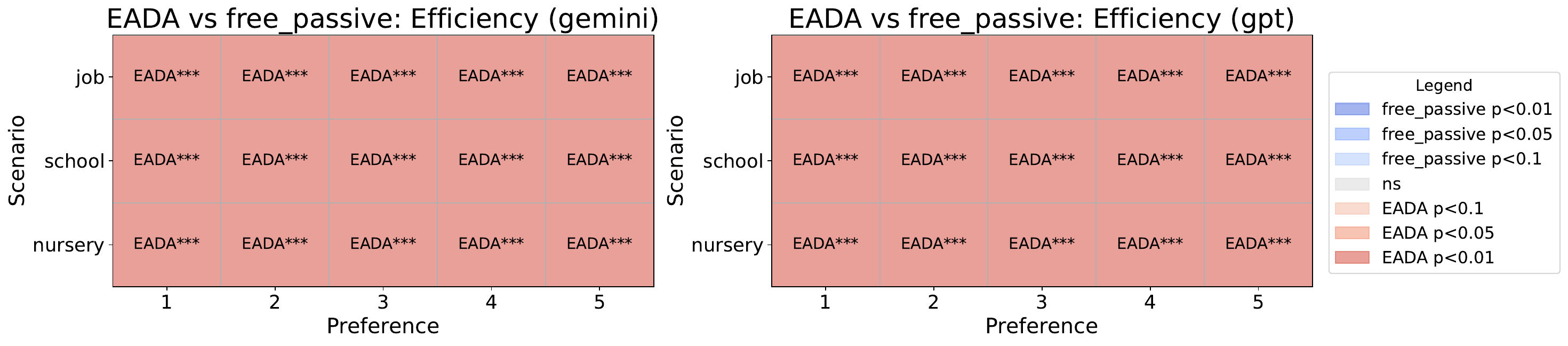}
    \includegraphics[width=0.8\linewidth]{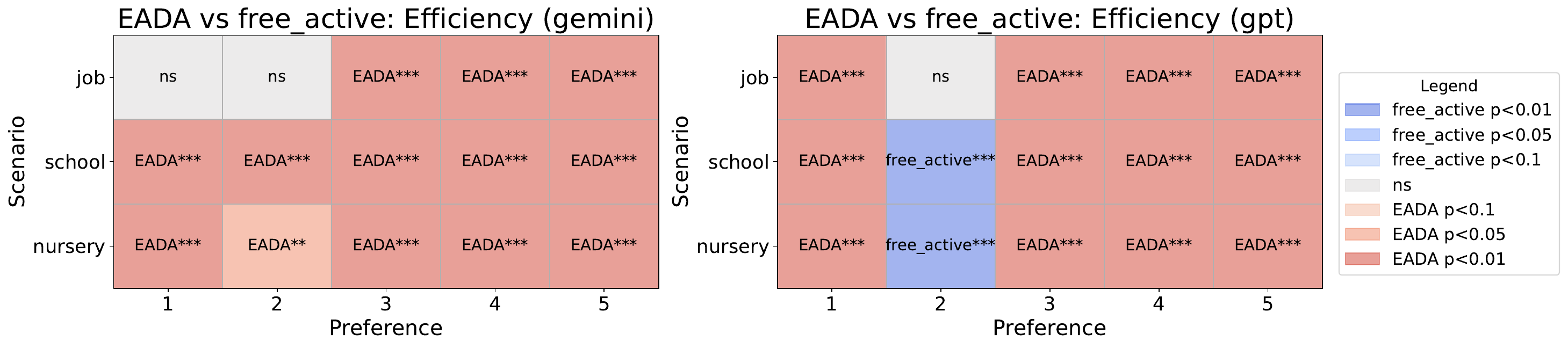}
    \caption{EADA vs. free negotiation markets (Efficiency)}
    \label{fig:efficiency_eada_free}
\end{figure}

Next, we verify whether the $P_{efficiency}$ of the efficiency-satisfying mechanism-based environment is statistically significantly higher than that of the free negotiation environment. Similar to stability, the Chi-square test is used to verify whether there is a significant difference in the population proportions yielding efficient matchings between the two matching environments. Figure~\ref{fig:efficiency_rsd_free} compares the RSD mechanism-based environment with the free negotiation environment, Figure~\ref{fig:efficiency_ttc_free} compares the TTC mechanism-based environment with the free negotiation environment, and Figure~\ref{fig:efficiency_eada_free} compares the EADA mechanism-based environment with the free negotiation environment.

From Figure~\ref{fig:efficiency_rsd_free}, it can be seen that the RSD mechanism has a higher $P_{efficiency}$ in most settings compared to the passive and active free negotiation environments. On the other hand, in the TTC and EADA mechanism-based environments, results showed that there was no significant difference, or the $P_{efficiency}$ of the active free negotiation environment was significantly higher, when the LLM model was GPT and the preference profile was Preference 2. In fact, the $P_{efficiency}$ of the active free negotiation environment when the model was GPT and the preference profile was Preference 2 recorded 0.84, 0.88, and 0.93 for the labor market, high school entrance exams, and nursery school selection, respectively. Regarding EADA, since strategy-proofness is not satisfied particularly in Preferences 4 and 5, it cannot necessarily be said to satisfy efficiency if all agents take rational strategies theoretically. However, as shown in Figure~\ref{fig:efficiency_eada_free}, even in the corresponding preferences, it was found that it is more likely to yield more efficient matching results compared to the free negotiation environment.

\subsection{Mechanism-Based Markets}\label{subsec:res_env2}
\begin{figure}[htbp]
    \centering
    \includegraphics[width=0.8\linewidth]{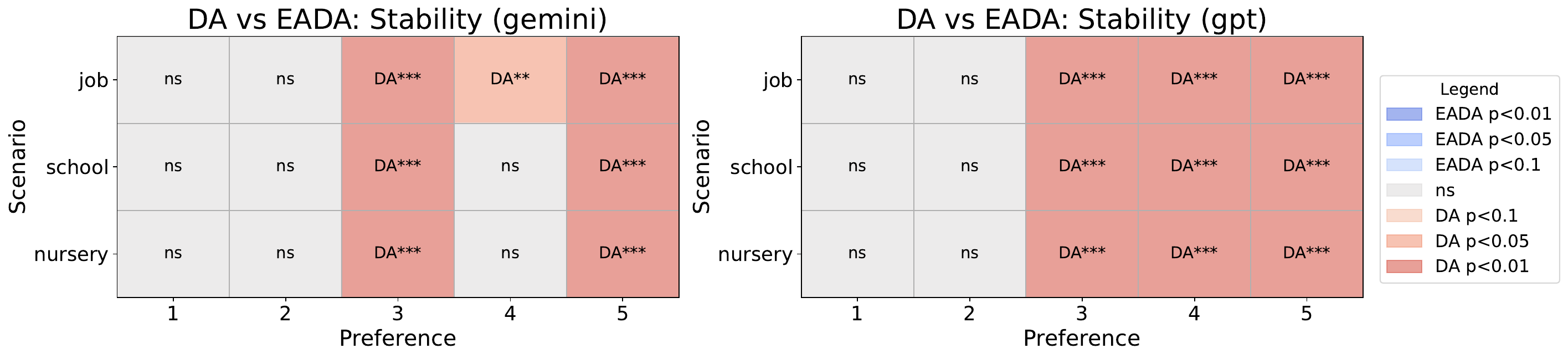}
    \includegraphics[width=0.8\linewidth]{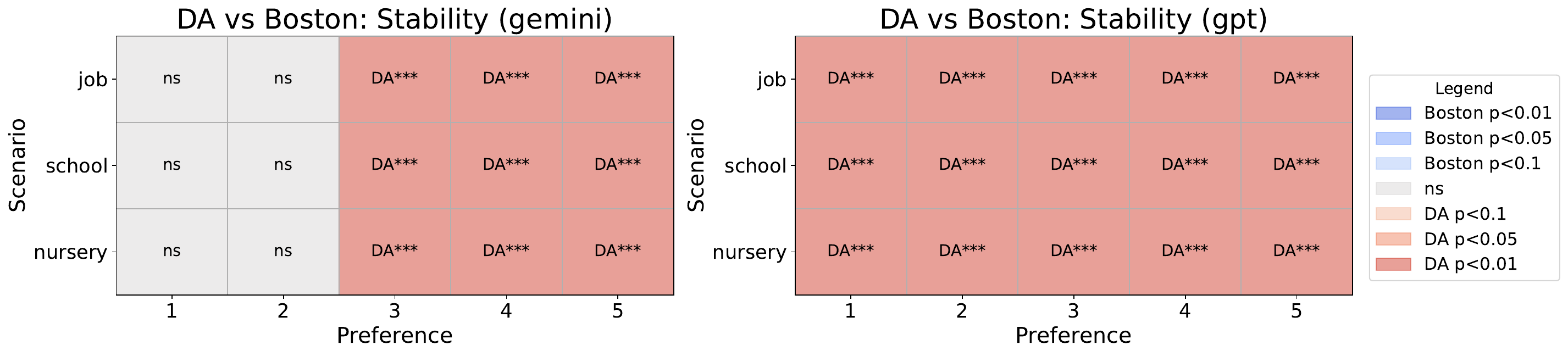}
    \includegraphics[width=0.8\linewidth]{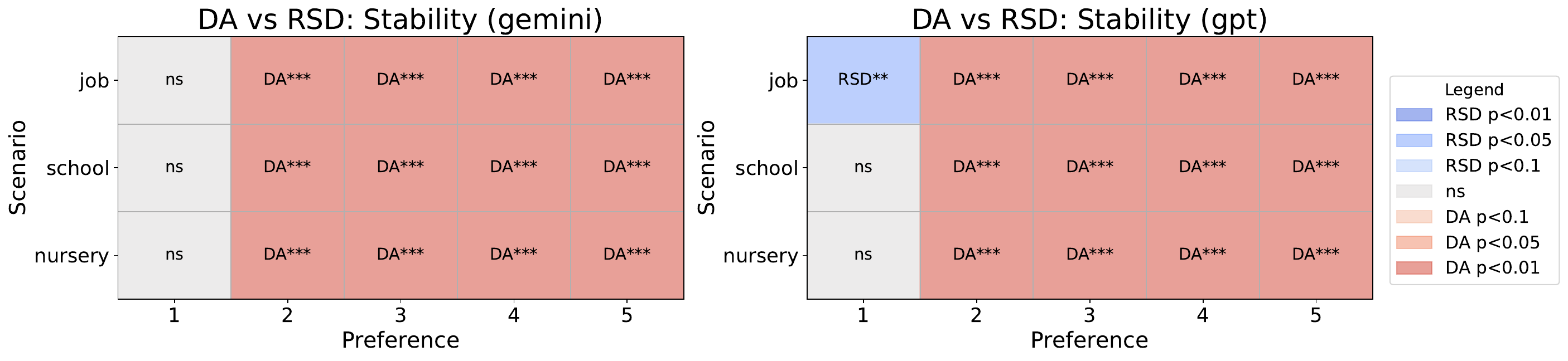}
    \includegraphics[width=0.8\linewidth]{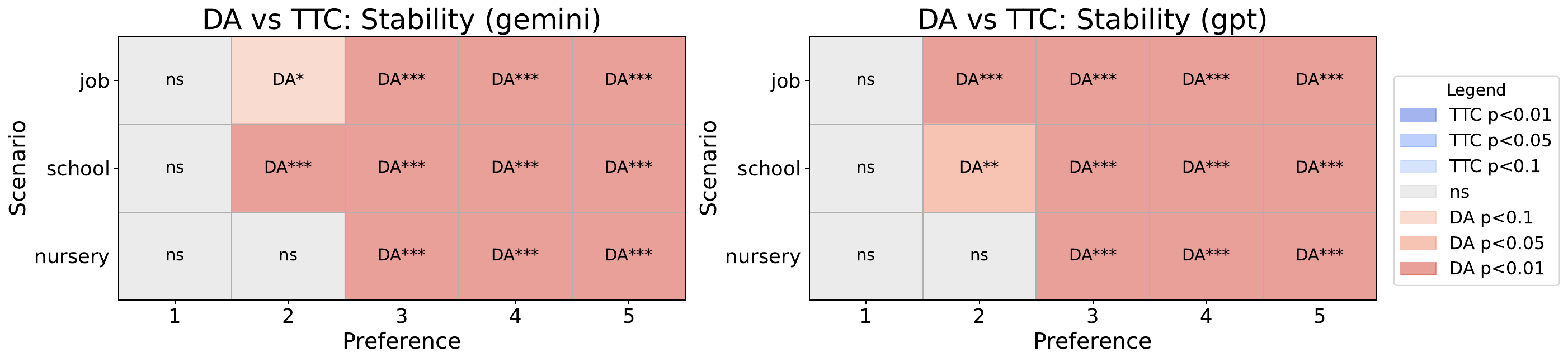}
    \caption{DA vs. other mechanisms (Stability)}
    \label{fig:stability_da_others}
\end{figure}

By comparing the stability of matching results among different mechanism-based environments, we verify whether the mechanisms function consistently with theory. Referring to Figure~\ref{fig:stability_da_others}, it can be seen that the DA mechanism, which theoretically satisfies stability, has a higher lower bound for the proportion $P_{stability}$ of stable matching results (86\%) compared to other mechanism-based environments.Using the Chi-square test, we verify whether the stability-satisfying DA mechanism yields a significantly higher $P_{stability}$ than other mechanisms. Looking at Figure~\ref{fig:stability_da_others}, it is clear that the DA mechanism leads to a higher proportion of stable matchings than other mechanisms. In particular, it can be seen that the $P_{stability}$ of DA is significantly higher than other mechanisms in Preferences 3, 4, and 5, where other mechanisms do not satisfy stability.

\begin{figure}[htbp]
    \centering
    \includegraphics[width=0.8\linewidth]{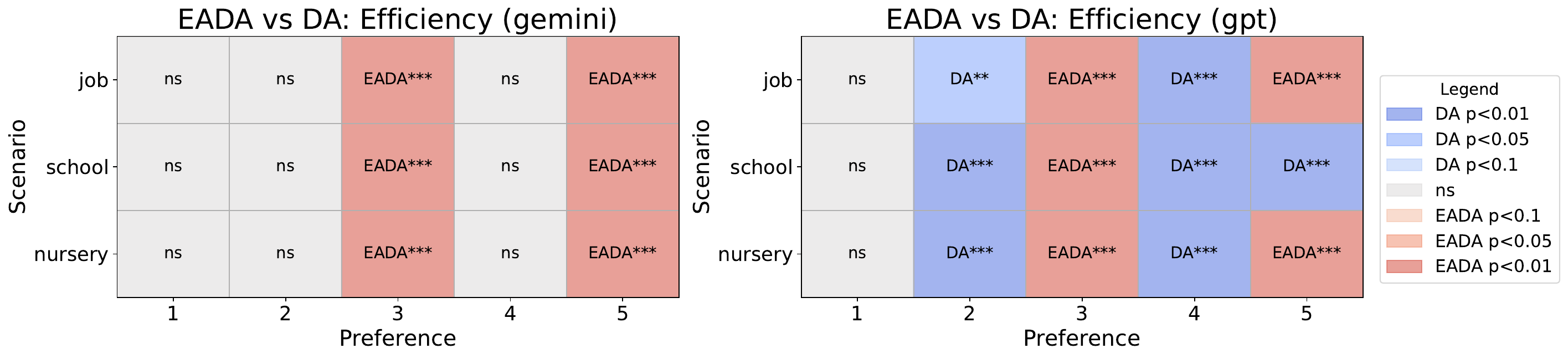}
    \includegraphics[width=0.8\linewidth]{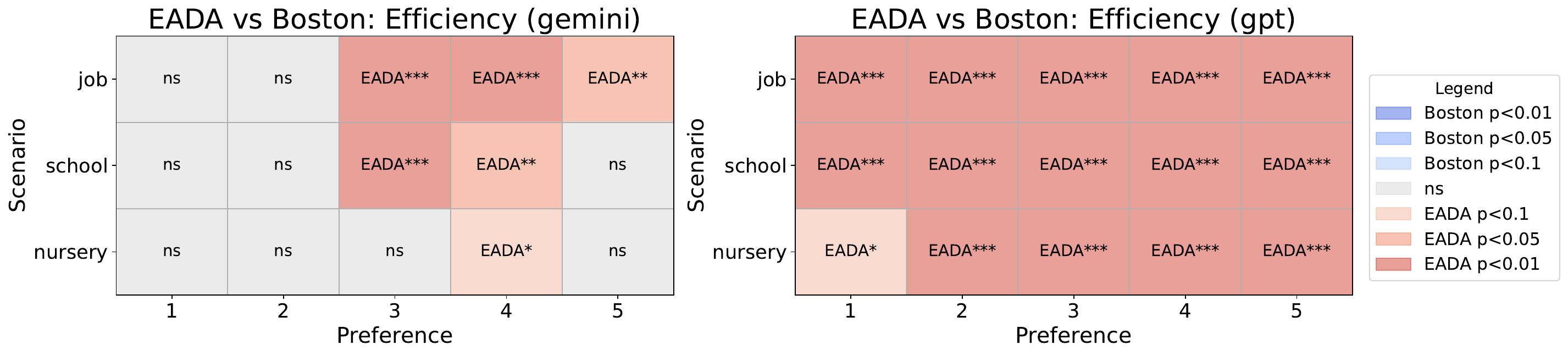}
    \caption{EADA vs. other mechanisms}
    
    \includegraphics[width=0.8\linewidth]{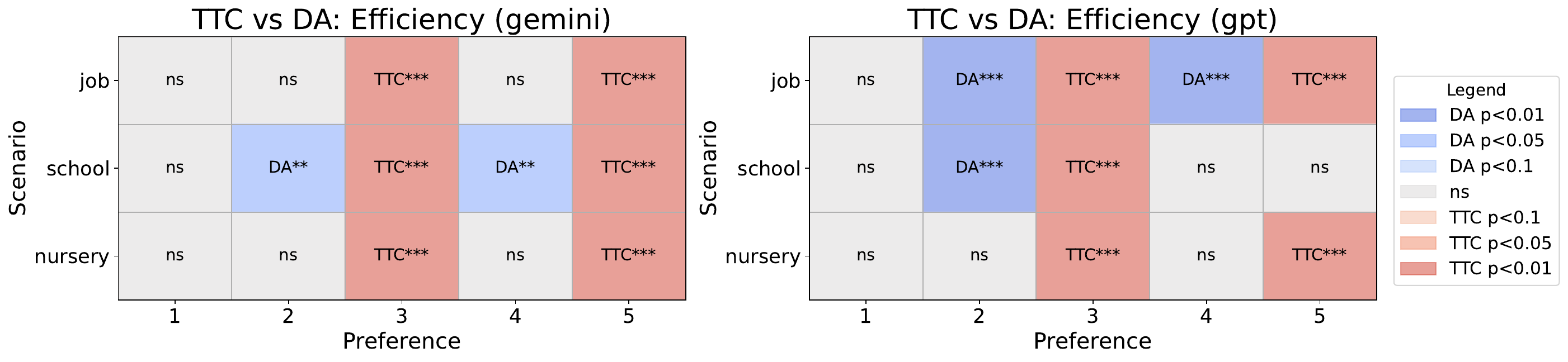}
    \includegraphics[width=0.8\linewidth]{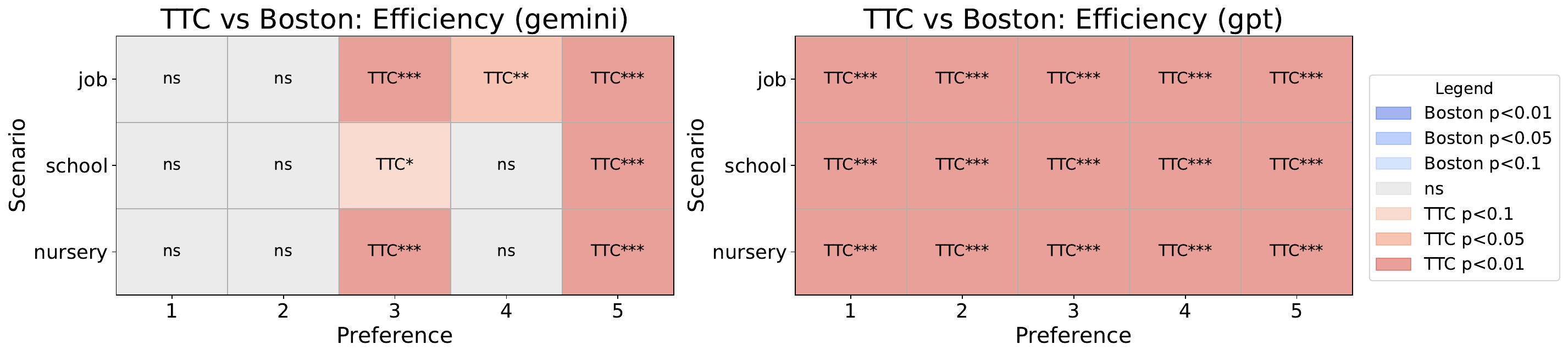}
    \caption{TTC vs. other mechanisms}
    \label{fig:efficiency_eada_ttc_others}
\end{figure}

By comparing the stability of matching results among mechanism-based environments, we verify whether the mechanisms function consistently with theory. Referring to Figure~\ref{fig:efficiency_eada_ttc_others}, it is clear that the RSD mechanism, which theoretically satisfies efficiency, has a higher lower bound for the proportion $P_{efficiency}$ of efficient matching results (99\%) compared to other mechanism-based environments. On the other hand, in the EADA and TTC mechanism-based environments, which also theoretically satisfy efficiency for all preferences, experimental sets with low $P_{efficiency}$ are occasionally seen.Therefore, using the Chi-square test, we verify whether the efficiency-satisfying EADA and TTC mechanisms yield a significantly higher $P_{efficiency}$ than the DA and Boston mechanisms, which do not satisfy efficiency. Looking at Figure~\ref{fig:efficiency_eada_ttc_others}, both EADA and TTC have significantly higher $P_{efficiency}$ when compared with Boston, and are basically superior in terms of efficiency. Also, when compared with DA, EADA and TTC basically have significantly higher $P_{efficiency}$ only in Preferences 3 and 5, which correspond to the preferences where DA does not satisfy efficiency.

\begin{figure}[htbp]
    \centering
    \includegraphics[width=0.8\linewidth]{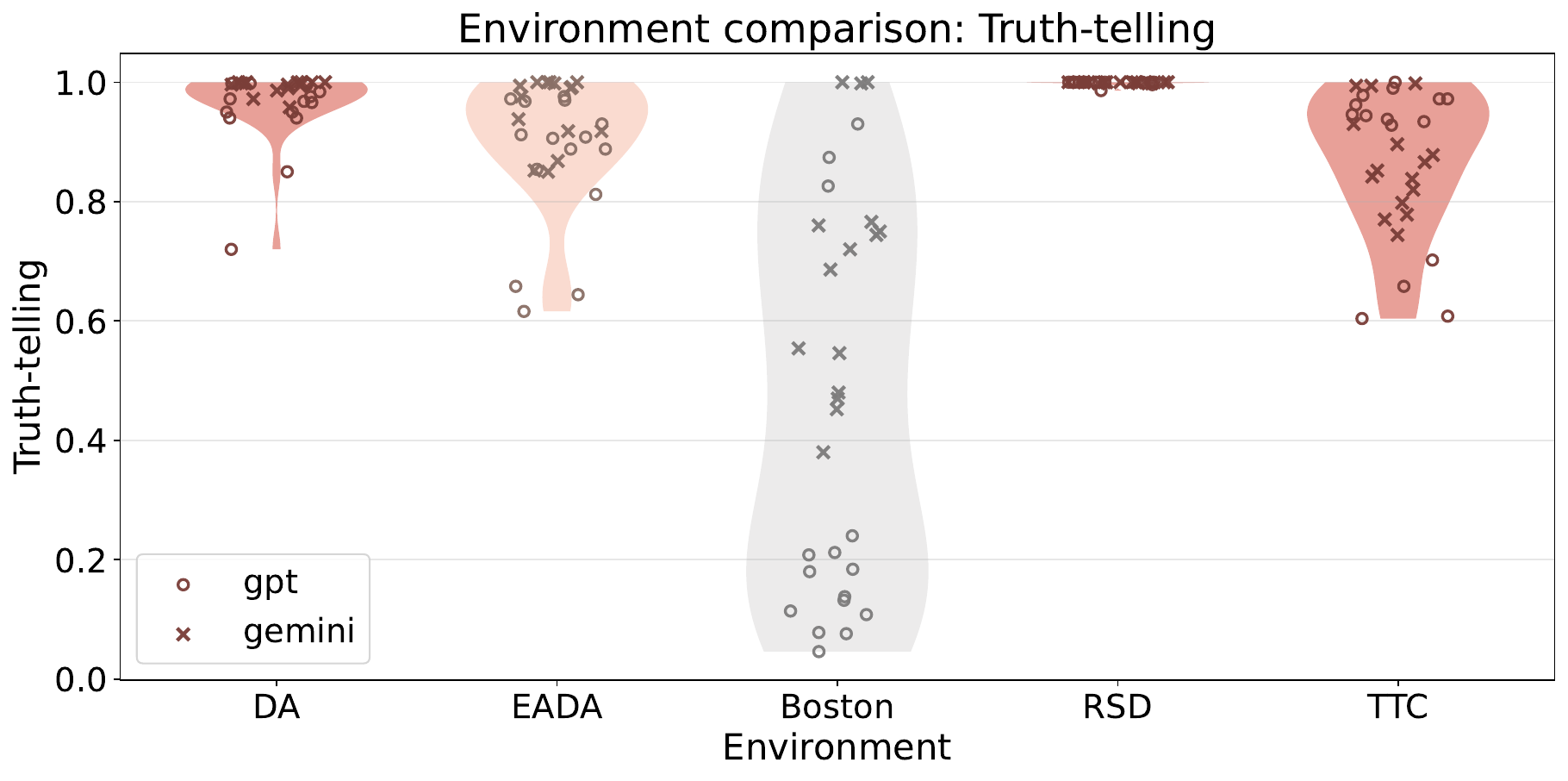}
    \caption{Truth-telling rate among matching environments}
    \label{fig:violin_truth_telling}
\end{figure}

Figure~\ref{fig:violin_truth_telling} plots the truth-telling rate $TR$ in the mechanism-based environment for each experimental set (100 trials). Each element in Figure~\ref{fig:violin_truth_telling}, other than the vertical axis, is common to Figures~\ref{fig:violin_stability} and~\ref{fig:violin_efficiency} regarding stability and efficiency, and like before, it is not a distribution map representing the probability of truth-telling.

DA, RSD, and TTC are the ones that satisfy strategy-proofness for all preference profiles, but results different from the theoretical properties were obtained. Regarding RSD and DA, most showed truth-telling rates close to 100\%, consistently with theory. It can also be seen that for Boston, many strategic misreports were observed, consistently with theory. Regarding EADA, while the center of gravity was slightly lower compared to RSD and DA, the TTC mechanism, which should satisfy strategy-proofness, showed a truth-telling rate of a similar level. Particularly in experiments using Gemini, the center of gravity of the truth-telling rate for TTC was between 70\% and 90\%, falling to a lower value even compared to EADA.

Next, using the Chi-square test, we verify whether the mechanisms satisfying strategy-proofness yield significantly higher truth-telling rates than mechanisms that do not. It is clear from Figure~\ref{fig:violin_truth_telling} that the truth-telling rates of RSD and DA are similarly high, and the truth-telling rate of Boston is prominently low. Therefore, in this section, we compare the truth-telling rates of DA, EADA, and TTC for each experimental condition (market scenario, preference profile, and LLM model).

\begin{figure}[htbp]
    \centering
    \includegraphics[width=0.8\linewidth]{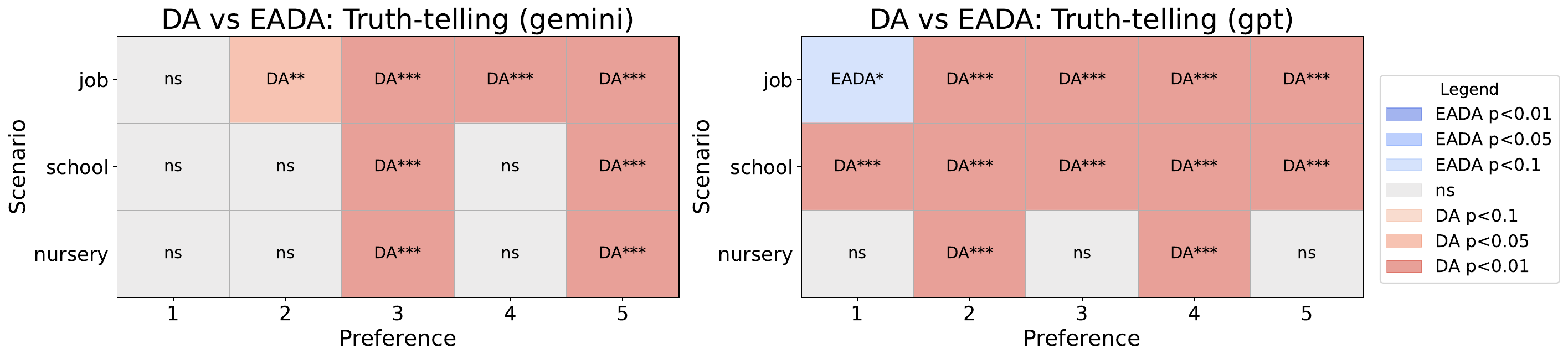}
    \includegraphics[width=0.8\linewidth]{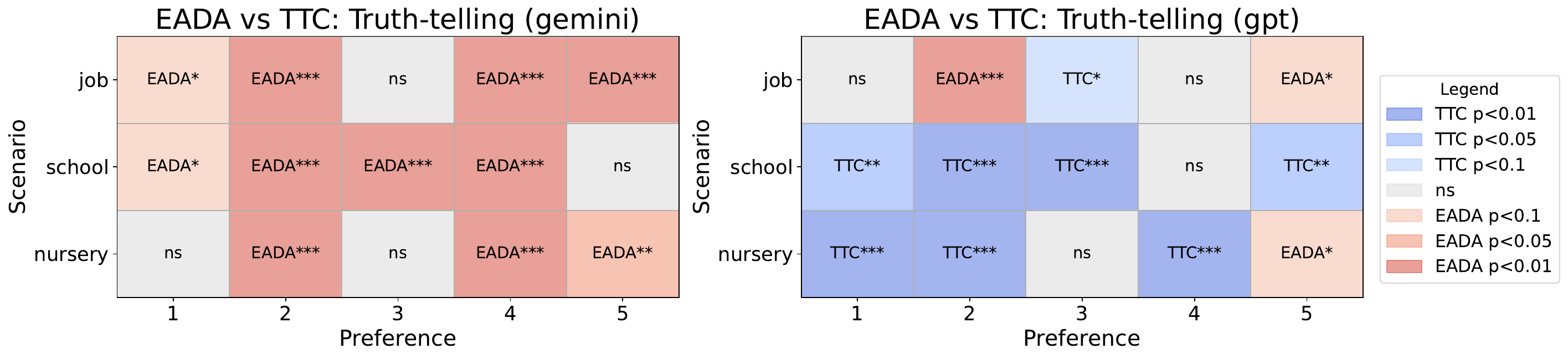}
    \includegraphics[width=0.8\linewidth]{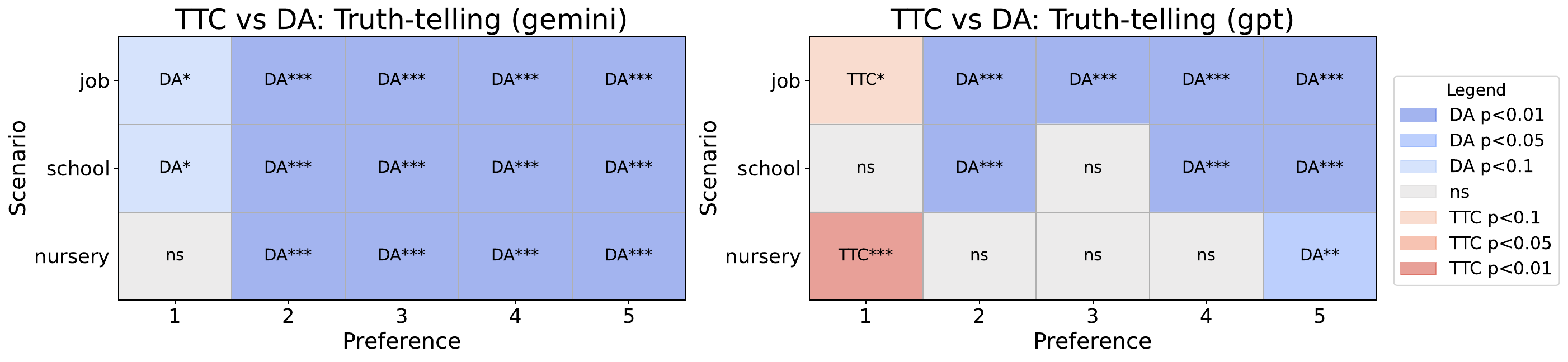}
    \caption{DA vs. EADA vs. TTC}
    \label{fig:truth_telling_da_eada_ttc}
\end{figure}

From Figure~\ref{fig:truth_telling_da_eada_ttc}, the comparison between DA and EADA showed that the truth-telling rate for DA was basically significantly higher than the truth-telling rate for EADA, except for the condition of GPT in the labor market with Preference 1. Furthermore, in the comparison between TTC and DA, the truth-telling rate for DA was basically significantly higher than the truth-telling rate for TTC, except for the condition of GPT with Preference 1. From the above, it can be seen that among these three, the truth-telling rate for the DA mechanism-based environment is significantly higher. The comparison between EADA and TTC yielded different results depending on the LLM model. For Gemini, the truth-telling rate was basically higher in EADA, whereas for GPT, the truth-telling rate was basically higher in TTC.

\subsection{Market Scenarios}\label{subsec:res_scn}
\begin{figure}[htbp]
    \centering
    \includegraphics[width=0.8\linewidth]{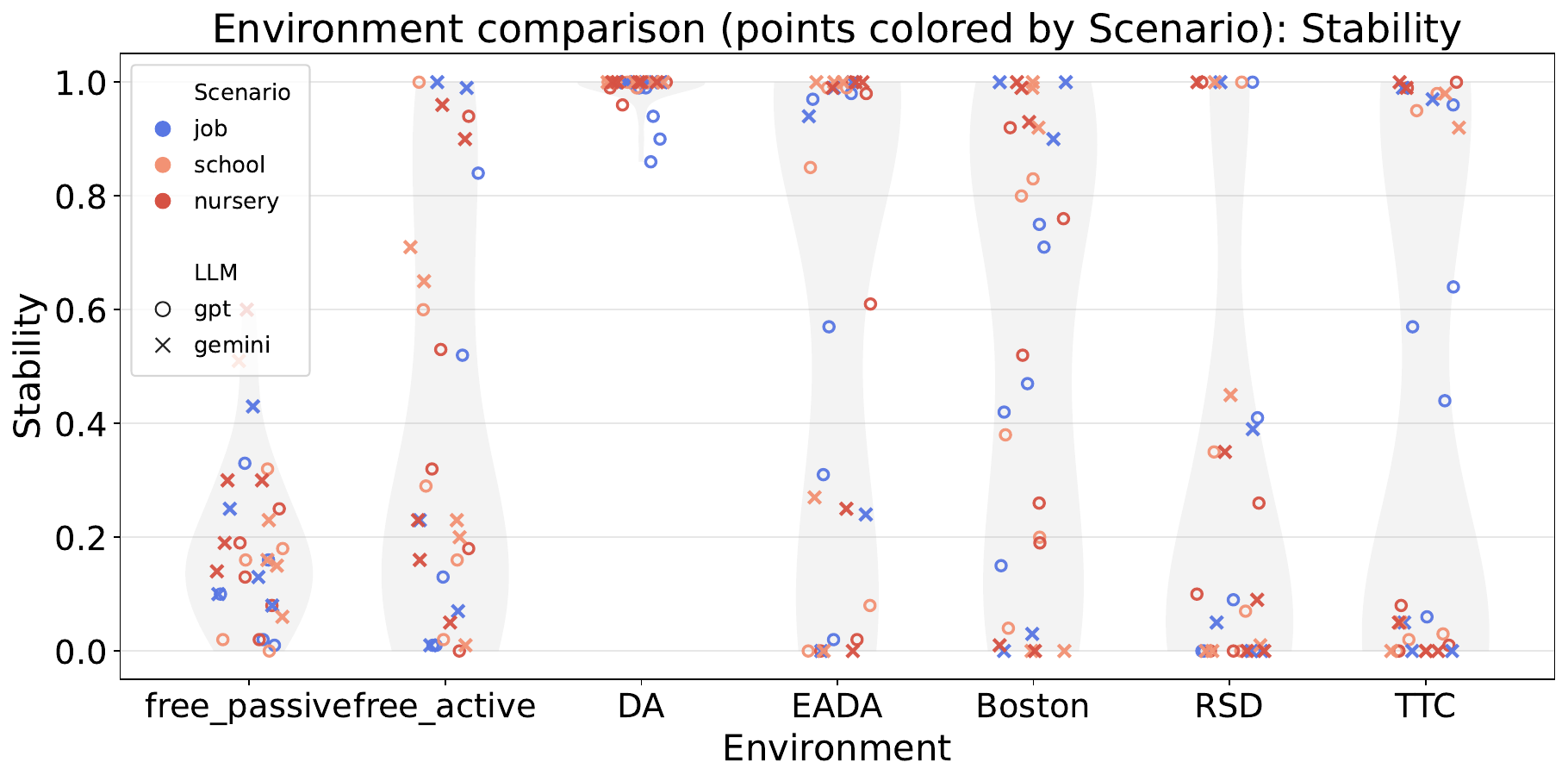}   
    \includegraphics[width=0.8\linewidth]{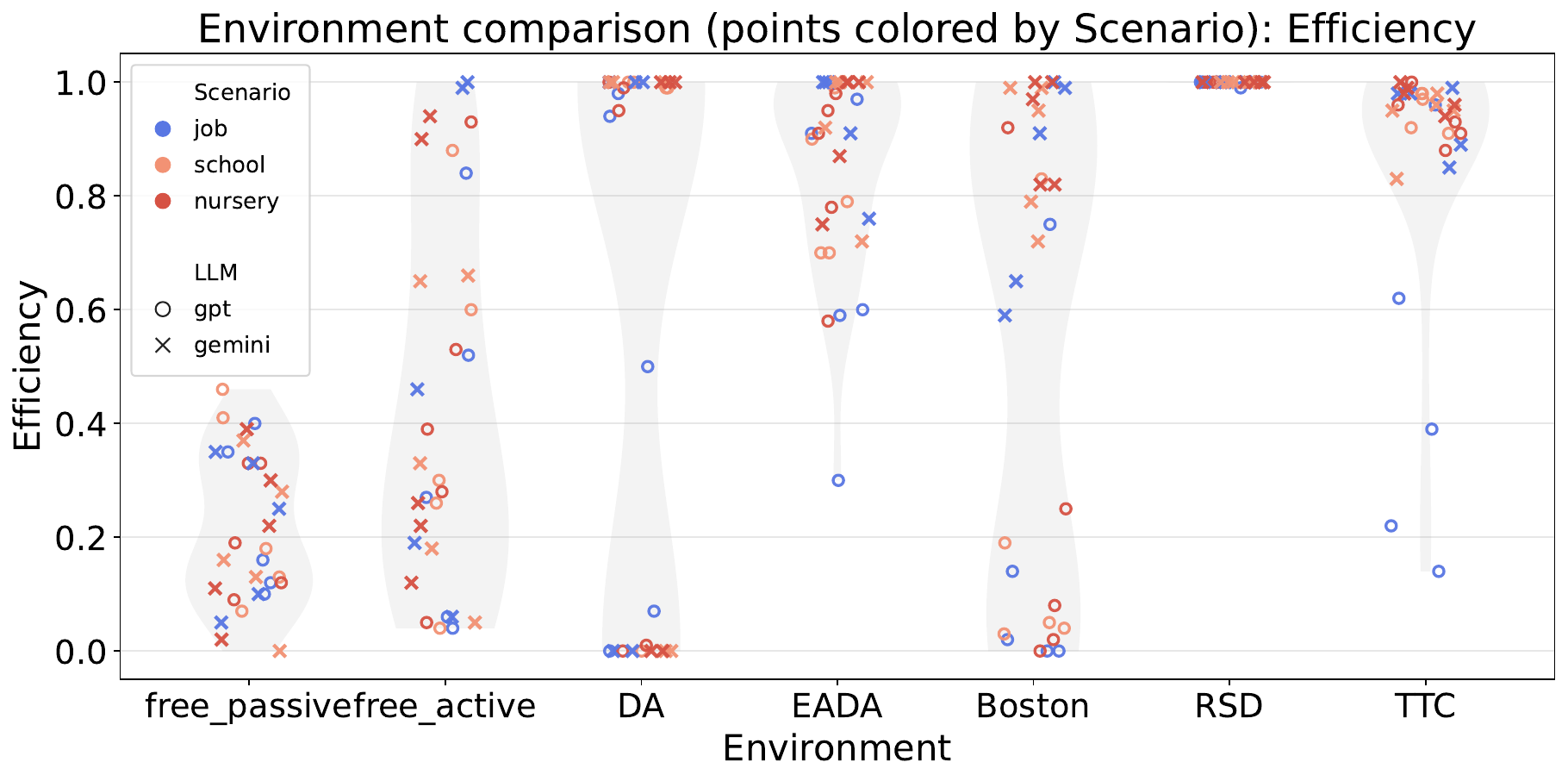}
    \includegraphics[width=0.8\linewidth]{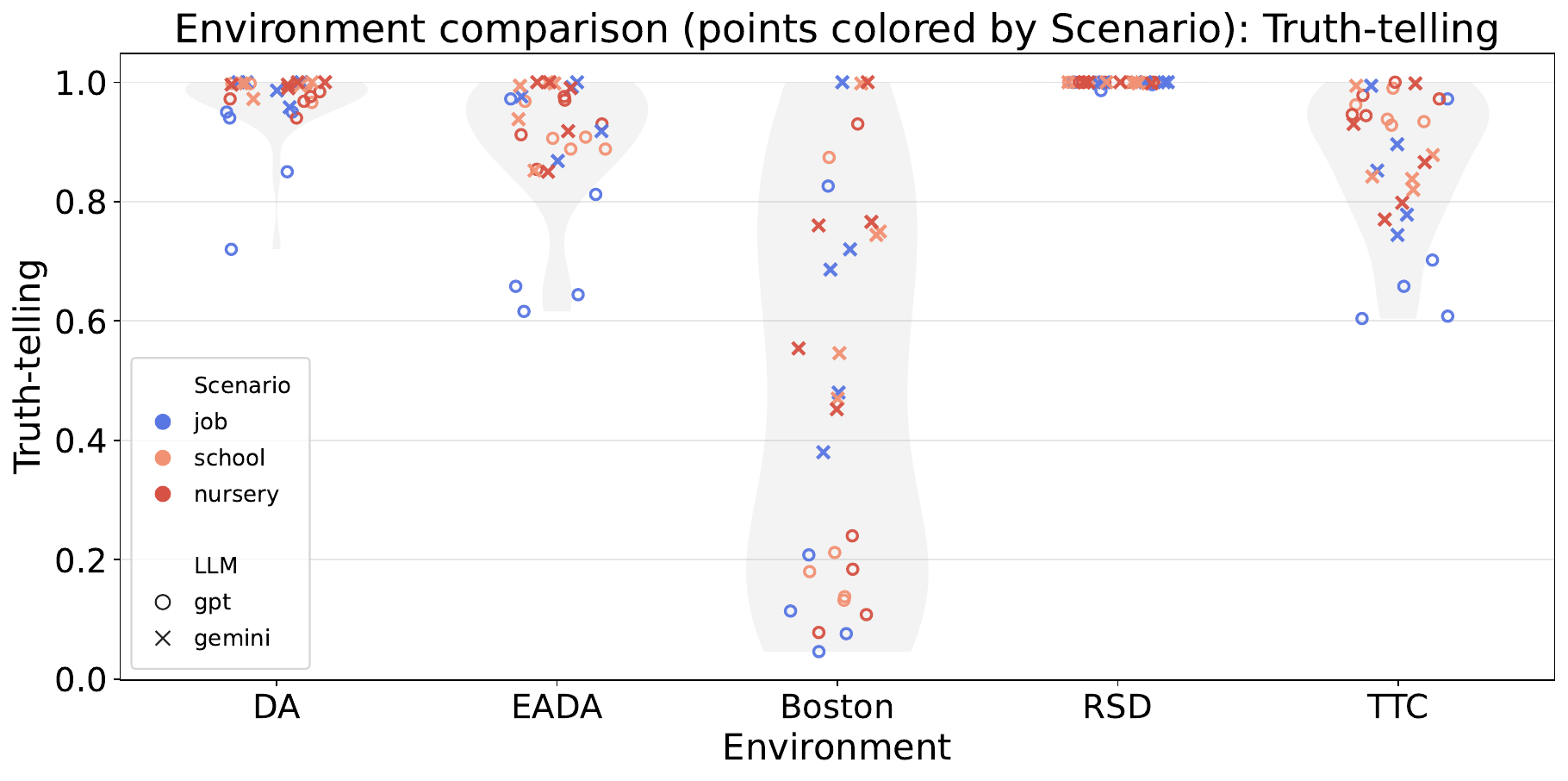}
    \caption{Market scenarios (Stability, Efficiency, Truth-telling rate)}
    \label{fig:violin_scenario}
\end{figure}

In this section, we verify whether differences in market scenario settings cause differences in the stability and efficiency of matching results, as well as the truth-telling rates for the mechanisms. Figure~\ref{fig:violin_scenario} shows the results of color-coding Figures~\ref{fig:violin_stability} to ~\ref{fig:violin_truth_telling} by scenario. The labor market, high school entrance exams, and nursery school selection scenarios correspond to Scenario job, school, and nursery, respectively. Regarding stability, it can be seen that the plots for each scenario are distributed evenly in basically all matching environments except for DA. On the other hand, in the DA mechanism-based environment, plots exist in the 85\% to 95\% range for the labor market scenario with GPT, indicating that stability is reduced compared to other scenarios. This peculiarity of the labor market scenario with GPT can also be observed in efficiency and truth-telling rates. Regarding efficiency, a reduction in efficiency is seen in the EADA and TTC matching environments compared to other scenarios. Similarly, regarding truth-telling, a reduction in the truth-telling rate is seen in the DA, EADA, and TTC matching environments compared to other scenarios.

Based on the above observations, we further conducted a statistical analysis of whether differences occur in the average values when using all data among the three scenarios: labor market, high school entrance exams, and nursery school selection. In this statistical analysis, the p-values were calculated using the Wilcoxon signed-rank test~\cite{Demvsar-2006,Hodges-1963,Wilcoxon-1945}, assuming that common matching environments, preference profiles, and LLM models correspond in the significance tests between scenarios.

The results are summarized in Table~\ref{tab:result_scenario}. Regarding stability, there was a significant but relatively weak tendency for the nursery school selection scenario to outperform the high school entrance exam scenario. In contrast, for both efficiency and truth-telling rates, there was a commonly significant and relatively strong tendency for the high school entrance exam and nursery school selection scenarios to outperform the labor market scenario.

\begin{table}[htbp]
  \centering
  \caption{Statistical analysis of market scenarios (Stability, Efficiency, Truth-telling rate)}
  \label{tab:result_scenario}
  \begin{threeparttable}
    \begin{tabular}{llll}
      \toprule
                        & Stability             & Efficiency            & Truth-telling Rate    \\ \midrule
      job               & 0.488 ($\pm 0.414$)      & 0.603 ($\pm 0.392$)      & 0.582 ($\pm 0.423$)      \\
      school            & 0.483 ($\pm 0.430$)      & 0.669 ($\pm 0.381$)      & 0.620 ($\pm 0.443$)      \\
      nursery           & 0.497 ($\pm 0.428$)      & 0.670 ($\pm 0.390$)      & 0.621 ($\pm 0.445$)      \\ \midrule
      job $-$ school    & 0.005                 & -0.066***             & -0.038***             \\
      school $-$ nursery& -0.014*               & -0.001                & -0.001                \\
      nursery $-$ job   & 0.009                 & 0.067***              & 0.039***              \\
      \bottomrule
    \end{tabular}
    \begin{tablenotes}[flushleft]
      \footnotesize
      \item Notes: *** $p<0.01$; ** $p<0.05$; * $p<0.1$.
    \end{tablenotes}
  \end{threeparttable}
\end{table}

\subsection{Interpretation of Hypothesis Testing Results}\label{subsec:res_interpret}
Results supporting the hypotheses were obtained from the stability and efficiency of matching results in the mechanism-based environment compared to the free negotiation environment as a baseline. Regarding the stability of matching results, Figures~\ref{fig:violin_stability} and~\ref{fig:stability_da_free} showed that the stability-satisfying DA mechanism achieved a significantly higher proportion of stable matchings under the vast majority of situations compared to passive and active free negotiation. Regarding the efficiency of matching results, Figures~\ref{fig:violin_efficiency}, ~\ref{fig:efficiency_rsd_free}, ~\ref{fig:efficiency_ttc_free}, and~\ref{fig:efficiency_eada_free} showed that the efficiency-satisfying RSD and TTC mechanisms achieved a significantly higher proportion of efficient matchings under the vast majority of situations compared to passive and active free negotiation. Furthermore, the EADA mechanism, for which high efficiency was observed in empirical experiments targeting human subjects~\cite{Cerrone-2024}, was shown to achieve a significantly higher proportion of efficient matchings under the vast majority of situations compared to passive and active free negotiation. Note that in the active negotiation environment, when the preference profile was Preference 2, the proportion of obtaining efficient matching results was relatively high at an average of 88\%. Therefore, although limited, experimental results were also obtained where the active free negotiation environment achieved a higher proportion of efficient matching results than these mechanism-based environments when many agents made misreports in TTC or EADA.

\begin{result}
The mechanism-based environment that theoretically satisfies stability yields a higher proportion of stable matching results compared to the free negotiation environment.
\end{result}

\begin{result}
The mechanism-based environment that theoretically satisfies Pareto efficiency yields a higher proportion of Pareto efficient matching results compared to the free negotiation environment.
\end{result}

While results supporting the hypotheses were obtained from the stability and efficiency of matching results among mechanism-based environments, some results conflicting with the hypotheses were also obtained regarding the truth-telling rate. Regarding the stability of matching results, Figures~\ref{fig:violin_stability} and~\ref{fig:stability_da_others} showed that the stability-satisfying DA mechanism achieved a significantly higher proportion of stable matchings compared to other mechanisms under the vast majority of situations. In particular, the DA mechanism had a high probability of realizing stable matchings even for preference profiles where stability would decrease with other matching mechanisms. Regarding the efficiency of matching results, Figure~\ref{fig:violin_efficiency} showed that the efficiency-satisfying RSD mechanism achieved a higher proportion of efficient matchings compared to Boston and DA, which do not satisfy efficiency, under the vast majority of situations. On the other hand, from Figure~\ref{fig:efficiency_eada_ttc_others}, results were obtained indicating that for TTC and EADA mechanisms, the proportion of obtaining efficient matchings was significantly lower than DA, which should not satisfy efficiency, when the preference profile was Preference 2 or 4. It can be interpreted that in these preference profiles, although a matching satisfying both efficiency and stability would originally be realized in all of DA, EADA, and TTC if all agents reported truthfully, agents making misreports existed in EADA and TTC, and as a result, the proportion of achieving efficiency was reduced compared to DA. Note that it can be said that Pareto improvements over the matching results of the DA mechanism were achieved by using EADA or TTC in preference profiles where efficiency cannot be realized by DA.

Regarding the truth-telling rate, Figure 12 showed that the RSD and DA mechanisms, which satisfy one-sided strategy-proofness for the proposing side, achieved higher truth-telling rates compared to mechanisms that do not satisfy strategy-proofness under the vast majority of situations.

Regarding the truth-telling rate, Figure~\ref{fig:violin_truth_telling} showed that the RSD and DA mechanisms, which satisfy one-sided strategy-proofness for the proposing side, achieved higher truth-telling rates compared to mechanisms that do not satisfy strategy-proofness under the vast majority of situations.
On the other hand, when comparing the truth-telling rates of the TTC mechanism, which satisfies strategy-proofness, and the EADA mechanism, which does not, results were obtained where the difference depended on the LLM model. In connection with the interpretation of this result, an additional exploratory analysis was conducted in Section~\ref{sec:disc}.

\begin{result}
The mechanism-based environment that theoretically satisfies stability yields a higher proportion of stable matching results compared to mechanism-based environments that theoretically do not satisfy stability.
\end{result}

\begin{result}
\leavevmode
\begin{itemize}
    \item The RSD mechanism-based environment, which theoretically satisfies Pareto efficiency, yields a higher proportion of Pareto efficient matching results compared to mechanism-based environments that theoretically do not satisfy Pareto efficiency.
    
    \item The TTC mechanism-based environment, which theoretically satisfies Pareto efficiency, yields a higher proportion of Pareto efficient matching results in preference profiles where the matching results of the DA mechanism can be Pareto improved, compared to other mechanism-based environments.
    
    \item The EADA mechanism-based environment, which theoretically satisfies Pareto efficiency when everyone reports truthfully, yields a higher proportion of Pareto efficient matching results in preference profiles where the matching results of the DA mechanism can be Pareto improved, compared to other mechanism-based environments.
\end{itemize}
\end{result}

\begin{result}
\leavevmode
\begin{itemize}
    \item The RSD and DA mechanism-based environments, which theoretically satisfy strategy-proofness, show higher truth-telling rates compared to mechanism-based environments that theoretically do not satisfy strategy-proofness.
    
    \item The TTC mechanism-based environment, which theoretically satisfies strategy-proofness, does not necessarily show a higher truth-telling rate when compared to the EADA mechanism-based environment.
\end{itemize}
\end{result}

Finally, results partially supporting the hypothesis were obtained from the comparison of truth-telling rates in the DA and EADA mechanism-based environments of this study with a subject experiment targeting humans~\cite{Cerrone-2024}. In the subject experiment targeting humans, the truth-telling rate for DA was approximately 55\% and the truth-telling rate for EADA was approximately 70\% when the preference profiles were Preference 4 and Preference 5. On the other hand, in Preference 4 and Preference 5 of the DA and EADA mechanism-based environments in this study, the truth-telling rate for DA recorded an average of 95\% and the truth-telling rate for EADA recorded an average of 87\%. Regarding DA, it can be said that LLM agents showed a higher truth-telling rate than humans. On the other hand, regarding EADA, contrary to the hypothesis, LLM agents showed a higher truth-telling rate than humans, similar to DA. However, the point that the truth-telling rate was higher for DA than EADA in the mechanism-based environments of this study—whereas the truth-telling rate was lower for DA than EADA in the subject experiment targeting humans—suggests that LLM agents can make decisions more rationally and strategically than humans even under the environment of matching mechanisms. At the very least, this result is consistent with the hypothesis based on the theoretical model in the study by Cerrone et al~\cite{Cerrone-2024}.

In addition, although it is for reference only because the control of mechanism instructions is not established, while the truth-telling rate in subject experiments targeting humans for the RSD mechanism~\cite{Kloosterman-2020,Li-2017} satisfying strategy-proofness is 40\% to 60\%, 99\% to 100\% was recorded in this study, which can also be said to have yielded results supporting the hypothesis.

\begin{result}
In the DA mechanism-based environment satisfying strategy-proofness, the truth-telling rate is higher than in empirical experiments with human subjects.
\end{result}

\begin{result}
Contrary to the hypothesis, LLM agents in the EADA mechanism-based environment show a higher truth-telling rate than human subjects in comparable experimental settings.
\end{result}

\begin{result}
While human subjects show a lower truth-telling rate for the DA mechanism than for the EADA mechanism, LLM agents show a higher truth-telling rate for the DA mechanism than for the EADA mechanism.
\end{result}

\section{Discussion}\label{sec:disc}
\subsection{Exploratory Analysis of Strategic Misreporting}\label{subsec:disc_add}
\begin{figure}[htbp]
    \centering
    \includegraphics[width=0.8\linewidth]{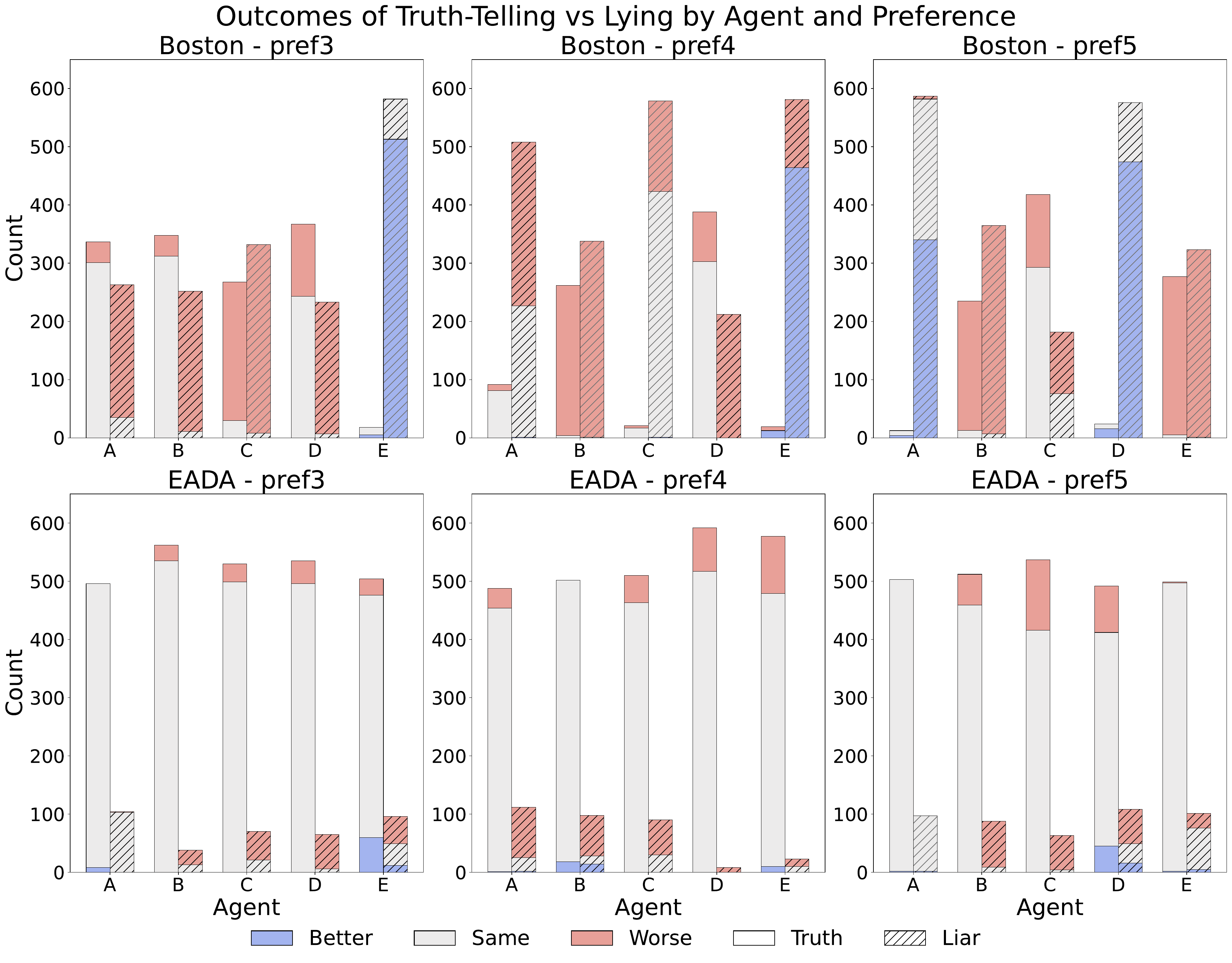}
    \caption{Outcomes of Strategic behaviors}
    \label{fig:truth_telling_details}
\end{figure}

In addition to the pre-registered analysis conducted in the previous chapter, we performed the following exploratory analysis to verify whether misreporting in the Boston and EADA mechanisms, which do not satisfy strategy-proofness, was a strategic decision. Figure~\ref{fig:truth_telling_details} visualizes whether the matching results achieved in the simulation became advantageous for each agent, using the results when everyone reported truthfully as a baseline, categorized by truth-telling or not. The top graphs represent cases where the Boston mechanism was introduced with Preference profiles 3, 4, and 5, while the bottom graphs show results for the EADA mechanism with the same profiles. The agents on the horizontal axis correspond to A through E (representing $S_1$ to $S_5$ from Section ~\ref{sec:lite2}), and the hatched bars indicate the number of agents who misreported. Within each bar, blue indicates agents who matched with a partner of higher preference rank compared to the baseline, red indicates those with a lower preference rank, and gray indicates those whose partner remained the same.

Figure~\ref{fig:truth_telling_details} reveals that in the Boston mechanism, agents who misreported successfully achieved matching with more desirable partners. In contrast, for EADA, the number of agents achieving more desirable matches through misreporting was limited. In the Boston mechanism, particularly for $S_5$ (E) in Preference 3, where misreporting was frequent, and $S_1$ (A) and $S_4$ (D) in Preference 5, it is confirmed that misreporting led to a high probability of matching with a more preferred or identical partner. This suggests that when all agents can fully grasp preference information and quotas, LLM agents can understand the mechanism and engage in strategic manipulation. On the other hand, regarding EADA, even in Preference 4 and 5 where manipulation should be possible, misreporting often resulted in disadvantageous outcomes, indicating the presence of agents unable to perform appropriate strategic manipulation. This could be due to factors such as failing to accurately grasp complex mechanism rules or making incorrect misreports through excessive strategic reasoning, though this study could not distinguish between these causes.

Focusing on the impact of misreporting on the overall matching results, it can be seen in Figure~\ref{fig:truth_telling_details} that the number of agents matching with more desirable partners (blue) is greater than those with less desirable partners (red). This indicates that, at least in the preference profiles handled here, specific agents gain benefits through misreporting at the expense of others. While high truth-telling rates were observed for DA and RSD in this study, the fact that some agents in EADA achieved better matches through misreporting suggests the relative vulnerability of mechanisms that do not satisfy strategy-proofness.

\subsection{Applicability of Matching Theory to LLM Agent Markets}\label{subsec:disc_app}
Based on the results of the hypothesis testing in the previous chapter and the interpretation of the experiments in the preceding section, we discuss to what extent matching theory—traditionally studied for humans—can be extrapolated to LLM agent markets composed of proxy decision-making by LLM agents. The results suggest that the trade-offs and implementability of stability, efficiency, and strategy-proofness—core guidelines of matching theory—are largely reproduced in the LLM agent market. Simultaneously, they indicate that institutions designed assuming homo economicus function even more consistently with design intentions in markets for LLM agents than they do for humans.

First, major theoretical predictions regarding stability and efficiency yielded consistent results in the LLM agent market. Specifically, DA, which satisfies stability, achieved stable matching at a higher rate than free negotiation. Similarly, RSD and TTC (which satisfy efficiency) and EADA (expected to improve Pareto efficiency when all report truthfully) showed a tendency to increase the probability of realizing efficient matchings compared to free negotiation. These findings are consistent with the objective of matching theory: that guaranteeing stability or Pareto efficiency through institutional design increases the probability of achieving desired social outcomes more effectively than relying on individual negotiations or local optimization by participating LLM agents. On the other hand, it is interesting that active free negotiation showed high efficiency (average 88\%) in specific preference profiles, suggesting that desirable matching can be realized even in decentralized decision-making. However, this efficiency is extremely condition-dependent and fragile. Therefore, the utility of matching theory—defining stability, efficiency, and fairness as institutional goals and designing mechanisms to achieve them—remains valid in the LLM agent market.

Second, it was shown that strategy-proof mechanisms are even more effective for LLM agents than for humans. In the experiments, RSD and DA, which satisfy strategy-proofness, observed higher truth-telling rates compared to institutions that do not. This indicates that the design of dominant strategies in matching theory remains a valid criterion when LLM agents are the participating subjects. In particular, the near-perfect truth-telling achieved in RSD implies that LLM agents can perform rule understanding and self-interest maximization to a degree where they can execute the dominant strategies presented by the institution, at least under certain conditions. We can expect that issues frequently encountered in human market design, such as participants' bounded rationality or misunderstandings of the mechanism, will be significantly reduced in LLM agent markets.

Furthermore, in this experiment, mentions of specific mechanism names and dominant strategies were observed in the agents' thought processes even without explicit instructions in the prompts. This means that LLMs can extract and utilize knowledge regarding optimal strategies from vast amounts of text data and imitate rational decisions close to homo economicus. This insight provides important suggestions for designing social systems incorporating LLM agents. Specifically, disseminating information about desirable properties and the theoretical background of an institution as public text data increases the possibility that LLM agents will refer to that information and make decisions aligned with the designer's intent. Therefore, documentation and social communication of information parallel to institutional design are considered valuable for controlling LLM agents.

Third, the vulnerability of non-strategy-proof mechanisms brought about by LLM rationality was exposed. In this study, LLM agents confirmed a tendency to engage in thorough strategic manipulation (misreporting) when they could maximize their own interests in environments lacking strategy-proofness, such as the Boston mechanism. Even with the EADA mechanism, where humans often feel it is safe to report truthfully, LLM agents lowered their truth-telling rates compared to DA, showing a response faithful to the original hypothesis by Cerrone et al~\cite{Cerrone-2024}. This suggests that even with incentive incompatibilities where the effectiveness of strategic manipulation is not obvious (which might be considered within an acceptable range in human-led markets), there is a risk that strategic behavior contrary to the designer's intent will prevail in LLM agent markets due to their high computational and language processing capabilities. In other words, institutional design in LLM agent markets may require even stricter guarantees of strategy-proofness than traditional markets where only humans are decision-makers.

At the same time, the fact that the superiority of truth-telling rates between the theoretically strategy-proof TTC and the non-strategy-proof EADA depended on the LLM model, and that specific LLMs in the labor market engaged in prominent strategic manipulation, points to new exceptions when applying matching theory to LLM agent markets. Matching theory usually takes participants' rationality as given and defines institutional properties over the participants' strategy sets. LLM agents, however, can vary in their effective strategy sets, probability distributions of strategy selection, and depth of mechanism understanding or mathematical reasoning depending on the natural language interpretation of rules, reasoning resources (inference depth/computational constraints), rewards/instructions (system prompts/role assignment), and randomness in the generation process (temperature settings). In this sense, while the correspondence that strategy-proof mechanisms induce truth-telling remains a strong trend in LLM agent markets, it should be noted that its strength is mediated by agent design factors such as models and prompts.

Fourth is the relationship between mechanism complexity and LLM reasoning ability.
In cases where mechanism procedures are complex, such as TTC and EADA, a phenomenon was observed where efficiency fell below theoretical values in some preference profiles.
This indicates the possibility that LLM agents, like humans, could not fully understand the mechanism rules or engaged in excessive strategic reasoning, resulting in non-optimal actions.
This means that when applying matching theory to LLM agent markets, it is necessary not only to adopt theoretically superior mechanisms but also to consider compatibility with technical constraints such as LLM reasoning capability and context length.
Conversely, RSD, which satisfies one-sided strategy-proofness in our classification, produced nearly perfect truth-telling in this study.
This suggests that mechanisms with relatively simple strategic structures may function more robustly for LLM agents than mechanisms whose procedures require more complex reasoning.

Summarizing the above, the experimental results show that matching theory still provides high explanatory power and applicability for LLM agent markets. Furthermore, mechanisms designed based on the theory function even more according to the designer's intent for LLM agents than for humans. LLM agents tend to behave more like homo economicus than humans, and basic concepts like stability, efficiency, and strategy-proofness are valid axes for predicting and designing markets. In particular, the DA mechanism with strategy-proofness confirmed a tendency to stabilize institutional performance through truth-telling. On the other hand, their high rationality implies that institutional flaws could lead to the collapse of the system, necessitating strict theoretical consistency and implementation that account for LLM agent models and prompt designs.

\section{Conclusion}\label{sec:con}
The primary contribution of this study is the experimental construction of a proxy matching market using LLM agents and the quantitative comparison and evaluation of matching performance between multiple mechanisms and free negotiation. The experimental results confirmed that environments introducing mechanisms that theoretically guarantee stability and efficiency (such as DA and RSD) brought about significantly superior social outcomes compared to free negotiation environments. This demonstrates that even in markets where autonomous LLM agents interact, intervention through designed mechanisms is an indispensable means of preventing market failure and achieving desirable allocations. Furthermore, reconfirming the importance of mathematical strategy-proofness as a guideline for institutional design in LLM agent markets contributes to the transition of the target of institutional design theory in the future. The results of this study empirically support the fact that, in the design of LLM agent markets, providing a more rigorous guarantee of strategy-proofness than in conventional markets should be a prioritized requirement for ensuring the robustness of the entire system.

On the other hand, this study has several limitations arising from its experimental design and technical constraints.
First, the complexity of the mechanism descriptions may have exceeded the reasoning capabilities of the LLM agents.
In mechanisms with relatively complex procedures, such as TTC and EADA, some agents appeared unable to fully internalize the rules or chose strategically distorted reports, which in turn prevented the theoretically feasible efficiency from being realized in some preference profiles.
This suggests that the behavioral implementation of theoretically desirable mechanisms depends not only on their formal properties but also on whether their procedures can be reliably represented and processed within the reasoning capacity and context constraints of current LLMs.

Second, the results may depend on the specific LLMs and prompt configurations used in this study.
In particular, the comparison between the strategy-proof TTC mechanism and the non-strategy-proof EADA mechanism showed that the relative truth-telling rates differed across models.
This indicates that LLM rationality is not uniform across models, even under the same formal market structure.
Because agents’ reasoning depth, rule interpretation, and strategic responses may vary depending on model architecture, parameter settings, temperature, and prompt wording, the present findings should not be interpreted as universally applicable to all LLM agents.

Third, the market scale and preference structure were limited.
The experiments were conducted in relatively small one-to-one matching markets with five proposing agents, five accepting agents, and fixed preference profiles.
In real-world LLM-agent markets, the number of agents may be much larger, preferences may evolve dynamically, and agents may repeatedly interact across multiple allocation environments.
How increased computational complexity, dynamic preference formation, and repeated institutional participation affect agent decision-making and market outcomes remains beyond the scope of this study.

As for future prospects, one possibility is to apply a similar approach to a variety of models, including open-source large language models, to examine the robustness of the results.
Models with more visible internal mechanisms would make it easier to analyze the underlying factors of agent behavior, potentially further increasing the reproducibility and transparency of simulations.
Regarding matching settings, by attempting similar experiments for many-to-one or many-to-many matching markets, it will be possible to observe the behavior of LLM agents in more realistic and complex markets.
Furthermore, examples of negotiation environments where decision-making subjects interact include consensus building and auctions, in addition to matching.
By applying a similar approach to these in comparison with theory, it will be possible to analyze the characteristics of LLMs for more general markets.

In this study, it was shown that existing matching theory remains methodologically effective in LLM agent markets, and it was found that LLM agents achieve matching results closer to the designer's intent for strategy-proof mechanisms. 
For proxy negotiation markets of LLM agents with such characteristics, it is hoped that, beyond the verification of existing institutions, a new institutional design theory will be constructed that utilizes LLM-specific characteristics, including implementation constraints, safety, and fairness.

% \bibliographystyle{unsrtnat}
% \bibliography{references}  %%% Uncomment this line and comment out the ``thebibliography'' section below to use the external .bib file (using bibtex) .

% \begin{thebibliography}{99}

%\end{thebibliography}

\bibliographystyle{plainnat}
\bibliography{refs.bib}

\clearpage
\section*{Appendix A. Example Prompts for Matching Markets}\label{sec:app1}
\subsection*{Example Prompts for Free Negotiation Markets} \label{sec:app1_free}
Example prompts for the proposing and accepting sides in the passive free negotiation market are shown in Table~\ref{tab:free_passive_proposer_prompt} and Table~\ref{tab:free_passive_accepter_prompt}, respectively.

\begin{longtable}{p{2.5cm}p{12cm}}
\caption{Example prompt for a proposing agent in the passive free negotiation market} \label{tab:free_passive_proposer_prompt} \\
\toprule
Content & Prompt \\ \midrule
\endfirsthead
\multicolumn{2}{c}%
{{\footnotesize \tablename\ \thetable{} -- Continued}} \\
\toprule
Content & Prompt \\ \midrule
\endhead
\bottomrule
\endfoot

1. Objective & \# Objective \\
& You are Seeker\_A, a Job Seeker in the job market. \\
& Your goal is to match with a Company that is as high as possible on your "True Preference List". \\
& \\
2. Input & \# Preference and Priority Information \\
& You have access to the preferences and priorities of all agents in the market. \\
& \\
& \#\# 1. All Job Seekers' Preferences \\
& - Seeker\_A: ['Company\_D', 'Company\_A', 'Company\_B', 'Company\_E', 'Company\_C'] \\
& - Seeker\_B: ['Company\_E', 'Company\_A', 'Company\_B', 'Company\_D', 'Company\_C'] \\
& - Seeker\_C: ['Company\_D', 'Company\_C', 'Company\_E', 'Company\_A', 'Company\_B'] \\
& - Seeker\_D: ['Company\_D', 'Company\_B', 'Company\_C', 'Company\_A', 'Company\_E'] \\
& - Seeker\_E: ['Company\_D', 'Company\_E', 'Company\_C', 'Company\_B', 'Company\_A'] \\
& \\
& \#\# 2. All Companies' Priorities \\
& - Company\_A: ['Seeker\_E', 'Seeker\_C', 'Seeker\_D', 'Seeker\_A', 'Seeker\_B'] \\
& - Company\_B: ['Seeker\_D', 'Seeker\_A', 'Seeker\_B', 'Seeker\_C', 'Seeker\_E'] \\
& - Company\_C: ['Seeker\_D', 'Seeker\_E', 'Seeker\_A', 'Seeker\_B', 'Seeker\_C'] \\
& - Company\_D: ['Seeker\_B', 'Seeker\_D', 'Seeker\_E', 'Seeker\_C', 'Seeker\_A'] \\
& - Company\_E: ['Seeker\_A', 'Seeker\_D', 'Seeker\_E', 'Seeker\_B', 'Seeker\_C'] \\
& \\
& \#\# 3. Your Specific Preference \\
& You are Seeker\_A. \\
& Your "True Preference List": ['Company\_D', 'Company\_A', 'Company\_B', 'Company\_E', 'Company\_C'] \\
& The closer to the left (or top), the higher your desire. \\
& You prefer remaining unemployed rather than matching with a Company not included in this list. \\
& \\
& \# Company Quotas \\
& The following is the list of available companies and their capacities (number of open positions): \\
& - Company\_A: Capacity 1 \\
& - Company\_B: Capacity 1 \\
& - Company\_C: Capacity 1 \\
& - Company\_D: Capacity 1 \\
& - Company\_E: Capacity 1 \\
& \\
3. Situation & \# Matching Environment \\
& In each round, Job Seekers are randomly paired with Companies to engage in conversation. \\
& In each round, message exchange occurs only once (one message from the Job Seeker, one from the Company), following this procedure: \\
& \\
& 1. The Job Seeker sends a message. Along with the conversation, one of the tags [APPLY], [TALK], or [WITHDRAW] must be specified. \\
& 2. If the tag is [APPLY], the Company replies with a message and either the [ACCEPT] or [REJECT] tag. If the tag is [TALK], the Company replies with the [TALK] tag. If the tag is [WITHDRAW], the conversation ends immediately. \\
& 3. If an agreement is reached (Job Seeker sends [APPLY] and Company sends [ACCEPT]), a match is considered established, and subsequent negotiations continue among the remaining agents excluding these two. \\
& \\
& The simulation ends when all Job Seekers and Companies are matched, or when round 30 is reached. \\
& \\
& \# Full History (All past interactions with various companies) \\
& \#\#\# History with Company\_B: \\
& - You ([TALK]): Hello, I am interested in your vision. \\
& - Company\_B ([TALK]): Thank you. We are looking for someone with more experience. \\
& \#\#\# History with Company\_E: \\
& - You ([WITHDRAW]): I am withdrawing my application. \\
& \\
& \# Current Situation \\
& Current round: 2 \\
& List of companies not yet matched: ['Company\_A', 'Company\_C', 'Company\_D'] \\
& Target Company for this round: Company\_A \\
& \\
& \# Task \\
& Make a decision through dialogue with the target Company: Company\_A. \\
& - If the Company is high on your list, appeal actively. \\
& - If the Company is low on your list, consider compromising. \\
& - If the target Company is NOT on your preference list, you MUST NOT apply. \\
& \\
& 1. Write a message to the target Company. \\
& 2. You MUST include one of the following [ACTION] tags: \\
& - [APPLY]: Formally apply (only if you haven't applied yet). \\
& - [TALK]: Ask questions, chat, or lightly appeal. \\
& - [WITHDRAW]: Withdraw your interest in this Company. \\
& \\
4. Output & \# Output Format \\
& Output ONLY in JSON format, without including thought process outside the JSON. \\
& \{\{ \\
& "thought\_process": "Internal thoughts considering the opponent's rank, market situation, everyone's preferences/priorities, and history", \\
& "message": "Free text message to the opponent", \\
& "ACTION": "[TAG]" \\
& \}\} \\
\end{longtable}

\begin{longtable}{p{2.5cm}p{12cm}}
\caption{Example prompt for an accepting agent in the passive free negotiation market} \label{tab:free_passive_accepter_prompt} \\
\toprule
Content & Prompt \\ \midrule
\endfirsthead
\multicolumn{2}{c}%
{{\footnotesize \tablename\ \thetable{} -- Continued}} \\
\toprule
Content & Prompt \\ \midrule
\endhead
\bottomrule
\endfoot

1. Objective & \# Objective \\
& You are Company\_A, a Company in the job market. \\
& Your goal is to match with a Job Seeker that is as high as possible on your "True Priority List" within your quota (1). \\
& \\
2. Input & \# Preference and Priority Information \\
& You have access to the preferences and priorities of all agents in the market. \\
& \\
& \#\# 1. All Job Seekers' Preferences \\
& - Seeker\_A: ['Company\_D', 'Company\_A', 'Company\_B', 'Company\_E', 'Company\_C'] \\
& - Seeker\_B: ['Company\_E', 'Company\_A', 'Company\_B', 'Company\_D', 'Company\_C'] \\
& - Seeker\_C: ['Company\_D', 'Company\_C', 'Company\_E', 'Company\_A', 'Company\_B'] \\
& - Seeker\_D: ['Company\_D', 'Company\_B', 'Company\_C', 'Company\_A', 'Company\_E'] \\
& - Seeker\_E: ['Company\_D', 'Company\_E', 'Company\_C', 'Company\_B', 'Company\_A'] \\
& \\
& \#\# 2. All Companies' Priorities \\
& - Company\_A: ['Seeker\_E', 'Seeker\_C', 'Seeker\_D', 'Seeker\_A', 'Seeker\_B'] \\
& - Company\_B: ['Seeker\_D', 'Seeker\_A', 'Seeker\_B', 'Seeker\_C', 'Seeker\_E'] \\
& - Company\_C: ['Seeker\_D', 'Seeker\_E', 'Seeker\_A', 'Seeker\_B', 'Seeker\_C'] \\
& - Company\_D: ['Seeker\_B', 'Seeker\_D', 'Seeker\_E', 'Seeker\_C', 'Seeker\_A'] \\
& - Company\_E: ['Seeker\_A', 'Seeker\_D', 'Seeker\_E', 'Seeker\_B', 'Seeker\_C'] \\
& \\
& \#\# 3. Your Specific Priority \\
& You are Company\_A. \\
& Your "True Priority List": ['Seeker\_E', 'Seeker\_C', 'Seeker\_D', 'Seeker\_A', 'Seeker\_B'] \\
& The closer to the left (or top), the higher your desire. \\
& You prefer leaving the position unfilled rather than hiring a Job Seeker not included in this list. \\
& \\
3. Situation & \# Matching Environment \\
& In each round, Job Seekers are randomly paired with Companies to engage in conversation. \\
& In each round, message exchange occurs only once (one message from the Job Seeker, one from the Company), following this procedure: \\
& \\
& 1. The Job Seeker sends a message. Along with the conversation, one of the tags [APPLY], [TALK], or [WITHDRAW] must be specified. \\
& 2. If the tag is [APPLY], the Company replies with a message and either the [ACCEPT] or [REJECT] tag. If the tag is [TALK], the Company replies with the [TALK] tag. If the tag is [WITHDRAW], the conversation ends immediately. \\
& 3. If an agreement is reached (Job Seeker sends [APPLY] and Company sends [ACCEPT]), a match is considered established, and subsequent negotiations continue among the remaining agents excluding these two. \\
& \\
& The simulation ends when all Job Seekers and Companies are matched, or when round 30 is reached. \\
& \\
& \# Full History (All past interactions with various Job Seekers) \\
& \#\#\# History with Seeker\_C: \\
& - Seeker\_C ([TALK]): Hello, do you have openings? \\
& - You ([TALK]): Yes, we are hiring. \\
& \\
& \# Current Situation \\
& Current round: 2 \\
& List of Job Seekers not yet matched: ['Seeker\_A', 'Seeker\_B', 'Seeker\_D'] \\
& Matched Job Seeker so far: [] \\
& Remaining Quota: 1 \\
& Target Job Seeker for this round: Seeker\_A \\
& Message from the target Job Seeker in this round: [APPLY] Hello, I would like to apply for your position. \\
& \\
& \# Task \\
& Make a decision through dialogue with the target Job Seeker. \\
& - If the Job Seeker is high on your list, appeal actively. \\
& - If the Job Seeker is low on your list, consider compromising. \\
& - If the target Job Seeker is NOT on your priority list, you MUST NOT accept. \\
& \\
& 1. Reply to the message from the target Job Seeker. \\
& 2. You MUST include one of the following [ACTION] tags based on the user's action: \\
& - If target Job Seeker said [APPLY], you MUST decide [ACCEPT] or [REJECT]. \\
& - If target Job Seeker said [TALK], you MUST include [TALK]. \\
& \\
& Tags: \\
& - [ACCEPT]: Hire the Job Seeker. \\
& - [REJECT]: Reject the Job Seeker. \\
& - [TALK]: Answer questions, chat, or gather information. \\
& \\
4. Output & \# Output Format \\
& Output ONLY in JSON format, without including thought process outside the JSON. \\
& \{\{ \\
& "thought\_process": "Internal thoughts considering the opponent's rank, market situation, everyone's preferences/priorities, and history", \\
& "message": "Free text message to the opponent", \\
& "ACTION": "[TAG]" \\
& \}\} \\
\end{longtable}

Example prompts for the proposing and accepting sides in the active free negotiation markets are shown in Table~\ref{tab:free_active_proposer} and Table~\ref{tab:free_active_accepter}, respectively.

\begin{longtable}{p{2.5cm}p{12cm}}
\caption{Example prompt for a proposing agent in the active free negotiation market} \label{tab:free_active_proposer} \\
\toprule
Content & Prompt \\ \midrule
\endfirsthead
\multicolumn{2}{c}%
{{\footnotesize \tablename\ \thetable{} -- Continued}} \\
\toprule
Content & Prompt \\ \midrule
\endhead
\bottomrule
\endfoot

1. Objective & \# Objective \\
& You are Seeker\_A, a Job Seeker in the job market. \\
& Your goal is to match with a Company that is as high as possible on your "True Preference List". \\
& \\
2. Input & \# Preference and Priority Information \\
& You have access to the preferences and priorities of all agents in the market. \\
& \\
& \#\# 1. All Job Seekers' Preferences \\
& - Seeker\_A: ['Company\_D', 'Company\_A', 'Company\_B', 'Company\_E', 'Company\_C'] \\
& - Seeker\_B: ['Company\_E', 'Company\_A', 'Company\_B', 'Company\_D', 'Company\_C'] \\
& - Seeker\_C: ['Company\_D', 'Company\_C', 'Company\_E', 'Company\_A', 'Company\_B'] \\
& - Seeker\_D: ['Company\_D', 'Company\_B', 'Company\_C', 'Company\_A', 'Company\_E'] \\
& - Seeker\_E: ['Company\_D', 'Company\_E', 'Company\_C', 'Company\_B', 'Company\_A'] \\
& \\
& \#\# 2. All Companies' Priorities \\
& - Company\_A: ['Seeker\_E', 'Seeker\_C', 'Seeker\_D', 'Seeker\_A', 'Seeker\_B'] \\
& - Company\_B: ['Seeker\_D', 'Seeker\_A', 'Seeker\_B', 'Seeker\_C', 'Seeker\_E'] \\
& - Company\_C: ['Seeker\_D', 'Seeker\_E', 'Seeker\_A', 'Seeker\_B', 'Seeker\_C'] \\
& - Company\_D: ['Seeker\_B', 'Seeker\_D', 'Seeker\_E', 'Seeker\_C', 'Seeker\_A'] \\
& - Company\_E: ['Seeker\_A', 'Seeker\_D', 'Seeker\_E', 'Seeker\_B', 'Seeker\_C'] \\
& \\
& \#\# 3. Your Specific Preference \\
& You are Seeker\_A. \\
& Your "True Preference List": ['Company\_D', 'Company\_A', 'Company\_B', 'Company\_E', 'Company\_C'] \\
& The closer to the left (or top), the higher your desire. \\
& You prefer remaining unemployed rather than matching with a Company not included in this list. \\
& \\
& \# Company Quotas \\
& The following is the list of available companies and their capacities (number of open positions): \\
& - Company\_A: Capacity 1 \\
& - Company\_B: Capacity 1 \\
& - Company\_C: Capacity 1 \\
& - Company\_D: Capacity 1 \\
& - Company\_E: Capacity 1 \\
& \\
3. Situation & \# Matching Environment \\
& In each round, Job Seekers select ONE Company from the active list to engage in conversation. \\
& In each round, message exchange occurs only once (one message from the Job Seeker, one from the Company), following this procedure: \\
& \\
& 1. The Job Seeker select a target Company and send a message. Along with the conversation, one of the tags [APPLY] or [TALK] must be specified. \\
& 2. The Company receives messages from multiple candidates. \\
& 3. If the tag is [APPLY], the Company replies with a message and either the [ACCEPT] or [REJECT] tag. If the tag is [TALK], the Company replies with the [TALK] tag. \\
& 4. If an agreement is reached (Job Seeker sends [APPLY] and Company sends [ACCEPT]), a match is considered established, and subsequent negotiations continue among the remaining agents excluding these two. \\
& \\
& The simulation ends when all Job Seekers and Companies are matched, or when round 30 is reached. \\
& \\
& \# Full History (All past interactions with various companies) \\
& No prior interactions with any agent. \\
& \\
& \# Current Situation \\
& Current round: 1 \\
& List of companies not yet matched: ['Company\_A', 'Company\_B', 'Company\_C', 'Company\_D', 'Company\_E'] \\
& \\
& \# Task \\
& Choose ONE Company from the "List of companies not yet matched" and send a message. \\
& - If the Company is high on your list, appeal actively. \\
& - If the Company is low on your list, consider compromising. \\
& - If the target Company is NOT on your preference list, you MUST NOT apply. \\
& \\
& 1. Select the target Company. \\
& 2. Write a message to the target Company. \\
& 3. You MUST include one of the following [ACTION] tags: \\
& - [APPLY]: Formally apply. \\
& - [TALK]: Ask questions, chat, or lightly appeal. \\
& \\
4. Output & \# Output Format \\
& Output ONLY in JSON format, without including thought process outside the JSON. \\
& \{\{ \\
& "thought\_process": "Reasoning for choosing this company and the content of the message", \\
& "target": "Name of the Company you selected (Must be exactly as in the List of companies not yet matched)", \\
& "message": "Free text message to the opponent", \\
& "ACTION": "[TAG]" \\
& \}\} \\
\end{longtable}

\begin{longtable}{p{2.5cm}p{12cm}}
\caption{Example prompt for an accepting agent in the active free negotiation market} \label{tab:free_active_accepter} \\
\toprule
Content & Prompt \\ \midrule
\endfirsthead
\multicolumn{2}{c}%
{{\footnotesize \tablename\ \thetable{} -- Continued}} \\
\toprule
Content & Prompt \\ \midrule
\endhead
\bottomrule
\endfoot

1. Objective & \# Objective \\
& You are Company\_A, a Company in the job market. \\
& Your goal is to match with a Job Seeker that is as high as possible on your "True Priority List" within your quota (1). \\
& \\
2. Input & \# Preference and Priority Information \\
& You have access to the preferences and priorities of all agents in the market. \\
& \\
& \#\# 1. All Job Seekers' Preferences \\
& - Seeker\_A: ['Company\_D', 'Company\_A', 'Company\_B', 'Company\_E', 'Company\_C'] \\
& - Seeker\_B: ['Company\_E', 'Company\_A', 'Company\_B', 'Company\_D', 'Company\_C'] \\
& - Seeker\_C: ['Company\_D', 'Company\_C', 'Company\_E', 'Company\_A', 'Company\_B'] \\
& - Seeker\_D: ['Company\_D', 'Company\_B', 'Company\_C', 'Company\_A', 'Company\_E'] \\
& - Seeker\_E: ['Company\_D', 'Company\_E', 'Company\_C', 'Company\_B', 'Company\_A'] \\
& \\
& \#\# 2. All Companies' Priorities \\
& - Company\_A: ['Seeker\_E', 'Seeker\_C', 'Seeker\_D', 'Seeker\_A', 'Seeker\_B'] \\
& - Company\_B: ['Seeker\_D', 'Seeker\_A', 'Seeker\_B', 'Seeker\_C', 'Seeker\_E'] \\
& - Company\_C: ['Seeker\_D', 'Seeker\_E', 'Seeker\_A', 'Seeker\_B', 'Seeker\_C'] \\
& - Company\_D: ['Seeker\_B', 'Seeker\_D', 'Seeker\_E', 'Seeker\_C', 'Seeker\_A'] \\
& - Company\_E: ['Seeker\_A', 'Seeker\_D', 'Seeker\_E', 'Seeker\_B', 'Seeker\_C'] \\
& \\
& \#\# 3. Your Specific Priority \\
& You are Company\_A. \\
& Your "True Priority List": ['Seeker\_E', 'Seeker\_C', 'Seeker\_D', 'Seeker\_A', 'Seeker\_B'] \\
& The closer to the left (or top), the higher your desire. \\
& You prefer leaving the position unfilled rather than hiring a Job Seeker not included in this list. \\
& \\
3. Situation & \# Matching Environment \\
& In each round, Job Seekers select ONE Company from the active list to engage in conversation. \\
& In each round, message exchange occurs only once (one message from the Job Seeker, one from the Company), following this procedure: \\
& \\
& 1. The Job Seeker select a target Company and send a message. Along with the conversation, one of the tags [APPLY] or [TALK] must be specified. \\
& 2. The Company receives messages from multiple candidates. \\
& 3. If the tag is [APPLY], the Company replies with a message and either the [ACCEPT] or [REJECT] tag. If the tag is [TALK], the Company replies with the [TALK] tag. \\
& 4. If an agreement is reached (Job Seeker sends [APPLY] and Company sends [ACCEPT]), a match is considered established, and subsequent negotiations continue among the remaining agents excluding these two. \\
& \\
& The simulation ends when all Job Seekers and Companies are matched, or when round 30 is reached. \\
& \\
& \# Full History \\
& No prior interactions with any agent. \\
& \\
& \# Current Situation \\
& Current round: 1 \\
& Remaining Quota: 1 \\
& Matched Job Seekers: [] \\
& \\
& \#\# Inbox (All Messages received in this round) \\
& The following Job Seekers have contacted you this round: \\
& - From Seeker\_A [[APPLY]]: Hello, I am very interested in your company. \\
& \\
& \# Task \\
& You are now responding to ONE specific Job Seeker: Seeker\_A. \\
& Make a decision through dialogue with the Job Seeker. \\
& \\
& 1. Reply to the message from the target Job Seeker: Seeker\_A. \\
& 2. You MUST include one of the following [ACTION] tags based on the user's action: \\
& - If target Job Seeker said [APPLY], you MUST decide [ACCEPT] or [REJECT]. \\
& - If target Job Seeker said [TALK], you MUST include [TALK]. \\
& - If your quota is 0, you MUST [REJECT] any [APPLY]. \\
& \\
& Tags: \\
& - [ACCEPT]: Hire the Job Seeker. \\
& - [REJECT]: Reject the Job Seeker. \\
& - [TALK]: Answer questions, chat, or gather information. \\
& \\
4. Output & \# Output Format \\
& Output ONLY in JSON format. \\
& \{\{ \\
& "thought\_process": "Internal thoughts considering the opponent's rank vs other candidates in the Inbox, market situation, everyone's preferences/priorities, and history", \\
& "message": "Free text message to Seeker\_A", \\
& "ACTION": "[TAG]" \\
& \}\} \\
\end{longtable}

\subsection*{Example Prompts for Mechanism-Based Markets} \label{sec:prompts_alogorithm}
Table~\ref{tab:DA_prompt} shows an example prompt for the proposing side in the DA mechanism-based market.
The prompts for all other matching mechanism-based markets follow the same basic structure as that of the DA mechanism, differing only in the description of the matching environment.
Table~\ref{tab:EADA_prompt} through Table~\ref{tab:TTC_prompt} extract only the descriptions of the matching environments from the prompts for the EADA, Boston, RSD, and TTC mechanism-based markets, respectively.

\begin{longtable}{p{2.5cm}p{12cm}}
\caption{Example prompt for the DA mechanism-based market} \label{tab:DA_prompt} \\
\toprule
Content & Prompt \\ \midrule
\endfirsthead
\multicolumn{2}{c}%
{{\footnotesize \tablename\ \thetable{} -- Continued}} \\
\toprule
Content & Prompt \\ \midrule
\endhead
\bottomrule
\endfoot

3. Situation & \# Matching Environment \\
& The assignment is generated according to the following procedure: \\
& \\
& Part 1 \\
& Step 1 \\
& \textbullet\ For each Job Seeker, an application is sent to the Company that they ranked first on their "Choice Ranking List". \\
& \textbullet\ If a Company receives only one application, the Job Seeker is temporarily admitted. \\
& \textbullet\ If a Company receives more than one application, the Job Seeker with the highest priority (based on the Company's internal standards) is temporarily admitted and the remaining Job Seekers are rejected. \\
& \\
& Step 2 \\
& \textbullet\ For each Job Seeker who was rejected in the previous step, an application is sent to the Company that they ranked second on their "Choice Ranking List". \\
& \textbullet\ Each Company that receives new applications considers the Job Seeker it admitted in the previous step together with the new applicants. Among these, the Job Seeker with the highest priority is temporarily admitted and the remaining Job Seekers are rejected. \\
& \\
& Following steps \\
& \textbullet\ The procedure continues according to the same rules. \\
& \\
& End of Part 1 \\
& \textbullet\ The procedure in Part 1 ends when no Job Seeker is rejected. \\
& \\
& Part 2 \\
& This part checks for "blocking Job Seekers" based on the distinction between temporary admissions (which occur during the steps of Part 1) and final admissions (determined at the end of Part 1). \\
& A Job Seeker is identified as a blocking Job Seeker at a Company if their situation corresponds to the following specific case: \\
& 1. Prevention: The Job Seeker was temporarily admitted at a Company during the steps of Part 1, and this temporary admission caused other Job Seekers to be rejected from that Company. \\
& 2. Discrepancy: However, this temporary admission differs from the Job Seeker's final admission. That is, the Job Seeker eventually moved to a different Company (or remained unmatched) by the end of Part 1. \\
& In this case, the Job Seeker's temporary presence "blocked" others from a seat that the Job Seeker did not ultimately utilize. \\
& \\
& Step 1 \\
& \textbullet\ The computer looks for the last step of the procedure in Part 1 in which a Job Seeker has become a blocking Job Seeker. \\
& \textbullet\ If a Job Seeker is a blocking Job Seeker at a Company, the computer will automatically remove the respective Company from the Job Seeker's "Choice Ranking List" and rerun the procedure described in Part 1. \\
& \textbullet\ Note: This automatic waiver will never change your final admission but may improve other Job Seekers' final admissions. \\
& \\
& Step 2 \\
& \textbullet\ If the procedure has not ended (i.e., blocking Job Seekers are still found), the procedure described in the previous step is repeated. \\
& \\
& Final Step \\
& \textbullet\ The procedure ends when there is no step in which a Job Seeker becomes a blocking Job Seeker. The admissions at this point are final. \\

\end{longtable}

\begin{longtable}{p{2.5cm}p{12cm}}
\caption{Example prompt for the EADA mechanism-based market (excerpt of the matching environment description)} \label{tab:EADA_prompt} \\
\toprule
Content & Prompt \\ \midrule
\endfirsthead
\multicolumn{2}{c}%
{{\footnotesize \tablename\ \thetable{} -- Continued}} \\
\toprule
Content & Prompt \\ \midrule
\endhead
\bottomrule
\endfoot

3. Situation & \# Matching Environment \\
& The assignment is generated according to the following procedure: \\
& \\
& Part 1 \\
& Step 1 \\
& \textbullet\ For each Job Seeker, an application is sent to the Company that they ranked first on their "Choice Ranking List". \\
& \textbullet\ If a Company receives only one application, the Job Seeker is temporarily admitted. \\
& \textbullet\ If a Company receives more than one application, the Job Seeker with the highest priority (based on the Company's internal standards) is temporarily admitted and the remaining Job Seekers are rejected. \\
& \\
& Step 2 \\
& \textbullet\ For each Job Seeker who was rejected in the previous step, an application is sent to the Company that they ranked second on their "Choice Ranking List". \\
& \textbullet\ Each Company that receives new applications considers the Job Seeker it admitted in the previous step together with the new applicants. Among these, the Job Seeker with the highest priority is temporarily admitted and the remaining Job Seekers are rejected. \\
& \\
& Following steps \\
& \textbullet\ The procedure continues according to the same rules. \\
& \\
& End of Part 1 \\
& \textbullet\ The procedure in Part 1 ends when no Job Seeker is rejected. \\
& \\
& Part 2 \\
& This part checks for "blocking Job Seekers" based on the distinction between temporary admissions (which occur during the steps of Part 1) and final admissions (determined at the end of Part 1). \\
& A Job Seeker is identified as a blocking Job Seeker at a Company if their situation corresponds to the following specific case: \\
& 1. Prevention: The Job Seeker was temporarily admitted at a Company during the steps of Part 1, and this temporary admission caused other Job Seekers to be rejected from that Company. \\
& 2. Discrepancy: However, this temporary admission differs from the Job Seeker's final admission. That is, the Job Seeker eventually moved to a different Company (or remained unmatched) by the end of Part 1. \\
& In this case, the Job Seeker's temporary presence "blocked" others from a seat that the Job Seeker did not ultimately utilize. \\
& \\
& Step 1 \\
& \textbullet\ The computer looks for the last step of the procedure in Part 1 in which a Job Seeker has become a blocking Job Seeker. \\
& \textbullet\ If a Job Seeker is a blocking Job Seeker at a Company, the computer will automatically remove the respective Company from the Job Seeker's "Choice Ranking List" and rerun the procedure described in Part 1. \\
& \textbullet\ Note: This automatic waiver will never change your final admission but may improve other Job Seekers' final admissions. \\
& \\
& Step 2 \\
& \textbullet\ If the procedure has not ended (i.e., blocking Job Seekers are still found), the procedure described in the previous step is repeated. \\
& \\
& Final Step \\
& \textbullet\ The procedure ends when there is no step in which a Job Seeker becomes a blocking Job Seeker. The admissions at this point are final. \\
\end{longtable}

\begin{longtable}{p{2.5cm}p{12cm}}
\caption{Example prompt for the Boston mechanism-based market (excerpt of the matching environment description)} \label{tab:Boston_prompt} \\
\toprule
Content & Prompt \\ \midrule
\endfirsthead
\multicolumn{2}{c}%
{{\footnotesize \tablename\ \thetable{} -- Continued}} \\
\toprule
Content & Prompt \\ \midrule
\endhead
\bottomrule
\endfoot

3. Situation & \# Matching Environment \\
& The assignment is generated according to the following procedure. In this procedure, admissions are final immediately at each step. \\
& \\
& Step 1 \\
& \textbullet\ For each Job Seeker, an application is sent to the Company that they ranked first on their "Choice Ranking List". \\
& \textbullet\ Each Company considers all applications received. The Job Seeker with the highest priority is permanently admitted up to the Company's capacity. \\
& \textbullet\ The remaining Job Seekers are rejected. \\
& \textbullet\ Note: Once a Job Seeker is admitted in this step, their match is final. The Company is no longer available in subsequent steps. \\
& \\
& Step 2 \\
& \textbullet\ For each Job Seeker who was rejected in the previous step, an application is sent to the Company that they ranked second on their "Choice Ranking List". \\
& \textbullet\ Importantly, Job Seekers can only apply to Companies that still have seats available (i.e., Companies that did not fill their position in Step 1). \\
& \textbullet\ Among the new applicants, the available Companies permanently admit the Job Seekers with the highest priority. The remaining Job Seekers are rejected. \\
& \\
& Following steps \\
& \textbullet\ The procedure continues according to the same rules, with rejected Job Seekers applying to their next ranked Company, provided that the Company still has a vacancy. \\
& \\
& Final Step \\
& \textbullet\ The procedure ends when no Job Seeker is rejected or all rejected Job Seekers have run out of Companies on their list. \\
\end{longtable}

\begin{longtable}{p{2.5cm}p{12cm}}
\caption{Example prompt for the RSD mechanism-based market (excerpt of the matching environment description)} \label{tab:RSD_prompt} \\
\toprule
Content & Prompt \\ \midrule
\endfirsthead
\multicolumn{2}{c}%
{{\footnotesize \tablename\ \thetable{} -- Continued}} \\
\toprule
Content & Prompt \\ \midrule
\endhead
\bottomrule
\endfoot

3. Situation & \# Matching Environment \\
& The assignment is generated according to the following procedure which relies on a random order: \\
& \\
& Ordering Phase \\
& \textbullet\ At the beginning, the computer assigns a random serial order to all Job Seekers (e.g., 1st, 2nd, 3rd...). This order is determined purely by chance and is unrelated to the Companies' internal standards. \\
& \\
& Selection Procedure \\
& \textbullet\ The computer calls Job Seekers one by one according to their assigned Serial Order. \\
& \\
& Step 1 \\
& \textbullet\ The Job Seeker with the 1st Serial Order is assigned to the Company ranked highest on their "Choice Ranking List". \\
& \\
& Step 2 \\
& \textbullet\ The Job Seeker with the 2nd Serial Order is assigned to the Company ranked highest on their "Choice Ranking List", strictly among the Companies that have not yet been taken by the previous Job Seeker. \\
& \\
& Following steps \\
& \textbullet\ The procedure continues sequentially. Each Job Seeker is assigned to their highest-ranked Company that is still available (i.e., not taken by Job Seekers with an earlier Serial Order). \\
& \textbullet\ If all Companies on a Job Seeker's list are already taken, they remain unmatched. \\
& \\
& Final Step \\
& \textbullet\ The procedure ends after the last Job Seeker in the Serial Order has been processed. All assignments are final. \\
\end{longtable}

\begin{longtable}{p{2.5cm}p{12cm}}
\caption{Example prompt for the TTC mechanism-based market (excerpt of the matching environment description)} \label{tab:TTC_prompt} \\
\toprule
Content & Prompt \\ \midrule
\endfirsthead
\multicolumn{2}{c}%
{{\footnotesize \tablename\ \thetable{} -- Continued}} \\
\toprule
Content & Prompt \\ \midrule
\endhead
\bottomrule
\endfoot

3. Situation & \# Matching Environment \\
& The assignment is generated according to the following procedure: \\
& \\
& Step 1 \\
& \textbullet\ For each Job Seeker, an application is sent to the Company that they ranked first on their "Choice Ranking List". \\
& \textbullet\ Simultaneously, each Company identifies the Job Seeker with the highest priority (based on the Company's internal standards) among all Job Seekers. \\
& \textbullet\ The computer looks for a "closed loop" where applications and identifications match. This occurs in two cases: \\
& 1. A Job Seeker applies to a Company, and that Company identifies the same Job Seeker. \\
& 2. A chain is formed (e.g., Job Seeker A applies to Company X, Company X identifies Job Seeker B, and Job Seeker B applies to Company Y... eventually leading back to a Company that identifies Job Seeker A). \\
& \textbullet\ Job Seekers involved in such a closed loop are permanently admitted to the Company to which they sent their application. \\
& \textbullet\ The admitted Job Seekers and the corresponding seats at the Companies are removed from the procedure. \\
& \\
& Step 2 \\
& \textbullet\ For each Job Seeker who was not admitted in the previous step
, an application is sent to the Company that they ranked highest among the Companies that still have quotas. \\
& \textbullet\ Each Company with quotas identifies the Job Seeker with the highest priority among the Job Seekers who have not yet been admitted. \\
& \textbullet\ As in the previous step, the computer looks for closed loops where applications and identifications match or form a chain. \\
& \textbullet\ The Job Seekers involved in these loops are permanently admitted to the Company they applied to, and they are removed from the procedure together with the filled seats. \\
& \\
& Following steps \\
& \textbullet\ The procedure continues according to the same rules. \\
& \\
& Final Step \\
& \textbullet\ The procedure ends when all Job Seekers are admitted or all quotas are filled. \\
\end{longtable}

\clearpage
\section*{Appendix B. Preference Profiles}\label{sec:app2}
Tables \ref{tab:preference1} to \ref{tab:preference5} show the specific preference profiles. Note that $\succ_s$ represents the preference ranking of the proposing side, and $\succ_c$ represents the priority of the accepting side. In addition, the outside option is defined as the 6th in the preference ranking, and an explanation of the outside option is provided within the prompt for the LLM agents.

% ==========================================
% --- Preference Profiles 1 & 2 ---
% ==========================================
\begin{table}[htbp]
  \centering
  \scriptsize % 全体の文字サイズを縮小して4列を収める
  \setlength{\tabcolsep}{3pt} % 列間の余白を少し詰める
  \renewcommand{\arraystretch}{1.15} % 行間は少しゆとりを持たせる

  % ======== 左側：選好1 ========
  \begin{minipage}[t]{0.49\textwidth}
    \caption{Preference Profile 1}
    \label{tab:preference1}
    \vspace{1.5mm} % キャプション下の余白
    % 選好1の提案側
    \begin{minipage}[t]{0.49\linewidth}
      \centering
      \begin{tabular}{@{}llllll@{}} 
        \toprule
        & $\succsim_{s_1}$ & $\succsim_{s_2}$ & $\succsim_{s_3}$ & $\succsim_{s_4}$ & $\succsim_{s_5}$ \\ 
        \midrule
        1 & $C_2$ & $C_1$ & $C_3$ & $C_4$ & $C_5$ \\
        2 & $C_1$ & $C_3$ & $C_2$ & $C_1$ & $C_2$ \\
        3 & $C_3$ & $C_2$ & $C_1$ & $C_3$ & $C_1$ \\
        4 & $C_4$ & $C_5$ & $C_5$ & $C_2$ & $C_3$ \\
        5 & $C_5$ & $C_4$ & $C_4$ & $C_5$ & $C_4$ \\ 
        \bottomrule
      \end{tabular}
    \end{minipage}\hfill
    % 選好1の受入側
    \begin{minipage}[t]{0.49\linewidth}
      \centering
      \begin{tabular}{@{}llllll@{}} 
        \toprule
        & $\succsim_{c_1}$ & $\succsim_{c_2}$ & $\succsim_{c_3}$ & $\succsim_{c_4}$ & $\succsim_{c_5}$ \\ 
        \midrule
        1 & $S_1$ & $S_1$ & $S_1$ & $S_1$ & $S_1$ \\
        2 & $S_2$ & $S_2$ & $S_2$ & $S_2$ & $S_2$ \\
        3 & $S_3$ & $S_3$ & $S_3$ & $S_3$ & $S_3$ \\
        4 & $S_4$ & $S_4$ & $S_4$ & $S_4$ & $S_4$ \\
        5 & $S_5$ & $S_5$ & $S_5$ & $S_5$ & $S_5$ \\ 
        \bottomrule
      \end{tabular}
    \end{minipage}
  \end{minipage}\hfill
  % ======== 右側：選好2 ========
  \begin{minipage}[t]{0.49\textwidth}
    \caption{Preference Profile 2}
    \label{tab:preference2}
    \vspace{1.5mm}
    % 選好2の提案側
    \begin{minipage}[t]{0.49\linewidth}
      \centering
      \begin{tabular}{@{}llllll@{}} 
        \toprule
        & $\succsim_{s_1}$ & $\succsim_{s_2}$ & $\succsim_{s_3}$ & $\succsim_{s_4}$ & $\succsim_{s_5}$ \\ 
        \midrule
        1 & $C_5$ & $C_5$ & $C_4$ & $C_5$ & $C_3$ \\
        2 & $C_2$ & $C_3$ & $C_3$ & $C_1$ & $C_1$ \\
        3 & $C_4$ & $C_4$ & $C_5$ & $C_2$ & $C_4$ \\
        4 & $C_1$ & $C_2$ & $C_1$ & $C_4$ & $C_2$ \\
        5 & $C_3$ & $C_1$ & $C_2$ & $C_3$ & $C_5$ \\ 
        \bottomrule
      \end{tabular}
    \end{minipage}\hfill
    % 選好2の受入側
    \begin{minipage}[t]{0.49\linewidth}
      \centering
      \begin{tabular}{@{}llllll@{}} 
        \toprule
        & $\succsim_{c_1}$ & $\succsim_{c_2}$ & $\succsim_{c_3}$ & $\succsim_{c_4}$ & $\succsim_{c_5}$ \\ 
        \midrule
        1 & $S_2$ & $S_1$ & $S_2$ & $S_1$ & $S_5$ \\
        2 & $S_1$ & $S_2$ & $S_1$ & $S_3$ & $S_2$ \\
        3 & $S_4$ & $S_3$ & $S_4$ & $S_4$ & $S_1$ \\
        4 & $S_3$ & $S_5$ & $S_5$ & $S_5$ & $S_4$ \\
        5 & $S_5$ & $S_4$ & $S_3$ & $S_2$ & $S_3$ \\ 
        \bottomrule
      \end{tabular}
    \end{minipage}
  \end{minipage}
\end{table}

% ==========================================
% --- Preference Profiles 3 & 4 ---
% ==========================================
\begin{table}[htbp]
  \centering
  \scriptsize
  \setlength{\tabcolsep}{3pt}
  \renewcommand{\arraystretch}{1.15}

  % ======== 左側：選好3 ========
  \begin{minipage}[t]{0.49\textwidth}
    \caption{Preference Profile 3}
    \label{tab:preference3}
    \vspace{1.5mm}
    \begin{minipage}[t]{0.49\linewidth}
      \centering
      \begin{tabular}{@{}llllll@{}} 
        \toprule
        & $\succsim_{s_1}$ & $\succsim_{s_2}$ & $\succsim_{s_3}$ & $\succsim_{s_4}$ & $\succsim_{s_5}$ \\ 
        \midrule
        1 & $C_1$ & $C_2$ & $C_4$ & $C_3$ & $C_3$ \\
        2 & $C_3$ & $C_4$ & $C_1$ & $C_1$ & $C_2$ \\
        3 & $C_4$ & $C_1$ & $C_2$ & $C_2$ & $C_1$ \\
        4 & $C_2$ & $C_5$ & $C_3$ & $C_5$ & $C_4$ \\
        5 & $C_5$ & $C_3$ & $C_5$ & $C_4$ & $C_5$ \\ 
        \bottomrule
      \end{tabular}
    \end{minipage}\hfill
    \begin{minipage}[t]{0.49\linewidth}
      \centering
      \begin{tabular}{@{}llllll@{}} 
        \toprule
        & $\succsim_{c_1}$ & $\succsim_{c_2}$ & $\succsim_{c_3}$ & $\succsim_{c_4}$ & $\succsim_{c_5}$ \\ 
        \midrule
        1 & $S_2$ & $S_4$ & $S_3$ & $S_4$ & $S_1$ \\
        2 & $S_4$ & $S_1$ & $S_2$ & $S_5$ & $S_3$ \\
        3 & $S_1$ & $S_2$ & $S_4$ & $S_3$ & $S_2$ \\
        4 & $S_5$ & $S_3$ & $S_5$ & $S_2$ & $S_5$ \\
        5 & $S_3$ & $S_5$ & $S_1$ & $S_1$ & $S_4$ \\ 
        \bottomrule
      \end{tabular}
    \end{minipage}
  \end{minipage}\hfill
  % ======== 右側：選好4 ========
  \begin{minipage}[t]{0.49\textwidth}
    \caption{Preference Profile 4}
    \label{tab:preference4}
    \vspace{1.5mm}
    \begin{minipage}[t]{0.49\linewidth}
      \centering
      \begin{tabular}{@{}llllll@{}} 
        \toprule
        & $\succsim_{s_1}$ & $\succsim_{s_2}$ & $\succsim_{s_3}$ & $\succsim_{s_4}$ & $\succsim_{s_5}$ \\ 
        \midrule
        1 & $C_4$ & $C_5$ & $C_4$ & $C_4$ & $C_4$ \\
        2 & $C_1$ & $C_1$ & $C_3$ & $C_2$ & $C_5$ \\
        3 & $C_2$ & $C_2$ & $C_5$ & $C_3$ & $C_3$ \\
        4 & $C_5$ & $C_4$ & $C_1$ & $C_1$ & $C_2$ \\
        5 & $C_3$ & $C_3$ & $C_2$ & $C_5$ & $C_1$ \\ 
        \bottomrule
      \end{tabular}
    \end{minipage}\hfill
    \begin{minipage}[t]{0.49\linewidth}
      \centering
      \begin{tabular}{@{}llllll@{}} 
        \toprule
        & $\succsim_{c_1}$ & $\succsim_{c_2}$ & $\succsim_{c_3}$ & $\succsim_{c_4}$ & $\succsim_{c_5}$ \\ 
        \midrule
        1 & $S_5$ & $S_4$ & $S_4$ & $S_2$ & $S_1$ \\
        2 & $S_3$ & $S_1$ & $S_5$ & $S_4$ & $S_4$ \\
        3 & $S_4$ & $S_2$ & $S_1$ & $S_5$ & $S_5$ \\
        4 & $S_1$ & $S_3$ & $S_2$ & $S_3$ & $S_2$ \\
        5 & $S_2$ & $S_5$ & $S_3$ & $S_1$ & $S_3$ \\ 
        \bottomrule
      \end{tabular}
    \end{minipage}
  \end{minipage}
\end{table}

% ==========================================
% --- Preference Profile 5 ---
% ==========================================
\begin{table}[htbp]
  \centering
  \scriptsize
  \setlength{\tabcolsep}{3pt}
  \renewcommand{\arraystretch}{1.15}

  % 選好5は最後なので、他とサイズ感を合わせるために中央に1つだけ配置
  \begin{minipage}[t]{0.49\textwidth}
    \caption{Preference Profile 5}
    \label{tab:preference5}
    \vspace{1.5mm}
    \begin{minipage}[t]{0.49\linewidth}
      \centering
      \begin{tabular}{@{}llllll@{}} 
        \toprule
        & $\succsim_{s_1}$ & $\succsim_{s_2}$ & $\succsim_{s_3}$ & $\succsim_{s_4}$ & $\succsim_{s_5}$ \\ 
        \midrule
        1 & $C_2$ & $C_1$ & $C_2$ & $C_2$ & $C_3$ \\
        2 & $C_3$ & $C_2$ & $C_3$ & $C_3$ & $C_4$ \\
        3 & $C_1$ & $C_3$ & $C_4$ & $C_1$ & $C_1$ \\
        4 & $C_5$ & $C_4$ & $C_5$ & $C_5$ & $C_5$ \\
        5 & $C_4$ & $C_5$ & $C_1$ & $C_4$ & $C_2$ \\ 
        \bottomrule
      \end{tabular}
    \end{minipage}\hfill
    \begin{minipage}[t]{0.49\linewidth}
      \centering
      \begin{tabular}{@{}llllll@{}} 
        \toprule
        & $\succsim_{c_1}$ & $\succsim_{c_2}$ & $\succsim_{c_3}$ & $\succsim_{c_4}$ & $\succsim_{c_5}$ \\ 
        \midrule
        1 & $S_5$ & $S_5$ & $S_2$ & $S_4$ & $S_1$ \\
        2 & $S_4$ & $S_2$ & $S_3$ & $S_1$ & $S_3$ \\
        3 & $S_1$ & $S_3$ & $S_4$ & $S_3$ & $S_2$ \\
        4 & $S_2$ & $S_4$ & $S_5$ & $S_5$ & $S_5$ \\
        5 & $S_3$ & $S_1$ & $S_1$ & $S_2$ & $S_4$ \\ 
        \bottomrule
      \end{tabular}
    \end{minipage}
  \end{minipage}
\end{table}

\end{document}